\documentclass[a4paper,12pt]{article}
\usepackage{jheppub} % for details on the use of the package, please see the JINST-author-manual
\usepackage{lineno}
\usepackage{epsfig}
\usepackage{graphicx}
\usepackage{bm}
\usepackage{times}
\usepackage{braket}
\usepackage{color}
\usepackage{slashed}
\usepackage{hyperref}
\usepackage{ulem}
\usepackage{amsmath}
\allowdisplaybreaks[4] 
\title{Form factors of light pseudoscalar mesons from the perturbative QCD approach}

\author{Jian Chai$^{a,b}$}
\author[1]{and Shan Cheng$^{b,c,d}$ \note{Corresponding author: scheng@hnu.edu.cn}}
\affiliation{$^a$Institute of Theoretical Physics, School of Sciences,Henan University of Technology, Zhengzhou, Henan 450001, China}
\affiliation{$^b$School of Physics and Electronics, Hunan University, 410082 Changsha, China}
\affiliation{$^c$School for Theoretical Physics, Hunan University, 410082 Changsha, China}
\affiliation{$^d$Hunan Provincial Key Laboratory of High-Energy Scale Physics and Applications, 410082 Changsha, China}

\emailAdd{scheng@hnu.edu.cn}

\abstract{We study the electromagnetic and meson-photon transition form factors (TFF) of light pseudoscalar mesons from the perturbative QCD (pQCD) approach. 
To comprehensively account for both the longitudinal and transverse nonperturbative dynamics of hadronic constituents, we incorpoarate intrinsic transverse momentum distributions (iTMDs) alongside the conventional light-cone distribution amplitudes (LCDAs). 
The main motivations of this work are the disjointedness of electromagnetic form factors between the theoretical predictions and the experimental measurements, 
and the BaBar-Belle tension of pion-photon transition form factor in the large momentum transfers. 
Our calculation is carried out at the next-to-leading-order for the contributions from leading and subleading twist LCDAs, and leading order for the twist four contributions. Notably, this work presents the first systematic evaluation of higher-twist contributions to meson-photon TFFs.
The key findings are: (a) iTMDs play a crucial role in describing form factor data, 
particularly in the small-to-intermediate momentum transfer region where they induce significant modifications to pQCD predictions. 
(b) The extracted transverse size parameters for valence quark states are found to be 
$\beta_\pi^2 = 0.51 \pm 0.04$ GeV$^{-2}$ and $\beta_K^2 = 0.30 \pm 0.05$ GeV$^{-2}$, 
the chiral mass of pion meson $m_0^\pi$ at $1$ GeV is determined to be $1.84 \pm 0.07$ GeV. 
(c) The meson-photon TFFs are predominantly governed by leading-twist LCDAs. 
The iTMDs-enhanced pQCD results show better agreement with Belle's pion TFF data across intermediate and large momentum transfers and  
favor a small $\eta-\eta^\prime$ mixing angle. 
(d) Remarkably, the inclusion of iTMDs extends the applicability of pQCD calculations down to a few GeV$^2$ for all considered form factors, 
significantly improving the theory-data consistency. }

\begin{document}

\newcommand{\beq}{\begin{eqnarray}}
\newcommand{\eeq}{\end{eqnarray}}
\newcommand{\ben}{\begin{enumerate}}
\newcommand{\een}{\end{enumerate}}
\newcommand{\non}{\nonumber\\ }
\newcommand{\jpsi}{J/\Psi}
\newcommand{\ppa}{\phi_\pi^{\rm A}}
\newcommand{\ppp}{\phi_\pi^{\rm P}}
\newcommand{\ppt}{\phi_\pi^{\rm T}}
\newcommand{\ov}{ \overline }
\newcommand{\zerot}{ {\textbf 0_{\rm T}} }
\newcommand{\kt}{k_{\rm T} }
\newcommand{\fb}{f_{\rm B} }
\newcommand{\fk}{f_{\rm K} }
\newcommand{\rk}{r_{\rm K} }
\newcommand{\mb}{m_{\rm B} }
\newcommand{\mw}{m_{\rm W} }
\newcommand{\im}{{\rm Im} }
\newcommand{\kks}{K^{(*)}}
\newcommand{\acp}{{\cal A}_{\rm CP}}
\newcommand{\pb}{\phi_{\rm B}}
\newcommand{\xeba}{\bar{x}_2}
\newcommand{\xsba}{\bar{x}_3}
\newcommand{\peas}{\phi^A}
\newcommand{\Dsl}{ D \hspace{-2truemm}/ }
\newcommand{\pvsl}{ p \hspace{-2.0truemm}/_{K^*} }
\newcommand{\esl}{ \epsilon \hspace{-2.1truemm}/ }
\newcommand{\psl}{ p \hspace{-2truemm}/ }
\newcommand{\ksl}{ k \hspace{-2.2truemm}/ }
\newcommand{\lsl}{ l \hspace{-2.2truemm}/ }
\newcommand{\nsl}{ n \hspace{-2.2truemm}/ }
\newcommand{\vsl}{ v \hspace{-2.2truemm}/ }
\newcommand{\zsl}{ z \hspace{-2.2truemm}/ }
\newcommand{\epsl}{\epsilon \hspace{-1.8truemm}/\,  }
\newcommand{\bfkk}{{\bf k} }
\newcommand{\calm}{ {\cal M} }
\newcommand{\calh}{ {\cal H} }
\newcommand{\calo}{ {\cal O} }

%%%%========================================================
\definecolor{Red}{rgb}{1.,0.,0.}
\newcommand{\Red}[1]{{\color{nicered}{#1}}}
\definecolor{Blue}{rgb}{0.,0.,1.}
\newcommand{\Blue}[1]{{\color{Blue}{#1}}}
\definecolor{Green}{rgb}{0.,1.,0.}
\newcommand{\Green}[1]{{\color{Green}{#1}}}
\definecolor{Gray}{rgb}{0.5,0.5,0.5}
\newcommand{\Gray}[1]{{\color{Gray}{#1}}}

\definecolor{nicered}{rgb}{0.7,0.1,0.1}
\definecolor{nicegreen}{rgb}{0.1,0.5,0.1}
\bibliographystyle{apsrev}
\newcommand{\SC}[1]{{\color{magenta}{#1}}}

\maketitle 

\section{Introduction} \label{sec:introduction}

Emergence refers to the phenomenon wherein a system manifests properties or behaviors that are absent in its isolated components, 
emerging only through their interactions within an integrated whole. 
In quantum chromodynamics (QCD), confinement is an emergent phenomenon, 
where physical states manifest as color singlets despite the underlying degrees of freedoms are quark and gluon.
This is the foundation ideas of the parton picture, which describe hadron structures \cite{Mueller:1981sg,PQCD},  
and the factorization theorem, which separates the short- and long-distance contributions in hadron amplitudes \cite{Efremov:1979qk,Lepage:1980fj}.

To characterize parton distributions within hadrons, several complementary approaches have been developed. 
The most fundamental is the parton distribution function (PDF), 
a one-dimensional function describing the longitudinal momentum fraction $x_i$ distribution of partons. 
For a more complete picture that includes transverse momentum dependence, particularly in the small-$x$ regime, 
transverse-momentum-dependent distributions (TMDs) were introduced, 
along with their Fourier conjugates, says, the generalized parton distributions (GPDs) \cite{Diehl:2003ny,Belitsky:2005qn}. 
These different distribution functions are experimentally probed through distinct scattering processes. 
PDFs and TMDs are primarily investigated in inclusive scattering measurements, 
while GPDs are typically accessed through exclusive processes such as deeply virtual Compton scattering (DVCS) and deeply virtual meson production (DVMP). Such studies are predominantly conducted at fixed-target facilities like Jefferson Lab \cite{Dudek:2012vr} 
and will be a major focus of the future Electron-Ion Collider (EIC) \cite{AbdulKhalek:2022hcn}.
For high-energy exclusive processes involving large momentum transfers, 
the theoretical framework of light-cone distribution amplitudes (LCDAs) becomes particularly valuable. 
LCDAs provide a systematic twist expansion of parton distributions in the infinite-momentum frame, 
hence offer crucial insights into the dynamics of hard exclusive reactions. 

To our best knowledge, the first systematic analysis of high-twist hadron LCDAs was conducted for the pion 
in the context of investigating potential corrections to the pion form factor \cite{Braun:1989iv}. 
This foundational work was later extended to the kaon, incorporating the structure of $SU(3)$-breaking corrections \cite{Ball:2006wn}, 
and subsequently to vector mesons \cite{Ball:1998sk,Ball:1998ff,Ball:2007rt,Ball:2007zt} and nucleons \cite{Braun:2000kw}. 
Owing to its unique advantage of providing a rigorous power expansion, 
the LCDA framework has become a widely adopted tool for studying hard exclusive QCD processes.
Taking the pion as a prime example, the in-depth understanding of leading-twist LCDAs has enabled significant advancements in QCD calculations. 
Notably, the photon-pion transition form factor \cite{Zhang:2015mxa,Shen:2019vdc,Zhou:2023ivj} 
has been computed up to next-to-next-to-leading order (NNLO) \cite{Mikhailov:2021znq,Gao:2021iqq}. 
These refined calculations have substantially reduced hadronic uncertainties in interpretations of the $g$-2 anomaly \cite{Gerardin:2016cqj}. 
Furthermore, substantial progress has been made in calculating other key observables. 
The electromagnetic pion form factor has been investigated extensively, 
with next-to-leading-order (NLO) analyses incorporating subleading-twist contributions from pion LCDAs \cite{Braun:1999uj,Bijnens:2002mg,Cheng:2020vwr,Li:2010nn,Cheng:2014gba,Hu:2012cp,Cheng:2015qra,Cheng:2019ruz,Chen:2023byr}. 
Similarly, NLO corrections to subleading-twist effects have been included in calculations of $B \to \pi$ transition form factors \cite{Li:2012nk,Cheng:2014fwa,Ball:2004ye,Ball:2004rg,Hambrock:2015wka}.

While perturbative QCD calculations of hard amplitudes have achieved increasing precision, the nonperturbative LCDAs remain less understood. 
Modern QCD studies typically explore LCDAs shapes through three complementary avenues: 
(1) the nonperturbative QCD approache including the sum rules (QCDSRs) \cite{Colangelo:2000dp}, 
the QCD Dyson-Schwinger equation (DSE) \cite{Chang:2013pq}, and the instanton vacuum model \cite{Petrov:1998kg}; 
(2) the first-principles lattice QCD (LQCD) computations \cite{RQCD:2019osh} 
and (3) the phenomenological data-driven analyses of experimental results \cite{Cheng:2020vwr,Chai:2022srx}. 
QCDSRs, starting from appropriate correlation functions, can predict the lowest Gegenbauer moment of leading-twist LCDAs. 
However, their predictive power diminishes for higher-order coefficients due to the introduction of nonlocal vacuum condensates, 
which introduce significant model dependence. 
Complementary to this approach, the DSE framework has independently confirmed both the broadness (relative to the asymptotic profile) 
and the unimodal shape of the leading-twist pion LCDA. 
In contrast, LQCD struggles with higher Gegenbauer moments too because of the technical challenges associated with high-order derivatives, 
which result in poor signal-to-noise ratios. 
The data-driven approach, while highly effective, depend critically on two key prerequisites: 
(i) precise theoretical predictions and experimental measurements as inputs, 
and (ii) a truncation of the Gegenbauer expansion under the assumption of convergence. 
This method, though widely used, inherently inherits uncertainties from both the truncation ansatz and the quality of the input data.

In this paper, we investigate the light pseudoscalar mesons by addressing three key questions to elucidate their fine structures.
First, we examine the discrepancy between QCD-based predictions and direct measurements of electromagnetic form factors (EMFFs). 
Theoretical predictions are robust in the intermediate- to large-momentum Euclidean region ($q^2 <0$), 
while experimental data are reliable only in the small spacelike region and physical timelike regions. 
This significant mismatch severely limits the effectiveness of data-driven methods for extracting LCDAs.
Second, we address the controversy surrounding the pion-photon transition form factor (TFF) at large momentum transfers. 
The BaBar collaboration reported a rapid rise in the TFF, which was not corroborated by Belle. 
Aside from the BESIII measurement in the $\rho-\omega$ resonance region, no additional data have emerged in the past decade. 
Theoretically, this process is highly sensitive to the leading-twist pion LCDAs, making it a critical benchmark for understanding the Goldstone nature of the pion.
Lastly, we explore the differences in the $\eta$ and $\eta^\prime$ TFFs observed at low and intermediate momentum transfers. 
These measurements provide a unique opportunity to test the mixing scheme of the isoscalar components ($\eta_q,\eta_s$ and $\eta_c$). 

The paper is organized as follows. 
In Section \ref{sec:emff}, we introduce the dispersion relations within the modular representation 
to establish a model-independent connection between timelike and spacelike EMFFs. 
Subsequently, we present a feasible data-driven analysis for pion and kaon mesons.
Section \ref{sec:transition-ff} focuses on the photon-pion transition form factor (TFF), which we analyze using the perturbative QCD (pQCD) approach. 
Here, we incorporate LCDAs and the intrinsic transverse momentum distribution function (iTMDs) extracted in Section \ref{sec:emff}. 
We further extend our investigation to the TFFs of the isoscalar light pseudoscalar mesons $\eta, \eta^\prime$, 
aiming to interpret experimental measurements under different mixing schemes.
Finally, a summary of our findings is provided in Section \ref{sec:summary}.

\section{Electromagnetic form factors}\label{sec:emff}

Form factors, being the simplest hadronic matrix elements, play a crucial role in testing factorization theorems and perturbative QCD calculations, 
provided the hadron structure is well understood. 
Conversely, with sufficiently precise QCD calculations, they also shed light on the dynamical structure of hadrons.
In hard exclusive QCD processes, two distinct types of form factors emerge. 
For heavy-to-light transitions, the short-distance nature of the interaction in perturbative calculations is ensured by an intrinsic scale, 
namely, the large masses of the  $W/Z$ bosons and heavy quarks. 
In contrast, for electromagnetic and meson-photon transitions, 
the small interaction distance arises instead from an external large momentum transfers.
This work focuses on the latter case, aiming to probe the fine structure of light pseudoscalar mesons 
by combining state-of-the-art perturbative QCD calculations with high-precision experimental measurements.
In this paper, we focus on the latter ones, trying to understand the fine structures of light pseudoscalar mesons 
from the state-of-the-art perturbative calculations and the precise experiment measurements. 

%------------------------------------------------------------------------
\begin{figure}[t] \vspace{-4mm}
\begin{center}
\includegraphics[width=0.8\textwidth]{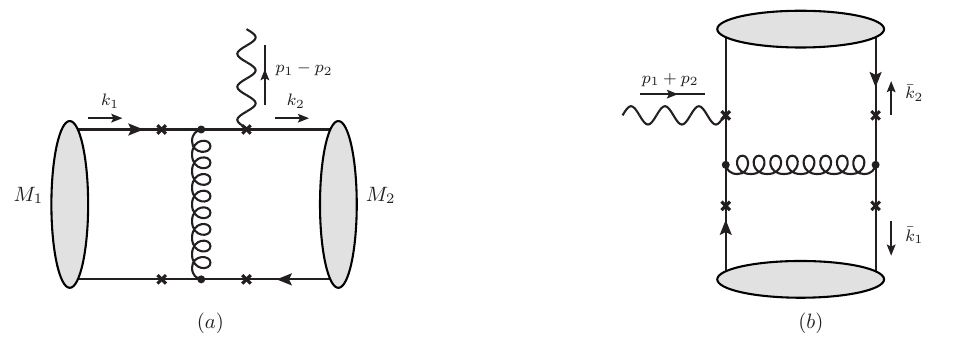}  
\end{center}
\vspace{-8mm}
\caption{Feynman diagrams of EMFFs at leading order, here $"\times"$ denotes the possible electromagnetic interaction vertexes.}
\label{fig:emff}
\end{figure} 
%------------------------------------------------------------------------

Here we discuss the EMFFs of pion and kaon mesons, 
which characterize the probability amplitude for a hadron to absorb a high-energy photon while remaining intact. 
From the perspective of perturbative QCD, when an energetic photon significantly alters the momentum of a struck parton, 
a hard gluon exchange becomes necessary to redistribute momentum between the active and spectator partons, 
thereby ensuring the formation of the final meson state.
Figure \ref{fig:emff} illustrates the leading-order Feynman diagrams for both spacelike (left) and timelike (right) form factors. 
These processes are mediated by the electromagnetic current 
$j^{\rm em}_{\mu} = e_u {\bar u} \gamma_\mu u + e_d {\bar d} \gamma_\mu d + e_s {\bar s} \gamma_\mu s$, 
where $e_q$ represents the fractional charge of the participating quark. 
For the changed pseudoscalar mesons $({\cal P}=\pi,K)$, the form factors are defined through the matrix elements 
\beq \langle {\cal P}^+(p_2) \vert J_{\mu}^{\rm em} \vert {\cal P}^+(p_1) \rangle = e_q (p_1+p_2)_\mu  {\cal F}_{\cal P}(Q^2), \non
\langle {\cal P}^+(p_2) {\cal P}^-(p_1) \vert J_{\mu}^{\rm em} \vert 0 \rangle = e_q (p_1-p_2)_\mu  {\cal G}_{\cal P}(Q^2). 
\label{eq:emff} \eeq

In QCD processes involving large momentum transfers, quark propagators with high virtuality can be effectively treated as free propagators. 
This treatment amounts to retaining only the large fourier components of the nonlocal quark operators 
while neglecting nonperturbative effects that appear as power corrections $\langle 0 \vert G_{\mu\nu}^2 \vert 0 \rangle / q^4$. 
Conversely, quark lines with low virtuality remain incorporated within the Heisenberg operator framework.
This approach constitutes a string operator representation of the factorization theorem, 
offering a covariant and systematic technique for disentangling short-distance (perturbative) 
and long-distance (nonperturbative) contributions in physical amplitudes.
The electromagnetic matrix element can be expressed through the following factorization formalism, 
\beq &~& \langle {\cal P}(p_2) \vert J_\mu^{\mathrm{em}} \vert {\cal P}(p_1) \rangle \non
&=& \oint dz_1 dz_2 \, \big\langle {\cal P}(p_2) \bigg\vert \left\{ \overline{q}_{1 \gamma}(0) \,
\mathrm{exp} \left( ig_s \int_{z_2}^0 d\sigma_{\nu^\prime} A^{\nu^\prime}(\sigma) \right) q_{2 \beta}(z_2) \right\}_{kj} \bigg\vert 0 \big\rangle_{\mu_t}  \non 
&~& \cdot H_{\gamma\beta\alpha\delta}^{ijkl}(z_1,0) \big\langle 0 \bigg\vert \left\{ \overline{q}_{2 \alpha}(z_2) \,
\mathrm{exp} \left(i g_s \int^{z_2}_{z_1} d\sigma_\nu A^\nu(\sigma) \right) q_{1 \delta}(z_1) \right\}_{il} \bigg\vert {\cal P}(p_1) \big\rangle_{\mu_t},  \quad 
\label{eq:ff-fact} \eeq
where $\gamma , \beta, \alpha, \delta$ are the spinor indices, and $i,j,k,l$ are the color indicators. 
The hard kernel associated with the lowest Fock state is
\beq H_{\gamma\beta\alpha\delta}^{ijkl}(z_1,z_2) = - i \left[ g_s \gamma_m \right]_{\alpha\beta} T^{ij}
\left[(e_q \gamma_\mu) S^{(0)}(z_2-z_1) (g_s \gamma_n) \right]_{\gamma\delta} T^{kl} D_{mn}^{(0)}(z_2-z_1).
\label{eq:hk-2p} \eeq
Here the quark and gluon free propagators read as 
\beq S^{(0)}(z) = \frac{1}{2\pi^2}\frac{\zsl}{z^4}, \qquad
D_{mn}^{(0)}(z) = \frac{1}{4\pi^2}\frac{g_{mn}}{z^2}. \label{eq:free-propagator} \eeq
The nonlocal matrix elements in the RHS of Eq. (\ref{eq:ff-fact}) imply the amplitudes of mesons breaking-up into a pair of soft quarks,
they receive contributions from different lorentz structures 
\beq 4 {\bar q}_{1 \alpha} q_{2 \delta} = \left\{ {\bar q}_1 q_2 + \gamma_5 \left( {\bar q}_{2} \gamma_5 q_1 \right) 
+ \gamma^\rho \left( \bar{q}_2 \gamma_\rho q_1 \right) + \gamma_5 \gamma^\rho \left( {\bar q}_1 \gamma_\rho \gamma_5 q_1 \right) 
+ \frac{\sigma^{\rho \tau} }{2} \left( {\bar q}_2 \sigma_{\rho \tau} q_1 \right) \right\}_{\delta \alpha},  \label{eq:fierz-indentity}\eeq
and define the light cone distribution amplitudes (LCDAs) of meson at different twists. 
The details of LCDAs definition are presented in appendix \ref{app:LCDAs-definition}. 

The state-of-the-art pQCD predictions of EMFFs are quoted \cite{Chai:2022srx,Cheng:2019ruz} as 
\beq
{\cal F}^{\rm em}_{\cal P}(Q^2) =&& {\cal F}^{2p}_{\cal P}(Q^2) + {\cal F}^{3p}_{\cal P}(Q^2) \label{eq:ff} \\
{\cal F}^{2p}_{\cal P}(Q^2) = 
&&\frac{8}{9} \pi \alpha_s f_{\cal P}^2 Q^2 \int_0^1 du_1 \int_0^1 du_2 \, \int_0^{1/\Lambda} \, b_1 db_1 b_2 db_2 \, e^{-S(x_i,y_i,b_1,b_2,\mu)} \non
&& \cdot S_t(u_1,t) \, S_t(u_2,t ) \, 
\Big\{ u_2 \varphi(u_1) \varphi(u_2) \, \left(1+ F_{t2}^{(1)} (u_1, u_2, t, Q^2) \right) \, \mathcal{H}\non
&& + \frac{2m_0^2}{Q^2} \Big[ \left( 1 - u_2 \right)  \varphi^p(u_1) \varphi^p(u_2) \left(1+F_{t3}^{(1)} (u_1, u_2, t, Q^2) \right)  \, \mathcal{H}_0 \non
&& + \frac{1}{3} \varphi^p(u_1) \varphi^\sigma(u_2) \left(1+\tilde{F}_{t3}^{(1)} (u_1, u_2, t, Q^2) \right) \non
&& \cdot \left( \mathcal{H}_0 - (1-u_2) Q^2 \mathcal{H}_1 - u_2(1+u_1) \mathcal{H}_2 \right) \Big] \non
&& - \Big[ 2 \left( u_2 \right)^2 \varphi(u_1) g_{2}(u_2) - 4 u_2 \varphi(u_1) \left( g_1(u_2) + \tilde{g}_2(u_2) \right) \Big] \, \mathcal{H}_1 \non
&& - \Big[ 2 u_1 u_2 \varphi(u_2) g_{2}(u_1) + 4 u_2 \varphi(u_1) \left( g_1(u_2) + \tilde{g}_2(u_2) \right) \Big] \, \mathcal{H}_2 \non
&& + \Big[ 4 \left( u_2 \right)^2 \left( 1 + u_1 \right) Q^2 \varphi(u_1) \left( g_1(u_2) + \tilde{g}_2(u_2) \right) \Big] \, \mathcal{H}_3 \Big\},
\label{eq:ff-2p} \\
{\cal F}^{3p}_{\cal P}(Q^2) = &&\frac{64}{9} \pi \alpha_s f_{\cal P}^2 Q^2 \int_0^1 \mathcal{D} u_{1i} \int_0^1 \mathcal{D} u_{2i} \,
\int_0^{1/\Lambda} \, b_1 db_1 b_2 db_2 \, e^{-S^3(u_{1i},u_{2i},b_1,b_2 \mu)}  \non
&& \, S_t(u_{1i}, Q) \, S_t(u_{2i}, Q) \, \tilde{\varphi}_\parallel(u_{2i}) \varphi_\parallel(u_{1i}) \, v \bar{v}  \non
&& \cdot \Big\{ \mathcal{H}^\prime_0 + u_{22} Q^2 \mathcal{H}^\prime_1 
- \left[ 3u_{11} u_{22} \left( u_{11} + u_{21} + u_{11} u_{22} \right) Q^4 \right] \mathcal{H}^\prime_2 \Big\}.
\label{eq:ff-3p} \eeq
The contribution from two-particle and three-particle LCDAs are written separately in equation (\ref{eq:ff-2p}) and \ref{eq:ff-3p}), respectively.  

The decay constant $f_{\cal P}$ is defined via 
\beq \langle 0 \vert {\bar q}_1 \gamma_\mu \gamma_5 q_2 \vert {\cal P}(p) = i f_{\cal P} p_\mu. \eeq
$\varphi$, $\varphi^{p,\sigma}$ and $g_{1,2}$ are the leading twist, twist three and twist four LCDAs 
associated to valence quark state (${\bar q}_1 q_2 $) of ${\cal P}$ meson, 
and $\varphi_{\parallel}$ is the twist four LCDAs associated to the three-particle state (${\bar q}_1 q_2 g$). 
See appendixes \ref{app:LCDAs-definition} and \ref{app:LCDAs-expression} for details. 
The auxiliary LCDAs  
\beq \tilde{g}_2(u_2) \equiv \int_0^{u_2} \, du'_2 \, g_2(u'_2),  \qquad
\tilde{\varphi}_\parallel(u_{2i}) \equiv \int_0^{u_{22}} \, du'_{22} \, \varphi_\parallel(u'_{2i}) \eeq
satisfy the bound conditions $\tilde{g}_2(u_2=0,1) = 0$ and $\tilde{\varphi}_\parallel(u_{22}=0,1)=0$. 
In the framework of LCDAs, we employ distinct momentum fraction parameterizations for different Fock state components:
(1) $u_1$ and $u_2$ are the longitudinal momentum fractions in the two-particle LCDAs 
carried by the anti-quark in the initial meson and the quark in the final meson, respectively, 
and (2) $u_{1i}$ and $u_{2i}$ are the momentum fractions in the three-particle LCDAs for initial-state and final-state partons, 
with index $i = 1, 2, 3$ corresponding to antiquarks, quarks and gluons, respectively. 
The gluon momentum fractions are effectively incorporated into the antiquark component through a weighted parameterization scheme, 
with the integration measure defined as
\beq v \equiv \frac{u_1 - u_{11}}{1- u _{11} - u_{12}}, \quad 
\int_0^1 {\cal D}u_{1i} \equiv \int_0^1 du_1 \int_0^{u_1} du_{11} \int_0^{1-u_1} \frac{d u_{12}}{1-u_{11} - u_{12}},
\label{eq:integral-degree} \eeq
where $v$ represents the effective momentum redistribution ratio and 
the measure ${\cal D}u_{1i}$ systematically accounts for all possible momentum configurations. 
$b_1$ and $b_2$ are the conjugated coordinates to the transversal momenta associated to the internal quark propagators. 
Here we show explicitly the result of spacelike form factor with $Q^2 = -q^2 = - (p_1+p_2)^2$,
The corresponding timelike form factors can be directly obtained through the analytic continuation $-Q^2 \to Q^2$. 

Within the framework of perturbative QCD, the transverse momentum $k_T$ (with its conjugate coordinate $b$ in transverse space) 
plays a vital role in regulating the end-point singularities that emerge in hard scattering kernels.
The characteristic scales of $k_T$ span three distinct regimes, 
saying the QCD scale $\lambda$, the hard-collinear scale $\sqrt{\Lambda Q}$ and the hard scale $Q$. 
The loop integration generates large logarithmic terms, particularly significant in the soft region ($k_T \sim \Lambda$). 
These logarithms are systematically resummed up to the well-known sudakov factors $S$ and $S^3$, 
which strongly suppress soft contributions, while enhance the dominance of hard scattering mechanisms, 
and hence provides crucial infrared protection. 
The functions $F^{(1)}_{t2}$ and $F^{(1)}_{t3}$ appearing in equations (\ref{eq:ff},\ref{eq:ff-2p},\ref{eq:ff-3p}) 
are the next-to-leading-order (NLO) QCD corrections to the hard scattering amplitudes 
contributed from the twist-two and twist-three LCDAs, respectively \cite{Li:2010nn,Cheng:2014gba,Hu:2012cp,Cheng:2015qra}. 
The hard functions ${\cal H}_{(i)}$ are written by means of the modified Bessel functions. 
\beq &&{\cal H}_0 = K_0(\beta b_1) \Big[ \theta(b_1 - b_2) I_0(\alpha b_2) K_0(\alpha b_1) - \theta(b_2 - b_1) I_0(\alpha b_1) K_0(\alpha b_2) \Big], \non
&&{\cal H}_1 = K_0(\beta b_1) 
\Big[ \theta(b_1 - b_2) \left( \frac{b_1}{2\alpha} I_0(\alpha b_2) K_1(\alpha b_1) - \frac{b_2}{2\alpha} I_1(\alpha b_2) K_0(\alpha b_1) \right) \non 
&& \hspace{1cm} - \{ b_1 \leftrightarrow b_2 \} \Big], \non
&&{\cal H}_2 = \frac{b_1 K_1(\beta b_1) }{2\beta} 
\Big[ \theta(b_1 - b_2) I_0(\alpha b_2) K_0(\alpha b_1) - \theta(b_2 - b_1) I_0(\alpha b_1) K_0(\alpha b_2) \Big], \non
&&{\cal H}_3 = \frac{b_1 K_1(\beta b_1) }{2\beta} 
\Big[ \theta(b_1 - b_2) \left( \frac{b_1}{2\alpha} I_0(\alpha b_2) K_1(\alpha b_1) - \frac{b_2}{2\alpha} I_1(\alpha b_2) K_0(\alpha b_1) \right) \non 
&& \hspace{1cm} - \{ b_1 \leftrightarrow b_2 \} \Big], \non
&&{\cal H}_0^\prime = \frac{\left(b_1 \right)^2 K_2(\beta^\prime b_1) }{8 \left( \beta^\prime \right)^2} 
\Big[ \theta(b_1 - b_2) I_0(\alpha^\prime b_2) K_0(\alpha^\prime b_1) - \theta(b_2 - b_1) I_0(\alpha^\prime b_1) K_0(\alpha^\prime b_2) \Big], \non
&&{\cal H}_1^\prime = \frac{\left(b_1 \right)^2 K_2(\beta^\prime b_1) }{8 \left( \beta^\prime \right)^2} 
\Big[ \theta(b_1 - b_2) \left( \frac{b_1}{2\alpha} I_0(\alpha^\prime b_2) K_1(\alpha^\prime b_1) - 
\frac{b_2}{2\alpha} I_1(\alpha^\prime b_2) K_0(\alpha^\prime b_1) \right) \non 
&& \hspace{1cm} - \{ b_1 \leftrightarrow b_2 \} \Big], \non
&&{\cal H}_2^\prime = \frac{1}{48} \left[ \frac{\left(b_1 \right)^3}{\left( \beta^\prime \right)^3} K_3(\beta^\prime b_1) 
+ \frac{b_1}{\left( \beta^\prime \right)^5} K_1(\beta^\prime b_1)  \right] \Big[ \theta(b_1 - b_2) \non
&& \hspace{1cm} \cdot \left( \frac{b_1}{2\alpha} I_0(\alpha^\prime b_2) K_1(\alpha^\prime b_1) - 
\frac{b_2}{2\alpha} I_1(\alpha^\prime b_2) K_0(\alpha^\prime b_1) \right) - \{ b_1 \leftrightarrow b_2 \} \Big].
\label{eq:hardfunctions} \eeq 
The characteristic hard scales emerging in the internal propagators are explicitly given by
\beq &&\alpha = \left[ u_2 Q^2 \right]^{1/2}, \quad \beta = \left[ u_1u_2 Q^2 \right]^{1/2}, \non
&&\alpha^\prime = \left[ \left(u_{21} + u_{23} \right) Q^2 \right]^{1/2}, \quad \beta^\prime = \left[ u_{11} u_{21} Q^2 \right]^{1/2}.
\label{eq:hardparameters}\eeq

We mark that the results in Eqs. (\ref{eq:ff-2p}, \ref{eq:ff-3p}) exhibit some differences compared to the previous pQCD calculation \cite{Cheng:2019ruz}. 
These discrepancies primarily manifest in the terms proportional to twist-four LCDAs and the contrbutions from three-particle Fock states, 
which are power suppressed by the transversal momentum ${\cal O}(k_T^2/Q^2)$ and quark mass. 
The origin of this discrepancy lies in the distinct treatments of hard-scattering amplitudes. 
Here we implement the factorization prescription by starting from the string operator representation in coordinate space, 
then systematically transforming to momentum space through light-cone coordinate integration. 
In contrast, the previous work \cite{Cheng:2019ruz} computed the hard amplitudes directly using Feynman rules in momentum space, 
which inadvertently neglected certain transverse momentum effects in the numerator algebra. 

\subsection{Intrinsic transversal momentum distribution functions}\label{subsec:iTMD}

The renormalization scale of nonperturbative LCDAs is conventionally chosen to match the factorization scale, 
typically set at the largest virtuality in the scattering process, says $\mu = {\rm max} \left( \sqrt{u_2} Q,1/b_1,1/b_2 \right)$. 
This choice implies that the LCDAs appearing in the pQCD calculations (Equations \ref{eq:ff-2p} and \ref{eq:ff-3p}) 
effectively represent the wave functions at zero transverse separation. 
Consequently, the soft transverse dynamics are not fully accounted for in the conventional LCDA framework, as discussed in \cite{Chai:2024tss,Li:2009pr}. 

%%%=========================================================
\begin{figure}[t] \begin{center} \vspace{-4mm}
\includegraphics[width=0.60\textwidth]{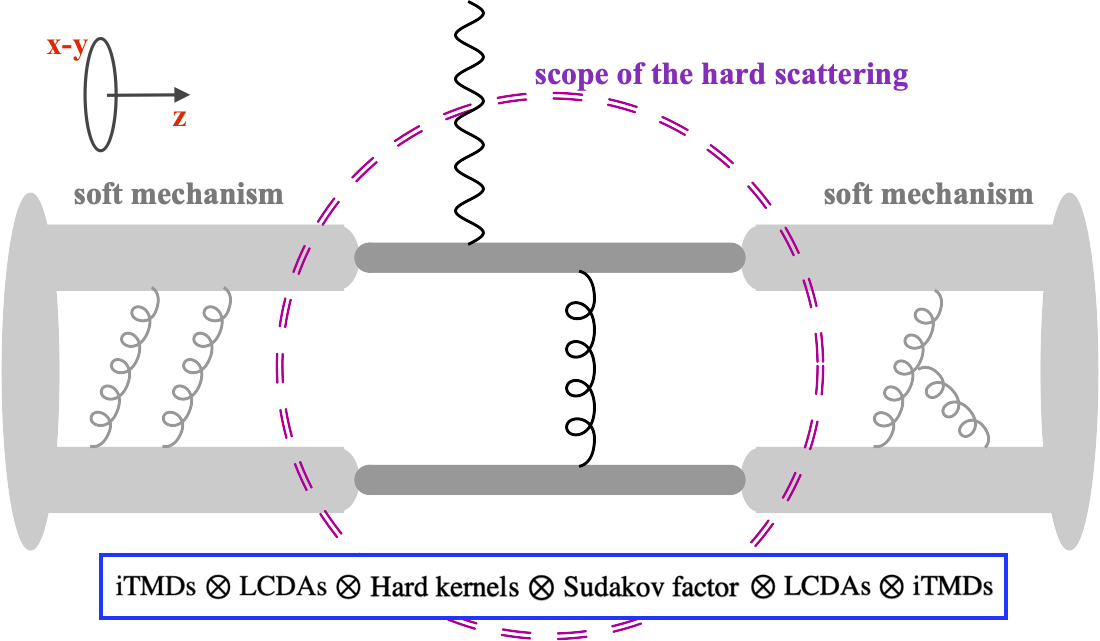} 
\end{center} \vspace{-6mm}
\caption{The sketch map of pion electromagnetic form factor \cite{Chai:2024tss}.}
\label{fig:kTfactorization}
\end{figure}
%%%=========================================================

For example, we show the sketch map of pion electromagnetic form factor in figure \ref{fig:kTfactorization}, 
where the electromagnetic interaction exhibites two distinct regimes separated by a double-dashed boundary.
The inner circle represents the hard scattering region (central electromagnetic potential field), 
described by a threshold-suppressed hard kernel $S_t H$ that captures the hard transverse plane parton radiation (thin gray stick). 
Externally, energetic pions propagate along the $z$-axis with soft bremsstrahlung radiation. 
Light-cone bremsstrahlung is incorporated into higher-twist LCDAs through a new coordinate $x^2$ 
that approximates but does not exactly coincide with the light-cone coordinate. 
In Eq. (\ref{eq:ff-fact}), the sudakov-multiplied LCDAs are defined at the hard transversal scale to match the potential field boundary. 
Hence, the transverse soft radiation (thick light-gray stick) has been frequently neglected in prior pQCD analyses. 

To systematically incorporate the soft transverse degrees of freedom in the parton distribution of hadrons, 
we introduce the iTMDs as complementary components to the standard LCDAs.  
From a physical perspective, the soft transverse oscillations within the pion, existing outside the electromagnetic potential field, 
become highly excited via electromagnetic interactions once the partons enter the domain of the potential field. 
This interaction energy is subsequently transferred to the spectator quark through hard gluon exchanges, 
thereby facilitating a transition of the Feynman diagram into a hard-scattering mechanism in a hard exclusive process.
 
The valence quark state wave function of the soft pion, denoted as $\psi(u, {\bf k}_T)$, satisfies the normalization condition 
\beq \int \frac{du d^2{\bf k}_T}{16\pi^3} \vert \psi(u, {\bf k}_T) \vert^2 = P_{q_1{\bar q}_2} \leqslant 1,  \label{eq:norm-valence-wf} \eeq 
where $P_{q_1{\bar q}_2}$ represents the probability of the valence quark-antiquark component.
Integrating over the transversal momenta, it deduces to the LCDAs 
\beq \frac{f_{\cal P}}{2\sqrt{6}} \varphi(u,\mu) \equiv \int \frac{d^2k_\perp}{16\pi^3} \psi(u,{\bf k}_T) \eeq
where $\varphi(u, \mu)$ satisfies the normalization condition $\int_0^1 du \varphi(u, \mu) = 1$. 
This definition ensures consistency with the $\pi \to \mu\nu$ decay amplitude \cite{LBHM}, yielding  
\beq \frac{f_\pi}{2\sqrt{6}} = \int \frac{du d^2k_\perp}{16\pi^3} \psi(u,{\bf k}_T), \quad
\psi(u,{\bf k}_T) = \frac{f_{\cal P}}{2\sqrt{6}} \varphi(u,\mu) \Sigma(u,{\bf k}_T).  \label{eq:pi-soft-DA} \eeq
Here, the transverse momentum profile function $\Sigma(u,{\bf k}_T)$ itself obeys the normalization 
\beq \int\frac{d^2k_\perp}{16\pi^3} \Sigma(u,{\bf k}_T) = 1. \label{eq:lt-twf-norm} \eeq
Taking into account the rotational symmetry along the $z$-axis, 
the iTMDs can be effectively modeled by a harmonic oscillator in the transversal plane with a transverse-size parameter $\beta^2$. 
Following \cite{Jakob:1993iw,tmd-LBHM}, we adopt a Gaussian ansatz for the transverse momentum dependence
\beq \Sigma(u,{\bf k}_T) = 16 \pi^2 \beta^2 g(u) \mathrm{Exp}\left[ - \beta^2 k_T^2g(u) \right], \label{eq:TMD-lt-kt} \eeq
where $g(u) = 1/(u {\bar u})$ preserves rotational invariance. 
%This form originates from a harmonic oscillator potential analysis in both the rest frame and the infinite momentum frame. 
The Fourier transform of the transverse wave function to impact parameter space yields 
\beq \hat{\Sigma}(u,{\bf b}_T) = 4 \pi \mathrm{Exp} \left[ - \frac{b_T^2 u(1-u)}{4\beta^2} \right], \label{eq:TMD-lt-bt} \eeq
leading to the modified soft wave function in coordinate space 
\beq \psi(u, {\bf b}_T) = \frac{f_\pi}{2\sqrt{6}} \varphi(u,\mu) \hat{\Sigma}(u,{\bf b}_T). \label{eq:pi-soft-DA-bt} \eeq

The transverse radius of the valence quark state must be smaller than the mean-square charge radius of the pion, 
$\langle r_\pi^2 \rangle = 0.45 \pm 0.01$ fm$^2$ \cite{PDG}. 
Translated into transverse momentum space, the condition imposes a lower bound on the mean-square transverse momentum 
\beq \langle {\bf k}_\pi^2 \rangle
= \frac{\int du d^2{\bf k}_T {\bf k}_T^2 \vert \psi(u, {\bf k}_T) \vert^2}{\int du d^2{\bf k}_T \vert \psi(u, {\bf k}_T) \vert^2} 
\geqslant 0.086 \, {\rm GeV}^2. \label{eq:iTMD-pi-average} \eeq
The double-photon transition \cite{LBHM, Li:2009pr} provides an additional constraint on the transverse-size parameter, 
requiring $\int du \psi(u, {\bf k}_T= {\bf 0}) =\sqrt{6}/f_\pi$, which can be expressed in terms of gegenbauer coefficients as 
\beq \beta^2_\pi = \frac{1}{8\pi^2f_\pi^2 \left(1 + a_2^\pi + a_4^\pi  + \cdots \right)}. \label{eq:TMD-t2-beta2} \eeq
With the coefficients $a_2^\pi(1 {\rm GeV}) = 0.28 \pm 0.05, a_4^\pi(1 {\rm GeV}) = 0.19 \pm 0.06$ 
extracted from the electromagnetic form factor \cite{Cheng:2020vwr}, along with the decay constant $f_\pi = 0.13$ GeV \cite{PDG}, 
we obtain $\beta^2_\pi = 0.51 \pm 0.04$ GeV$^{-2}$. 
This value corresponds to a mean-square transverse momentum $\langle {\bf k}_\pi^2 \rangle = 0.13 \pm 0.02$ GeV$^2$ 
and a mean-square transverse charge radius $\langle b_\pi^2 \rangle  = 0.30 \pm 0.03$ fm$^2$. 
The result satisfies Eq. (\ref{eq:iTMD-pi-average}) and exhibits excellent agreement with the puley dimensional relation 
$\langle b_\pi^2 \rangle = 2/3 \langle r_\pi^2 \rangle $ \cite{Kumano:2017lhr,Burkardt:2000za,Raya:2021zrz}. 

As shown in equations (\ref{eq:DA-t3-p}-\ref{eq:DA-t3-3p}), there are three sources for the high (power) twist LCDAs. 
They are the "bad" components with "wrong" spin projection in the wave functions, 
the transversal motion of valence quark Fock state in the leading twist components, 
and the higher Fock states with an additional gluon or quark-antiquark pair. 
The first two are defined to the genuine two-particle twist three LCDAs and their scale evolutions, 
the third one gives the quark mass correction terms in the two-particle twist three LCDAs. 
Note that the second source is partially related to the third one by the equation of motion. 
We also note that the transversal motion in the high twist LCDAs is definitely independent of the iTMDs introduced in this work, 
since they transfer different dynamics (hard/hard-collinear and soft) in the factorization formula. 

To clearly separate the contributions from the valence quark state and the three-particle state, 
we express the twist-three LCDAs in the following compact form
\beq \varphi^{p,\sigma}(u,\mu) = \varphi^{p,\sigma}_{2p}(u,\mu) + \varphi_{3p}^{p,\sigma}(u,\mu).  \label{eq:2pt3}\eeq
where $\varphi^{p,\sigma}_{2p}$ and $\varphi_{3p}^{p,\sigma}$ denote the two-particle (valence) and three-particle contributions, respectively. 
The three-particle term can be further decomposed as
\beq \varphi_{3p}^{p,\sigma}(u,\mu) = \rho_{\pm} \varphi_{\rm EOM}^{p,\sigma}(u,\mu) + \eta_3 \varphi_{{\bar q}qg}^{p,\sigma}(u,\mu), \eeq
with $\rho_\pm \equiv (m_{q_1} \pm m_{q_2})/m_0$ encoding quark mass effects and $\eta_3$ representing a three-particle coupling parameter. 

The soft transverse function associated with the valence quark state is given in Eqs. (\ref{eq:TMD-lt-kt}) and (\ref{eq:TMD-lt-bt}). 
For the three-particle state, we introduce a distinct iTMD function, denoted as $\Sigma^\prime$, to describe the corresponding soft transverse dynamics.
\beq &&\frac{f_\pi m_0^{\cal P}}{2\sqrt{6}} \varphi^{p,\sigma}(u, \mu) = \int \frac{d^2\vec{k}_T}{16\pi^3} \varphi_{2p}^{p,\sigma}(u, {\bf k}_T) 
+ \int \frac{d^2 {\bf k}_{1T}}{16\pi^3} \frac{d^2 {\bf k}_{2T}}{4\pi^2} \rho_{\pm} \varphi_{3p}^{p,\sigma}(u, {\bf k}_{1T}, {\bf k}_{2T}). \non
&&\psi_{2p}^{p,\sigma}(u, {\bf k}_T) = \frac{f_\pi m_0^{\cal P}}{2\sqrt{6}} \varphi_{2p}^{p,\sigma}(u, \mu) \Sigma(u, {\bf k}_T), \non
&&\psi_{3p}^{p,\sigma}(u, {\bf k}_{1T}, {\bf k}_{2T}) = \frac{f_\pi m_0^{\cal P}}{2\sqrt{6}}  \varphi_{3p}^{p,\sigma}(u, \mu) 
\int_0^u d \alpha_1 \int_0^{\bar u} d \alpha_2 \frac{\Sigma^\prime(\alpha_i, {\bf k}_{1T}, {\bf k}_{2T})}{1-\alpha_1-\alpha_2}.  \label{eq:iTMD} \eeq
The normalization $\int_0^1 du \varphi^p(u) = 1$ is automatically satisfied through the normalization constraints imposed on the iTMDs 
\beq &&\int\frac{d^2k_\perp}{16\pi^3} \Sigma(u,{\bf k}_T) = 1, \quad 
\int_0^u d \alpha_1 \int_0^{\bar u} d \alpha_2  \int \frac{d^2 {\bf k}_{1T}}{16\pi^3} \frac{d^2 {\bf k}_{2T}}{4\pi^2} 
\frac{\Sigma^\prime(\alpha_i, {\bf k}_{1T}, {\bf k}_{2T})}{1-\alpha_1-\alpha_2} = 1, \non
&&\int_0^1 du \varphi_{2p}^{p,\sigma}(u, \mu) = 1, \quad 
\int_0^1 du \varphi_{3p}^{p,\sigma}(u, \mu) = 0.  \label{eq:norm-Sigma-prime}\eeq
The newly introduced iTMD function is similarly expressed in a harmonic oscillator form 
\beq \Sigma^\prime(\alpha_i, {\bf k}_{1T}, {\bf k}_{2T}) = \frac{64 \pi^3 \beta'^4 }{\alpha_1 \alpha_2 (1- \alpha_1 -\alpha_2)}
\mathrm{Exp} \left[-\beta'^2 \left( \frac{k_{1T}^2}{\alpha_1} + \frac{k_{2T}^2}{\alpha_2} 
+ \frac{\left(k_{1T} + k_{2T} \right)^2}{1- \alpha_1 -\alpha_2} \right) \right]. \label{eq:TMD-t3-kt-1}\eeq
The Fourier transformation results in a Gaussian distribution as a function of the transverse distance. 
\beq \hat{\Sigma}^\prime(\alpha_i, {\bf b}_{1}, {\bf b}_{2}) = 4 \pi 
\mathrm{Exp} \left[- \frac{2 \alpha_3 (b_{1}^2 + b_{2}^2 ) + (\alpha_1 + \alpha_2) (b_{1} - b_{2})^2 }{16\beta'^2} \right]. \label{eq:TMD-t3-bt} \eeq
We notice that the crossing symmetry between $\alpha_1$ and $\alpha_2$ is properly accounted for in the Fourier transformation.
There are two iTMDs associated with twist-three LCDAs, corresponding to the two-particle and three-particle contributions, respectively.
Consequently, two transverse-size parameters are introduced, where $\beta'^2$ is expected to be smaller than $\beta^2$ 
due to the color-charged soft gluon reducing the transverse extent of the hadron. 

\subsection{Electromagnetic form factor of pion}\label{subsec:pi-em-ff}

The EMFF of the pion has been extensively investigated within QCD and the factorization theorem framework over several decades. 
Theoretically, most QCD-based calculations primarily concentrate on spacelike form factors, 
with distinct methodologies being applicable to specific momentum transfer regimes.
For instance, the Dyson-Schwinger equation (DSE) approach \cite{Roberts:1994dr,Roberts:2021nhw}, 
which unifies the parton distribution amplitude (PDA) at the hadronic scale with perturbative QCD (pQCD) predictions in the hard photon limit, 
provides a self-consistent description across the entire spacelike momentum domain \cite{Chang:2013nia,Raya:2015gva}. 
Alternatively, light-cone sum rules (LCSRs) yield reliable predictions for low-to-intermediate momentum transfers 
by systematically incorporating contributions from various twist terms \cite{Braun:1999uj,Cheng:2020vwr}. 
In contrast, the pQCD becomes particularly effective for processes involving large momentum transfers \cite{Jain:1999xc,Chai:2023htt}.
Notably, recent advancements in lattice QCD (LQCD) have significantly expanded the calculable momentum range, 
progressing from earlier studies covering $- 1 \, {\rm GeV}^2 \leq q^2 \leq 0$ \cite{Wang:2020nbf} 
to more recent computations extending to $- 10 \, {\rm GeV}^2 \leq q^2 \leq 0$ \cite{Ding:2024lfj}. 

From the experimental side, precise measurements of the pion EMFF have thus far been limited to the timelike region. 
The BaBar collaboration reported results for the invariant mass range $4 m_\pi^2 \leq s \lesssim 8.7 \, {\rm GeV}^2$ \cite{BaBar:2012bdw}, 
while the Belle collaboration measured the isospin-vector form factor in $4m_\pi^2 \leq s \leq 3.125 \,{\rm GeV}^2$ vai $\tau$ decays \cite{Belle:2008xpe}. 
Additionally, the BESIII collaboration provided data in the $\rho-\omega$ resonance region using the initial-state radiation (ISR) technique \cite{BESIII:2015equ}.
For spacelike momentum transfers, existing measurements are restricted to relatively low momentum transfers. 
Early electron-nucleon elastic scattering experiments \cite{NA7:1986vav} and studies of the $^1H(e,e'\pi^+)n$ process \cite{JeffersonLabFpi-2:2006ysh,JeffersonLab:2008jve} have yielded data in the range $-2.50 \, {\rm GeV}^2 \leq q^2 \leq -0.25 \,{\rm GeV}^2$. 
However, with the completion of Jefferson Lab's $12$ GeV Upgrade, which enables photon energies up to $9$ GeV \cite{Dudek:2012vr}, 
new high-precision measurements in previously inaccessible spacelike regions are anticipated. 

A common approach involves utilizing dispersion relations to bridge QCD predictions of spacelike form factors 
with experimental measurements in the timelike region. 
\beq {\cal F}^{\rm pQCD}_{\pi}(q^2) = \frac{1}{\pi} \int_{s_0}^\infty ds \frac{{\rm Im}{\cal F}_\pi(s)}{s - q^2 - i \epsilon}, \,\quad q^2 < s_0. 
\label{eq:Fpi_DR1} \eeq
The conventional dispersion relation requires the imaginary part of the timelike form factor as its integrand, 
necessitating prior parameterization of experimental data using resonance models, hence inevitably introduces additional theoretical uncertainties.
To circumvent this limitation, we propose employing a modular representation of the dispersion relation 
where the imaginary part is replaced by the absolute value of the form factor \cite{Cheng:2020vwr,Chai:2023htt}. 
This approach enables direct incorporation of experimental measurements, 
thereby eliminating model-dependent assumptions and their associated uncertainties.
\beq \mathcal{F}_\pi^{\rm pQCD} (q^2) = \exp \left[ \frac{q^2 \sqrt{s_0 - q^2}}{2 \pi} \int\limits_{s_0}^\infty d s 
\frac{ \ln |\mathcal{F}_\pi (s)|^2}{s\,\sqrt{s - s_0}  \, (s -q^2)} \right], \quad q^2 < s_0. \label{eq:Fpi_DR2} \eeq
The derivation of the modular dispersion relation can be found in appendix \ref{app:dr}. 
The modulus square in the integrand is written in terms of heavy theta functions 
\beq |\mathcal{F}_\pi(s)|^2 = \Theta(s_{\rm max} - s) \, \vert \mathcal{F}^{\rm data}_{\pi, {\rm Inter.}}(s) \vert^2
+ \Theta(s - s_{\rm max}) \,  \vert \mathcal{F}_\pi^{\rm pQCD}(s) \vert^2, \label{eq:Fpi_DR2_modular} \eeq 
in which the available data ($4m_\pi^2 \leqslant s \leqslant s_{\rm max} \simeq 8.7 $ GeV$^2$) are interpolated 
with evenly distribution with the interval $0.01 \, {\rm GeV}$, 
and the high energy tail is well calculated by the pQCD approach\footnote{Timelike form factor 
described by the resonant models \cite{BaBar:2012bdw} deviates from the asymptotic 
behavior ${\cal F}_\pi(q^2) \sim 1/q^2$ at large momentum transfers, 
this is the other reason to take pQCD calculation to describe the high energy tails.}. 
We mark that the modular dispersion relation shown in Eq. (\ref{eq:Fpi_DR2}) 
strengthens the role of high energy tail in the dispersion relation with the logarithm of the timelike form factor, 
we will discuss again this in the numerics.  

In the pQCD calculations of pion and kaon EMFFs, the dominant contributions originate from twist-three LCDAs rather than leading-twist one. 
This contrasts sharply with the LCSRs predictions, where only even-twist LCDAs contribute, 
making the EMFFs predominantly sensitive to leading-twist effects. 
The enhanced role of twist-three contributions in pQCD does not imply a breakdown of the power expansion, 
but rather reflects the chiral enhancement mechanism characteristic of processes involving pseudoscalar mesons. 
As evident from Eqs. (\ref{eq:ff-2p},\ref{eq:ff-3p}), the various LCDA contributions, 
leading twist, two-parton twist-3, twist-2 $\otimes$ twist-4, three-parton twist-3 and twist-4 LCDAs, 
all maintain consistent power-counting behavior: 
\beq &~&{\cal F}_{\cal P}^{t2}(Q^2) : {\cal F}_{\cal P}^{2p,t3}(Q^2) :{\cal F}_{\cal P}^{2p, t2 \otimes t4}(Q^2) : {\cal F}_{\cal P}^{3p, t3}(Q^2) : 
{\cal F}_{\cal P}^{3p, t4}(Q^2) \non
&=& {\cal O}(1) : {\cal O}(\frac{m_0^2}{\tilde{ Q^2 }}) : {\cal O}(\frac{\delta_P^2}{\tilde{Q^2}}) :
{\cal O}(\frac{f_{3 {\cal P }}^2}{f_{\cal P}^2 \tilde{Q^2}}) : {\cal O}(\frac{\delta_{\cal P}^4}{\tilde{Q^4}}).
\label{eq:pQCD-power} \eeq  

Here $f_{\cal P}, f_{3 {\cal P}}$ represent decay constants, while $\delta_{\cal P}$ is a parameter in the two-particle twist four LCDAs. 
The chiral mass is defined through the relation $m_0^{\cal P} \equiv m_{\cal P}^2/(m_{q_1} + m_{q_2})$.
Our analysis demonstrates that $k_T$ and threshold resummations generate Sudakov suppression factors 
that effectively eliminate contributions from both the small-$k_T$ region and endpoint configurations. 
These resummation effects preferentially select configurations where the momentum fraction $x \sim {\cal O}(0.1)$ 
and transverse momentum $k_T \sim {\cal O}(0.1 Q^2)$ in the multidimensional integration. 
The effective longitudinal virtuality $\tilde{Q^2}$ in the hard scattering process grows significantly more slowly than the momentum transfer $Q^2$, 
This behavior gives rise to chiral enhancement effects that become particularly important in the intermediate to large momentum transfers\footnote{The leading twist contribution begins to be dominate at an ultraviolet scale like $m_Z^2$ \cite{Cheng:2018khi}.}. 
While the two-particle twist-3 LCDAs benefit from this chiral enhancement, 
we have verified that all other LCDA contributions maintain the expected hierarchy dictated by the power counting rules \cite{Cheng:2020vwr}.

We immediately conclude that the chiral mass $m_0^\pi$ and the lowest oder gegenbauer coefficient $a_2^\pi$ 
constitute the two most crucial parameters in the pQCD calculation of the pion EMFF. 
Motivated by this observation, we rewrite the pQCD expressions in Eqs. (\ref{eq:ff-2p}, (\ref{eq:ff-3p}) explicitly in terms of these key parameters. 
\beq {\cal F}^{\rm em}_{\pi}(Q^2) 
&~& =(m_0^\pi)^2 {\cal F}^{\rm em}_{\pi, 1}(Q^2) + m_0^\pi m_\pi {\cal F}^{\rm em}_{\pi, 2}(Q^2) + m_0^\pi m_\pi a_2^\pi  {\cal F}^{\rm em}_{\pi, 3}(Q^2)  \non
&~& + a_2^\pi {\cal F}^{\rm pQCD}_{\pi, 4}(Q^2) + \left( a_2^\pi \right)^2 {\cal F}^{\rm em}_{\pi, 5}(Q^2) + {\cal F}^{\rm em}_{\pi, 6}(Q^2).
\label{eq:pi-ff-a2m0} \eeq
The first row shows the contributions with twist three LCDAs $\phi^{p,\sigma}$ both in the initial and final states, 
where the first term gives the purely asymptotic terms, 
the second and third terms indicate the convolution of asymptotic term and quark mass suppressed term. 
In the second line, the first two terms show the contributions from the lowest gegenbeur polynomials in the leading twist LCDAs, 
and the last term ${\cal F}^{\rm em}_{\pi, 6}$ is the summation of asymptotic contribution from leading twist LCDAs 
and the retaining higher twist LCDAs, like two-particle twist four and three-particle LCDAs. 
Additionly, ${\cal F}^{\rm em}_{\pi, i}$ with $i=1, \cdots, 6$ are invariant functions which are not dependent on $m_0$ and $a_2^\pi$. 

%%-----------------------------------------------------------------------
\begin{table}[t]\begin{center}
\caption{Input parameters of pion meson in the pQCD calculations.} 
\label{tab:parameter_pi}
\begin{tabular}{c c c c c c} \hline
$f_{3\pi} (10^{-2})$ \cite{Ball:2006wn} & $a_2^\pi$ \cite{Cheng:2020vwr} & $a_4^\pi$ \cite{Cheng:2020vwr} & $\omega_{3\pi}$ \cite{Ball:2006wn} & 
$\delta_{\pi}^2$ \cite{Ball:2006wn} & $\omega_{4\pi}$ \cite{Ball:2006wn} \\ \hline
$0.45 \pm 0.15$ & $0.28\pm 0.05$ & $0.19 \pm 0.06$ & $-1.50 \pm 0.70$ & $0.18 \pm 0.06$ & $0.20 \pm 0.10$ \\ 
\hline \end{tabular} \vspace{-4mm}
\end{center}\end{table}
%%-----------------------------------------------------------------------
%-----------------------------------------------------------------------
\begin{table}[t]\begin{center}
\caption{$m_0^\pi$ obtained by matching the pQCD calculations to the modular dispersion relation.}
\label{tab:m0pi}\vspace{2mm}
\begin{tabular}{c | c  c | c c} \hline
{\rm Set} \quad & \quad ${\rm IA } $ \quad & \quad ${\rm IB } $ \quad & \quad ${\rm IIA}$ \quad & \quad ${\rm IIB}$ \quad  \\
\hline 
$m_0^{\pi}$(GeV) \quad  & \quad $1.37 \pm 0.10$ \quad  &\quad $1.30 \pm 0.10$ \quad  
& \quad $1.54 \pm 0.06$ \quad & \quad $1.84 \pm 0.07$ \quad  \\ 
\hline \end{tabular}\end{center} \vspace{-6mm} \end{table}
%-----------------------------------------------------------------------

Substituting the pQCD expression in equation (\ref{eq:pi-ff-a2m0}) into the dispersion relation in equation (\ref{eq:Fpi_DR2_modular}), 
we can extract out the $m_0^\pi$ and $a_2^\pi$ with the input data in the physical regions. 
While, in practice, we are only able to obtain a reliable result for $m_0^\pi$, 
the fit result of $a_2^\pi$ is out of control with a large uncertainty which is traced to the non-leading role of leading twist contribution. 
Since the LQCD evaluations only hold well for the lowest gegenbauer coefficient $a_2$ so far, 
we take the result obtained from the joint analysis of electromagnetic form factor 
with modular dispersion relation and precise LCSRs calculation \cite{Cheng:2020vwr}, 
they are shown in table \ref{tab:parameter_pi} as well as other parameters of pion LCDAs. 
We take $m_\pi = 0.14$ GeV and $f_\pi = 0.13$ GeV, $f_{3\pi}$ and $\delta_{\pi}^2$ are in units of GeV$^2$. 

We consider two versions of pQCD calculations in the form factor fitting. 
The conventional framework employs a Sudakov-dominated formalism \cite{Li:1992nu} 
where only hard transverse degrees of freedom are considered through the Sudakov exponential suppression. 
Alternatively, an improved Sudakov-plus-Gaussian formulation that additionally incorporates soft transverse momentum dynamics via iTMDs \cite{Jakob:1993iw}. For the combined approach, we determine the transverse size parameter as $\beta^2_\pi = 0.511^{+0.040}_{-0.035}$. 
The three-particle configurations contribute to two-particle twist-three LCDAs through equations of motion, 
as manifested in the $\rho_{\pm}^{\cal P}$ terms of Eqs. (\ref{eq:DA-t3-p},\ref{eq:DA-t3-t}). 
Since these contributions scale with quark masses, they become negligible for the pion case. 
Consequently, we do not consider the iTMD function $\Sigma^\prime$ in Eq. (\ref{eq:TMD-t3-bt}) in our pQCD calculation of the pion form factor.

Table \ref{tab:m0pi} presents our determinations of the chiral mass parameter. 
Scenario A incorporates scale evolution effects for all nonperturbative parameters in the LCDAs, 
while Scenario B maintains the parameters at the default scale. 
The Set-I results are obtained within the conventional Sudakov-dominated framework, 
fitted to deep spacelike momenta in the range $-30 \leqslant q^2 \leqslant -10$ GeV$^2$. 
In contrast, Set-II employs our improved Sudakov-plus-Gaussian formulation and extends the fitting range to $-30 \leqslant q^2 \leqslant -5$ GeV$^2$.  
The pQCD predictions for the timelike form factor enter our analysis through their role 
as the high-energy tail contribution in the dispersion integral of Eq. (\ref{eq:Fpi_DR2}). 
The chiral mass terms appear in a non-trivial logarithmic form that cannot be simply isolated. 
To handle this complexity, we firstly take an initial value of chiral mass ($1.6 \pm 0.4$ GeV) for the high energy tail contribution in the integrand, 
and do the numerical iteration to find the optional value of $m_0^\pi$ with the modular dispersion relation. 

%-----------------------------------------------------------------------
\begin{figure*}[t]\begin{center}
\vspace{-4mm}
\includegraphics[width=0.45\textwidth]{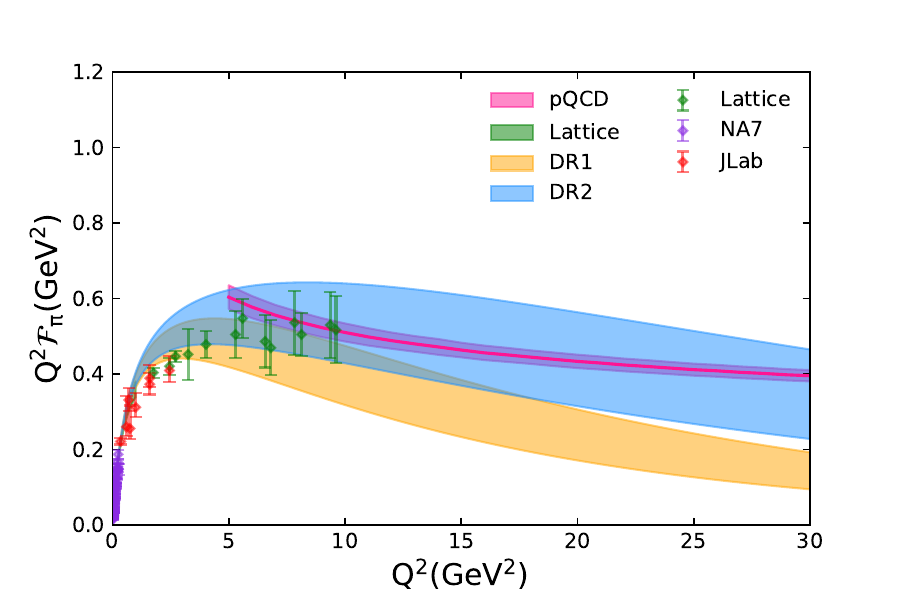}  
\hspace{1cm}
\includegraphics[width=0.45\textwidth]{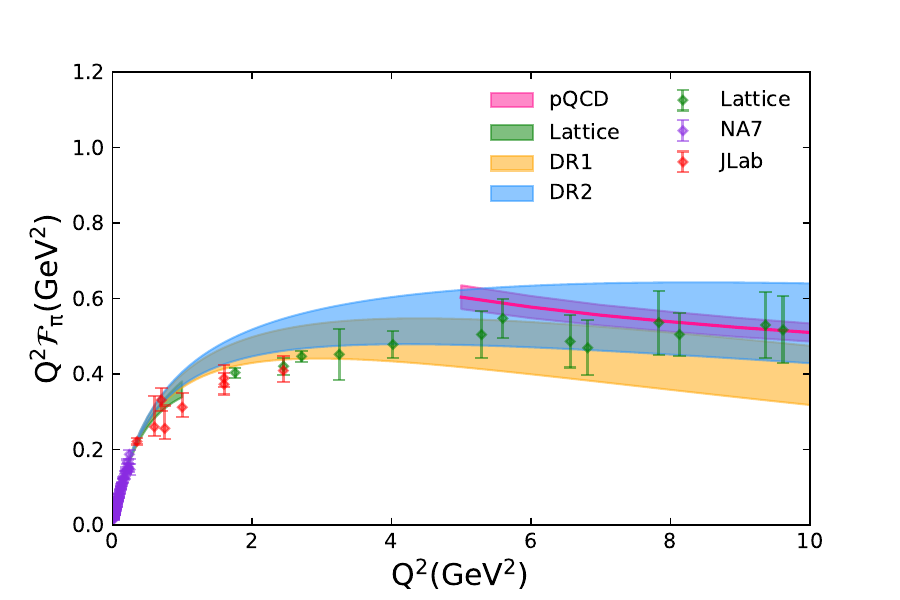}  \non
\vspace{8mm}
\includegraphics[width=0.45\textwidth]{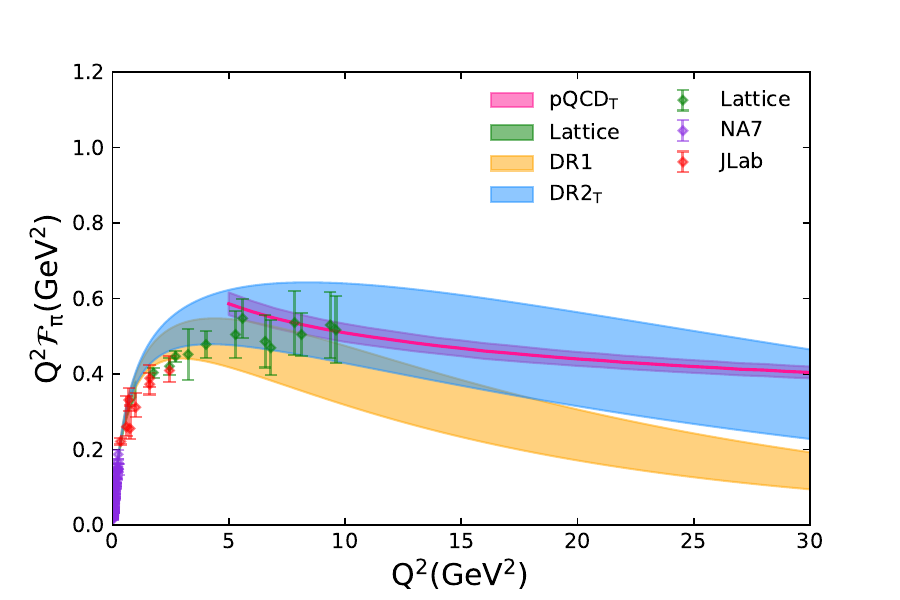}  
\hspace{1cm}
\includegraphics[width=0.45\textwidth]{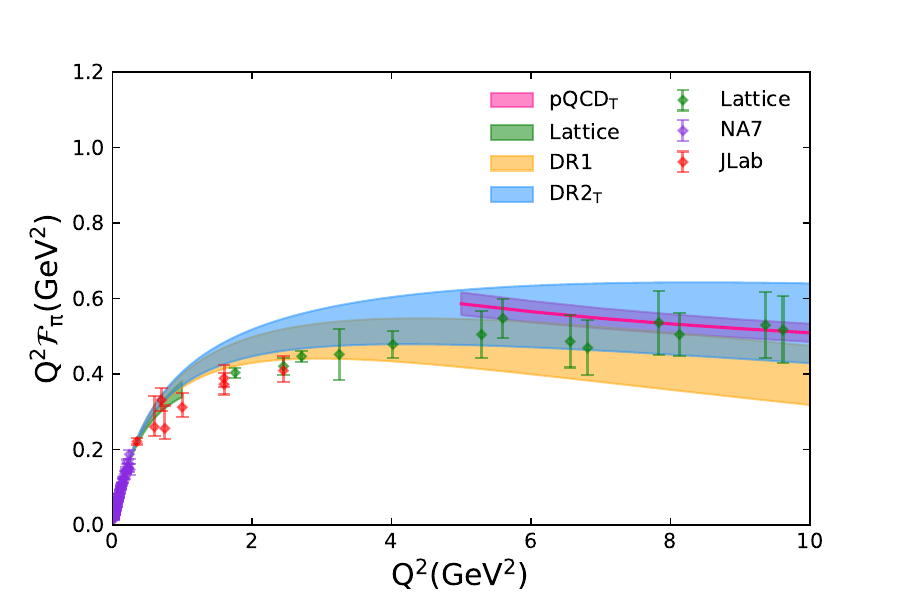}  
\end{center}\vspace{-6mm}
\caption{Up: $Q^2{\cal F}_\pi(Q^2)$ obtained under the sudokov factor-predominated picture in $5 \leqslant Q^2 \leqslant 30$ GeV$^2$.  
Low: Result obtained under the sudakov plus gaussian picture in $5 \leqslant Q^2 \leqslant 30$ GeV$^2$.  }
\label{fig:Fpi-fit-sl}
\end{figure*} 
%-----------------------------------------------------------------------

\begin{figure}[b]\begin{center} \vspace{-2mm}
\includegraphics[width=0.75\textwidth]{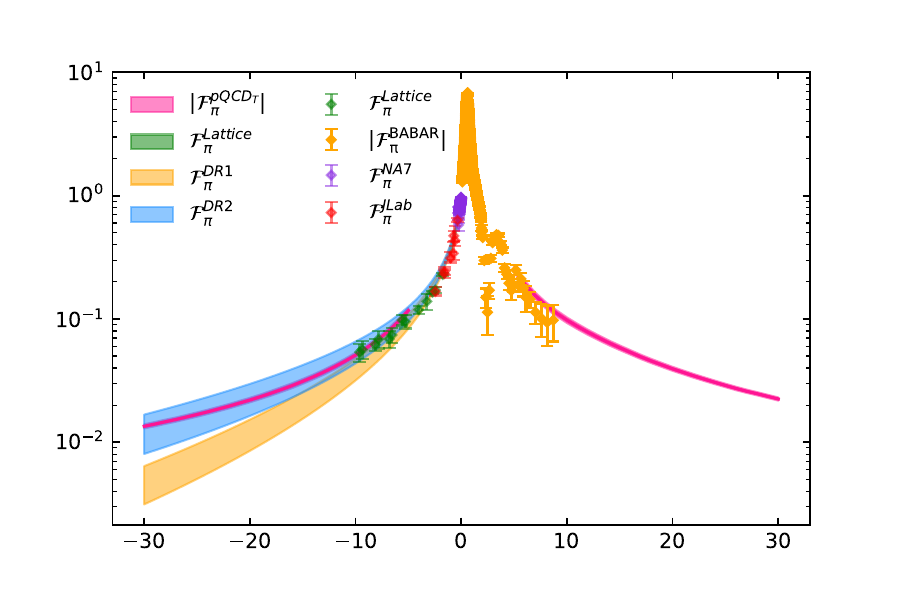}  
\end{center}\vspace{-10mm}
\caption{Pion EMFFs in the whole kinematical region.} 
\label{fig:ff-pion}
\end{figure}

Using the chiral mass obtained in Table \ref{tab:m0pi}, we plot the spacelike form factor $Q^2{\cal F}_\pi(Q^2)$ in figure \ref{fig:Fpi-fit-sl}. 
The upper (lower) row displays the results for scenario IB (IIB), comparing the predictions from the modular dispersion relation. 
The blue (DR2) and yellow (DR1) bands represent the dispersion relation results with and without the high-energy tail included in the integrand, respectively. 
The right panel provides an enlarged view for the momentum transfer range $0 \leqslant Q^2 \leqslant 10$ GeV$^2$. 
We also include available experimental data from NA7 \cite{NA7:1986vav} and Jefferson Lab \cite{JeffersonLabFpi-2:2006ysh,JeffersonLab:2008jve}, 
along with recent lattice QCD results \cite{Wang:2020nbf,Ding:2024lfj}. 
Additionally, Figure \ref{fig:ff-pion} shows the pion EMFF across the full kinematic range, 
where the iTMDs-improved pQCD predictions are indicated by magenta bands.

%-----------------------------------------------------------------------
%-----------------------------------------------------------------------

We give some intermediate conclusions. 
\begin{itemize}
\item[(a)] In the modular dispersion relation (\ref{eq:Fpi_DR2}), 
the contribution from the high-energy tail is not as strongly suppressed as in the standard dispersion relation (\ref{eq:Fpi_DR1}). 
This is evident from the gap between the blue (DR2) and yellow (DR1) bands, 
which arises from the logarithmic expression and the subtraction factor $q^2 \sqrt{s_0 - q^2}$ in the numerator. 
\item[(b)] The spacelike form factor obtained from the modular dispersion relation is insensitive to the iTMD function. 
his is expected because the iTMD function primarily modifies the form factor at small momentum transfers, 
whereas the modular dispersion relation relies on pQCD calculations at large momentum transfers.
\item[(c)] For the direct pQCD calculation of the spacelike form factor (left-hand side of Eq. (\ref{eq:Fpi_DR2})), 
the iTMD functions suppress the result in the small and intermediate momentum transfer regions, improving predictive accuracy up to a few GeV$^2$. 
Their effect becomes negligible at large momentum transfers. 
Conversely, as noted in (b), the dispersion relation result remains insensitive to the iTMD function. 
This explains why the chiral mass derived from the Sudakov-plus-Gaussian picture (Set-II) 
is significantly larger than that from the Sudakov-dominated picture (Set-I).
\item[(d)] The chiral mass obtained by fitting the iTMDs-improved pQCD prediction to the dispersion relation is 
$m_0^\pi(1 {\rm GeV}) = 1.84 \pm 0.07$ GeV. 
This value is thirty percents larger than the previous pQCD result $1.30 \pm 0.10$ GeV, 
but consistent with ChPT \cite{Leutwyler:1996qg}, indicating a significant suppression of the form factor due to soft transverse dynamics, 
particularly at small and intermediate momentum transfers. 
\item[(e)] We also perform a fit using the first-principles result for the Gegenbauer coefficient from lattice QCD,
says $a_2^\pi(1 {\rm GeV}) = 0.16 \pm 0.03$ \cite{RQCD:2019osh}, instead of the value $0.28 \pm 0.05$ given in table \ref{tab:parameter_pi}. 
The smaller Gegenbauer coefficient implies a larger transverse-size parameter $\beta^2 = 0.65 \pm 0.06$ GeV$^{-2}$. 
Refitting the chiral mass yields $m_0^\pi(1 {\rm GeV}) = 1.83 \pm0.06$ GeV, which is consistent with our central result of $1.84 \pm 0.07$ GeV. 
\item[(f)] The asymptotic behavior is confirmed to be identical in the timelike and spacelike regions. 
The pQCD predictions approach the asymptotic form slowly \cite{Gousset:1994yh}, 
with a roughly constant enhancement expected in the timelike region at measurable $Q^2 \sim {\cal O}(10)$ GeV$^2$. 
This trend agrees with experimental data and lattice results \cite{Ding:2024lfj}. 
\end{itemize}
%-----------------------------------------------------------------------

\subsection{Electromagnetic form factor of kaon}\label{subsec:K-em-ff}

We now turn to the kaon EMFF. 
While experimental measurements of the spacelike kaon form factor remain scarce, extensive data exist for the timelike region. 
Early measurements were made by the DM1 detector in the energy range $1.40-2.18$ GeV \cite{Delcourt:1980eq,Mane:1980ep}  
and by the DM2 detector covering $1.35-2.40$ GeV \cite{DM2:1988obi} at the Orsay storage rings DCI. 
Additional results were obtained by the OLYA detector ($1.0-1.4$ GeV) \cite{Ivanov:1981wf,Ivanov:1982cr}, 
including precise measurements at the $\phi(1020)$ resonance \cite{Achasov:2000am}, 
and by the SND detector in the energy region $\sqrt{q^2} = 1.04-1.38$ GeV \cite{Achasov:2007kg}.  
The CMD2 detector at VEPP-2M provided data in the $\phi(1020)$ region ($1.01-1.034$ GeV) \cite{CMD-2:2008fsu}. 
Higher-energy measurements include a single point at $q^2=13.48$ GeV$^2$ from CLEO-c \cite{CLEO:2005tiu,Seth:2012nn}, 
as well as comprehensive results from BaBar using the ISR method: from threshold to $5$ GeV \cite{BaBar:2013jqz} and
 from $2.6$ to $8.0$ GeV \cite{BaBar:2015lgl}. 
Most recently, BESIII at BEPCII reported the most precise measurements to date in the $\sqrt{q^2}=2.00-3.08$ GeV region \cite{BESIII:2018ldc}.

%-----------------------------------------------------------------------
\begin{figure}[t]
\begin{center} \vspace{-2mm}
\includegraphics[width=0.45\textwidth]{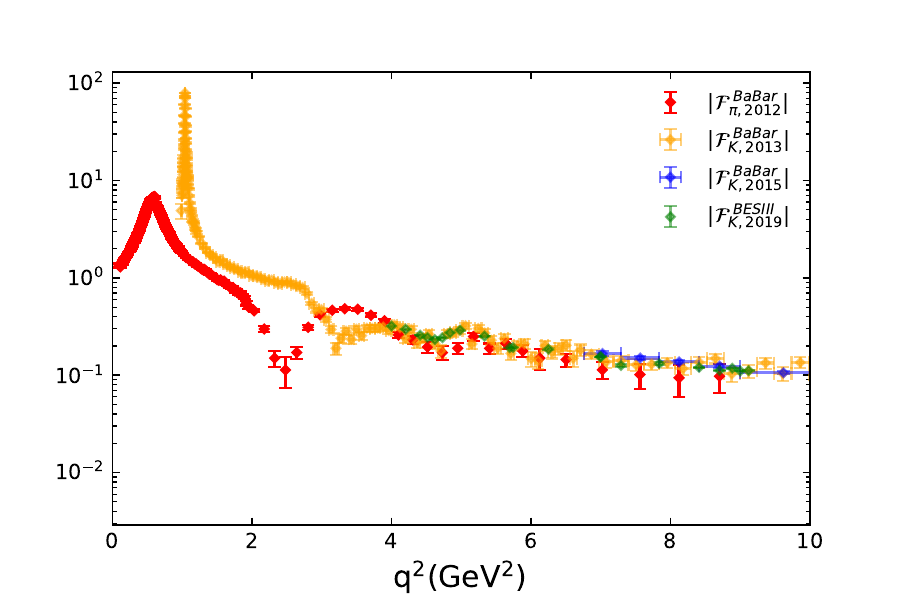}  
\hspace{1cm} 
\includegraphics[width=0.45\textwidth]{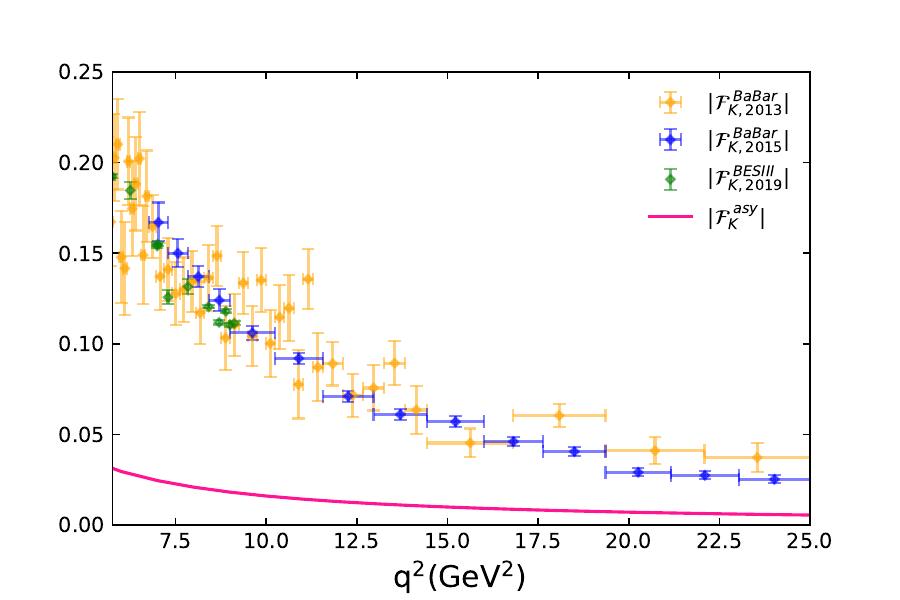}    
\end{center}
\vspace{-6mm}
\caption{Left: Timelike EMFFs of pion and kaon from production threshold up to $10$ GeV$^2$. 
Right: Kaon EMFF data in the large momentum transfers.}
\label{fig:FF-tl-data}
\end{figure} 
%-----------------------------------------------------------------------

In figure \ref{fig:FF-tl-data}, we present the recent measurements of kaon EMFFs from the BaBar and BESIII collaborations. 
For the sake of direct comparison to the data of pion EMFF, the $q^2$ is truncated up to the energy $10$ GeV$^2$. 
The pion form factor analysis \cite{BaBar:2012bdw} employed a superposition of four $\rho$ resonances 
(the ground state $\rho(770)$ followed by three radial excitations $\rho(1450), \rho''(1700), \rho'''(2250)$)
with taking into account the $\rho-\omega$ mixing effects. 
These resonant contributions were modeled using the Gounaris-Sakurai (GS) \cite{Gounaris:1968mw} parameterization 
for $\rho$ states and the K\"ohn-Santamaria (KS) \cite{Kuhn:1990ad} formalism for the $\omega$ resonance, 
both representing variants of the Breit-Wigner function.
The kaon form factor measurement \cite{BaBar:2013jqz} adopted a similar approach 
but incorporated eleven resonances spanning the energy range $4m_K^2 \leqslant q^2 \leqslant 2.4^2$ GeV$^2$, 
including three $\phi$ ($1020,1680, 2170$), four $\rho$ ($770, 1450, 1700, 2250$) and four $\omega$ ($782, 1420, 1650, 2200$). 
In the perturbative regime $2.4^2 \leqslant q^2 \leqslant 5^2$ GeV$^2$, 
the squared form factor was parameterized as $A \alpha_s(q^2)/s^n$, 
with the fitted exponent $n = 2.04 \pm 0.22$ \cite{BaBar:2012bdw,BaBar:2013jqz} matching the QCD asymptotic evolution \cite{Efremov:1979qk,Lepage:1980fj} ${\cal F}^{\rm asy}_K(q^2) = 8 \pi \alpha_s(q^2) f_K^2/q^2$. 
However, using the lattice-QCD derived decay constant $f_K = 0.156$ GeV \cite{PDG} 
reveals a fourfold enhancement in magnitude of the data relative to the asymptotic expectation. 
As evident in the right panel of figure \ref{fig:FF-tl-data}, 
this enhancement underscores significant non-asymptotic contributions predominantly arising from chiral enhancement effects 
in the two-particle twist-three LCDAs.

%-----------------------------------------------------------------------
\begin{figure}[t]
\begin{center} \vspace{-2mm}
\includegraphics[width=0.45\textwidth]{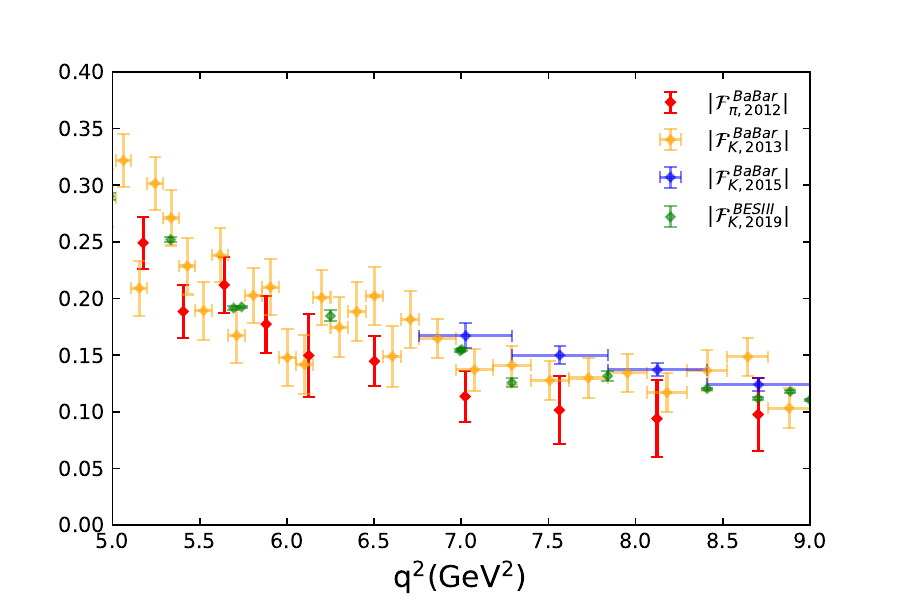}  
\hspace{1cm} 
\includegraphics[width=0.45\textwidth]{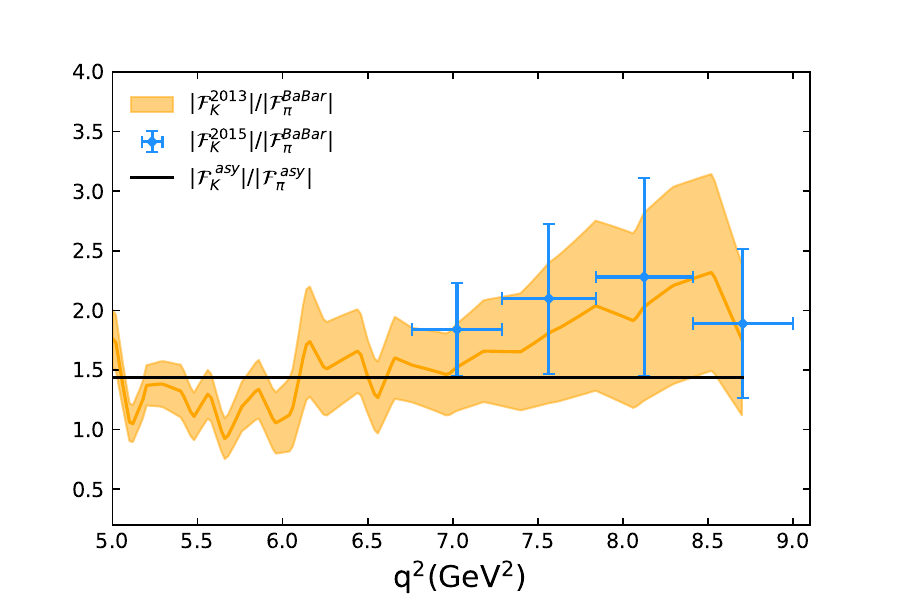}    
\end{center}
\vspace{-6mm}
\caption{Left: The Babar data of kaon and pion EMFFs in the perturbative region $5 \leqslant q^2 \leqslant 8.7$ GeV$^2$. 
Right: The data ratios of kaon and pion EMFFs. }
\label{fig:FF-ratio-data}
\end{figure} 
%-----------------------------------------------------------------------

In figure \ref{fig:FF-ratio-data}, we compare the BaBar data of kaon and pion EMFFs 
in the perturbative working region $5 \leqslant q^2 \leqslant 8.7$ GeV$^2$. 
The choice of this region is motivated by two considerations. First, the BaBar data for the pion form factor extends up to $8.7$ GeV$^2$. 
Second, this region lies sufficiently far from the physical resonance domain, ensuring a clean measurement free from resonance backgrounds.
We note that the 2013 data of kaon form factor were measured at different energy points and with different bins compared to the pion form factor.
To facilitate a direct comparison, we plot the ratio $\vert {\cal F}^{\rm BaBar}_{K,2013}\vert/\vert {\cal F}^{\rm BaBar}_{\pi,2012}\vert$ 
(yellow data points) by interpolating the kaon and pion data in the overlapping energy regions. 
For the 2015 data, however, the energy points and bins of kaon from factor match those of pion. 
Thus, we directly compute the ratio $\vert {\cal F}^{\rm BaBar}_{K,2015}\vert/\vert {\cal F}^{\rm BaBar}_{\pi,2012}\vert$ 
at four energy points (blue data points), where the dominant uncertainties originate from the pion form factor.
For comparison, we include the asymptotic pQCD prediction $\vert F^{\rm asy}_K(q^2) \vert/\vert F^{\rm asy}_\pi(q^2) \vert = f_K^2/f_\pi^2 = 1.44$ 
in black line. 
The measured ratios are found to approach the asymptotic value in the intermediate momentum-transfer region.
However, at higher momentum transfers ( $7 \leqslant q^2 \leqslant 9$ GeV$^2$), 
the central values of the ratios deviate noticeably from the asymptotic prediction, particularly in the 2015 measurement \cite{BaBar:2015lgl}.
This deviation suggests significant $SU(3)$-flavor symmetry breaking between the kaon and pion LCDAs, 
beyond what is accounted for by their decay constants alone. 

\begin{figure}[t]
\begin{center}  \vspace{-2mm}
\includegraphics[width=0.55\textwidth]{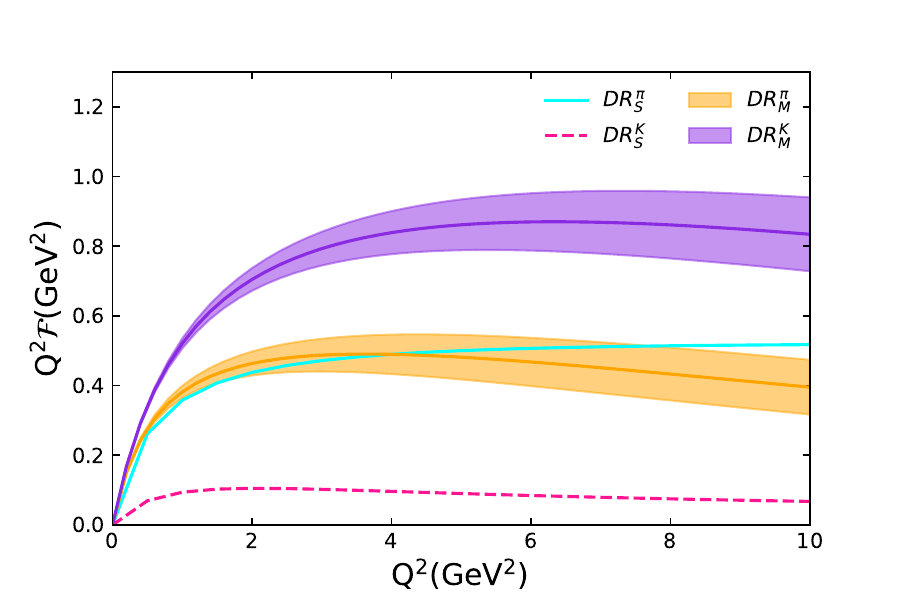}    
\end{center} \vspace{-6mm}
\caption{EMFFs obtained from the standard (curves) and modular (bands) dispersion relations. }
\label{fig:FF-K-pi-sl-DR1}
\end{figure}

With precise measurements of the kaon EMFFs in the timelike region, we can extract the spacelike form factors using dispersion relations.
In Figure \ref{fig:FF-K-pi-sl-DR1}, we present the kaon form factors derived from the standard dispersion relation (magenta curve) 
and the modular dispersion relation (purple band), where high-energy tail contributions are not included. 
For comparison, we also show the corresponding pion form factor results (cyan curve and yellow band).
We observe that the pion EMFF obtained from the standard dispersion relation closely matches the result from the modular approach. 
This consistency provides important validation in two respects. 
First, it confirms that the GS and KS resonance models \cite{BaBar:2012bdw} properly account for the imaginary part of the form factor. 
Second, it supports the assumption in the derivation of the modular dispersion relation \cite{Cheng:2020vwr,Chai:2023htt} 
that the pion form factor ${\cal F}_\pi(q^2)$ possesses no zeros in the complex plane. 

For the kaon form factor, however, 
the result obtained from the modular dispersion relation is significantly larger than that derived from the standard approach. 
Interestingly, the modular result becomes comparable to the pion form factor when accounting for a reasonably large  $SU(3)$ breaking effect. 
This discrepancy suggests two key issues. First, the resonance model employed by the BaBar collaboration \cite{BaBar:2013jqz}, 
while successfully describing the absolute magnitude of the timelike kaon form factor, may not correctly capture its imaginary part. 
Second, the modular dispersion relation itself might not be fully applicable to the kaon case.
The charged kaon form factor measured in BaBar via $e^+e^-$ annihilation receives contributions from both  $P$- and $S$-wave resonances, 
with the latter being phase-space suppressed. 
Crucially, the isospin-scalar $S$-wave component could exhibit a sharp dip near the $f_0$ resonance region, 
potentially invalidating the assumptions underlying the modular dispersion relation.
Given these complications, we do not employ the dispersion relations in our analysis of the kaon EMFF. 
Instead, we directly fit the pQCD calculations to the experimental measurements across the broad timelike region ($10 \leqslant q^2 \leqslant 60$ GeV$^2$).

%-----------------------------------------------------------------------
%----------------------------------------------------------------------- 

In terms of chiral mass, the pQCD result in equations (\ref{eq:ff-2p},\ref{eq:ff-3p}) is modified to  
\beq {\cal F}^{\rm em}_{K}(q^2) 
= (m_0^K)^2 {\cal F}^{\rm em}_{K ,1}(q^2) + m_0^K m_K {\cal F}^{\rm em}_{K, 2}(q^2) + m_0^K m_K a_1^K  {\cal F}^{\rm em}_{K, 3}(q^2) 
+ {\cal F}^{\rm em}_{K, 4}(q^2). \label{eq:K-ff-a1m0}  \eeq
Here the first term arises from the two-particle twist-three LCDAs, 
the second and third terms primarily stem from the interplay between the leading twist LCDAs 
(specifically, the asymptotic term and the first Gegenbauer expansion term) and the two-particle twist-three LCDAs, 
the last term collects the remaining contributions from higher Gegenbauer terms in leading twist LCDAs and from the high twist LCDAs. 
The well-known chiral perturbative theory (ChPT) relations \cite{Leutwyler:1996qg} 
\beq \mathcal{R} \equiv \frac{2m_s}{m_u+m_d} = 24.4 \pm 1.5, \quad 
\mathcal{Q}^2 \equiv \frac{m_s^2-(m_u+m_d)^2/4}{m_d^2-m_u^2} = (22.7 \pm 0.8)^2 \label{eq:ChPT-relation} \eeq
allow us to determine the chiral mass without explicit dependence on the individual light quark masses ($m_u$ and $m_d$):   
\beq 
m_0^K(1 {\rm GeV}) = \frac{m_K^2}{m_s \left[1 + \frac{1}{\mathcal{R}} \left( 1- \frac{\mathcal{R}^2-1}{4\mathcal{Q}^2}\right)\right]} = 1.90 \, {\rm GeV}. \eeq
We adopt the current quark mass $\overline{m}_s(2 \, \mathrm{GeV}) = 96^{+8}_{-4} \, \mathrm{MeV}$ from PDG \cite{PDG}. 
As previously noted, the chiral enhancement effect renders the determination of $a_n^K$ through the EMFFs unreliable. 
Therefore, we treat these parameters as inputs along with other relevant quantities, which are summarized in table \ref{tab:parameter_kaon}.

%-----------------------------------------------------------------------
\begin{table}[t]\begin{center}
\caption{Input parameters of kaon meson in the pQCD calculations.} \vspace{2mm}
\label{tab:parameter_kaon}
\begin{tabular}{c c c  c c c } \hline
$m_K$ \cite{PDG} & $m_0^K$ \cite{Leutwyler:1996qg} & $f_K$ \cite{PDG} & ~ & 
$a_1^K$ \cite{LatticeParton:2022zqc} & $a_2^K$ \cite{LatticeParton:2022zqc}  \non
$0.50$ & $1.90$ & $0.16$ &  ~ & $0.05\pm 0.00$  & $0.11 \pm 0.02$  \non \hline
$f_{3K} (10^{-2})$ \cite{Ball:2006wn} & $\omega_{3K}$ \cite{Ball:2006wn} & $\lambda_{3K}$ \cite{Ball:2006wn} & $\delta_{K}^2$ \cite{Ball:2006wn} & 
$\omega_{4K}$ \cite{Ball:2006wn} & $\kappa_{4K}$ \cite{Ball:2006wn} \non
$0.45 \pm 0.15$ & $-1.2 \pm 0.7$ & $1.6 \pm 0.4$ & $0.20 \pm 0.06$ & $0.20 \pm 0.10$ & $-0.12 \pm 0.01$ \non
\hline
\end{tabular}\end{center}\vspace{-4mm}\end{table}
%-----------------------------------------------------------------------

Using the twist-three LCDAs $\phi^{s,\sigma}(u, {\bf k}_T)$ defined in Eq. (\ref{eq:2pt3}), 
along with the associated iTMD functions given in Eq. (\ref{eq:iTMD}) and Eqs. (\ref{eq:TMD-lt-bt}, \ref{eq:TMD-t3-bt}), 
we separate the contributions from the valence quark state and the $q{\bar q}g$ state.
The invariant functions ${\cal F}^{\rm em}_{K ,1}$ and ${\cal F}^{\rm em}_{K ,2}$ in equation (\ref{eq:K-ff-a1m0}) 
can be expressed in terms of the dimensionless parameter $\rho_K = m_s/m_0^K$, 
\beq &&{\cal F}^{\rm em}_{K ,1}(q^2) = {\cal F}^{\rm a}_{K ,1}(q^2) + \rho_K {\cal F}^{b}_{K ,1}(q^2) + \rho_K^2 {\cal F}^{c}_{K ,1}(q^2), \non
&&{\cal F}^{\rm em}_{K ,2}(q^2) = {\cal F}^{\rm a}_{K ,2}(q^2) + \rho_K {\cal F}^{b}_{K ,2}(q^2). \label{eq:K-ff-a1m0rho}\eeq
In the first equation, ${\cal F}^{\rm a}_{K ,1}$ and ${\cal F}^{c}_{K ,1}$ denote the contribution from purely valence quark state $\phi^{p,\sigma}_{K, 2p}$ 
and purely $q{\bar q}g$ states $\phi^{p,\sigma}_{K, 3p}$, respectively, and ${\cal F}^{b}_{K ,1}(q^2)$ stands for the interplay between them. 
In the second equation, ${\cal F}^{\rm a}_{K ,2}$ combines contributions from both the leading twist LCDAs $\phi_K$ 
and the twist-three LCDAs $\phi^{p,\sigma}_{2p}$ of valence quark state. 

%-----------------------------------------------------------------------
\begin{table}[b]\begin{center}
\caption{The mean transversal momenta and the conjugated distances of $\pi, K$ mesons associated to leading twist and twist-three LCDAs.} \vspace{2mm}
\label{tab:transversal_meanvalue}
\begin{tabular}{c | c c c } \hline
{\rm mean value} \quad & \quad $\phi$ \quad & \quad $\phi^p$  \quad & \quad $\phi^\sigma$ \quad  \non \hline
$\langle k_T^2 \rangle^{1/2}_\pi$ ({\rm GeV}) \quad & \quad $0.36 \pm 0.02$ \quad & \quad $0.40 \pm 0.02$ \quad & \quad $0.40 \pm 0.02$ \quad  \non 
$\langle b_T^2 \rangle^{1/2}_\pi$ ({\rm fm}) \quad & \quad $0.56 \pm 0.02$ \quad & \quad $0.50 \pm 0.02$ \quad & \quad $0.50 \pm 0.02$ \quad   \non \hline
$\langle k_T^2 \rangle^{1/2}_K$ ({\rm GeV}) \quad & \quad $0.55 \pm 0.07$ \quad & \quad $0.53 \pm 0.07$ \quad & \quad $0.52 \pm 0.07$ \quad    \non 
$\langle b_T^2 \rangle^{1/2}_K$ ({\rm fm}) \quad & \quad $0.37 \pm 0.05$ \quad & \quad $0.38 \pm 0.05$ \quad & \quad $0.39 \pm 0.05$ \quad   \non 
\hline
\end{tabular}\end{center}\vspace{-4mm} \end{table}
%-----------------------------------------------------------------------

We extract the transverse-size parameter $\beta_K^2$ by fitting pQCD calculations to precise BaBar measurements 
in the high-momentum transfer region ($q^2 \geqslant 7.0$ GeV$^2$). 
The obtained value $\beta_K^2 = 0.30 \pm 0.05$ GeV$^{-2}$ corresponds to a mean transverse momentum 
$\left[ \langle {\bf k}_T^2 \rangle \right]^{1/2} = 0.55 \pm 0.07$ GeV for the valence quark state at leading twist.
For the twist-three LCDAs $\phi_{2p}^{p,\sigma}$, the three-particle contribution is suppressed by ${\cal O}(m_s/m_0^K)$. 
Given the current precision of both experimental measurements and pQCD calculations, 
we cannot reliably determine the transverse-size parameters associated with the ${\bar q}qg$ state. 
In our numerical analysis, we conservatively constrain this parameter to be no larger than the corresponding pion meson value 
($\beta_K^{' 2} < 0.511 $ GeV$^{-2}$).
Table \ref{tab:transversal_meanvalue} presents the mean transverse momenta defined in Eq. \ref{eq:iTMD-pi-average} 
and the conjugate distances for $\pi, K$ mesons associated with two-particle LCDAs at different twists. 
Key observations include: (1) both pion and kaon exhibit soft-scale transverse momenta, 
(2) their conjugate distances are smaller than the respective charge radii\footnote{The mean square electric charge radius of the kaon is 
$\langle r_K^2 \rangle^{1/2} = 0.56 \pm 0.03$ fm, 
this value is smaller than the charge radius of pion $\langle r_\pi^2 \rangle^{1/2} = 0.67 \pm 0.08$ fm \cite{PDG}. }, 
(3) The minor differences between leading-twist and twist-three LCDAs mean values originate from three-particle iTMD contributions. 

\begin{figure}[t]\begin{center}\vspace{-2mm}
\includegraphics[width=0.45\textwidth]{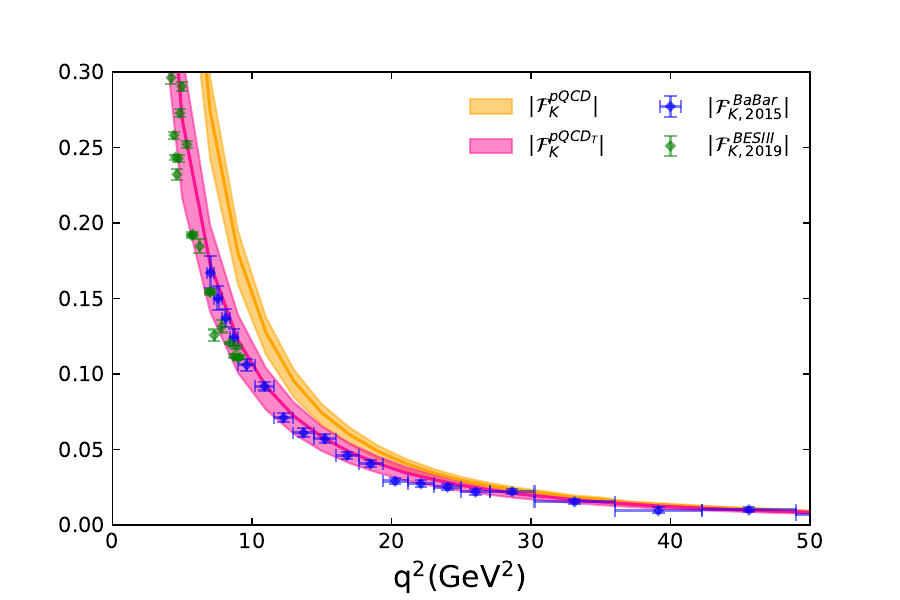}  
\hspace{1cm}
\includegraphics[width=0.45\textwidth]{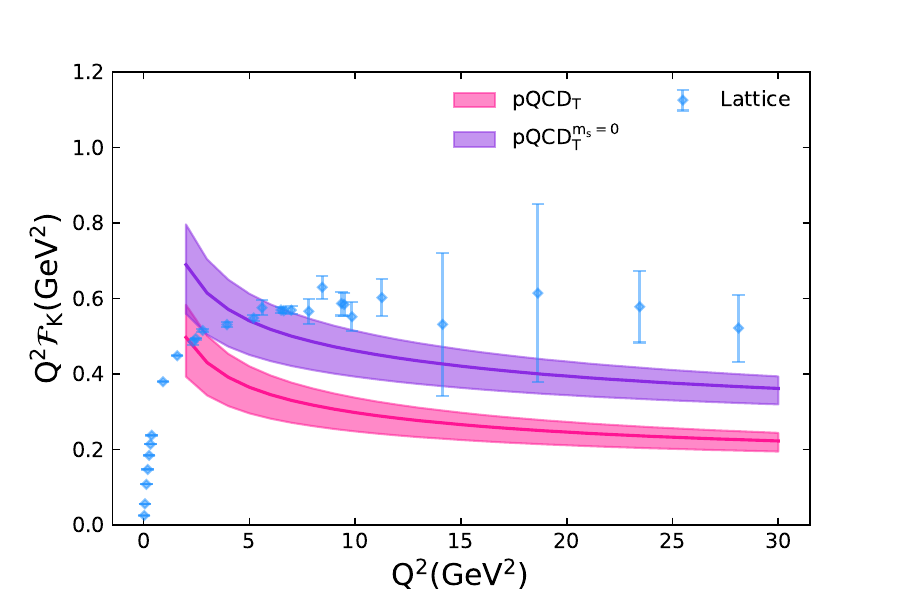}  
\end{center}\vspace{-6mm}
\caption{The pQCD predictions of timelike (left) and spacelike (right) kaon EMFFs. }
\label{fig:ff-kaon-pQCD}
\end{figure}

With the fitted iTMD functions, we show in figure \ref{fig:ff-kaon-pQCD} the pQCD predictions for kaon EMFFs 
across the momentum transfer range $\vert q^2 \vert \leqslant 10$ GeV$^2$. 
The magenta and yellow bands respectively represent calculations with and without iTMD effects, 
compared against the BaBar measurements \cite{BaBar:2013jqz,BaBar:2015lgl} 
and the recent lattice QCD evaluations \cite{Koponen:2017fvm,Ding:2024lfj}. 
The analysis reveals that iTMD contributions are crucial for explaining the timelike form factor data $\vert {\cal F}_K(q^2) \vert$ 
at intermediate and large  $q^2$.
The iTMD-improved pQCD prediction for $Q^2 {\cal F}_K(q^2)$ (magenta band) shows a small result in comparing to the lattice evaluation \cite{Ding:2024lfj}, 
which attributable to significant $SU(3)$ flavor symmetry breaking in kaons. 
This effect manifests through additional $s$-quark mass dependent terms in the twist-three LCDAs, 
whose importance becomes evident when comparing to the $m_s = 0$ case (yellow band). 
We found that the strange quark mass terms reduce the pQCD prediction by $30\%$ to $35 \%$. 
We mark that the lattice result still have large uncertainty in the large $q^2$ region, 
our findings show good agreement with both Dyson-Schwinger equation  \cite{Yao:2024drm} and the collinear QCD factorization \cite{Chen:2024oem}.
Notably, the inclusion of iTMDs extends the applicability of pQCD predictions down to lower momentum transfers (a few GeV$^2$), 
mirroring the improvement previously observed in pion studies. 
This demonstrates the universal importance of iTMDs effect in light meson form factor calculations.

%-----------------------------------------------------------------------
%-----------------------------------------------------------------------

%-----------------------------------------------------------------------
\section{Meson-Photon transition form factors}\label{sec:transition-ff}
%-----------------------------------------------------------------------

Meson-photon TFFs serve as another excellent platform for investigating LCDAs. 
The electromagnetic current conservation, 
expressed as $j_\mu^{\rm em} = e_u \bar{u} \gamma_\mu u + e_d \bar{d} \gamma_\mu d + e_s \bar{s} \gamma_\mu s$, preserves chiral symmetry. 
Consequently, only the chiral-even LCDAs defined by the quark operator ${\bar q}(z_2) \gamma_\rho \gamma_5 q(z_1)$ in Eq. (\ref{eq:DA-lt}) 
contribute to the TFFs, while the chiral-odd LCDAs from Eqs. (\ref{eq:DAs-sigma}, \ref{eq:DAs-p}) make no contribution \cite{Chai:2023htt}. 

The study of meson-photon TFFs through perturbative QCD was first proposed by Lepage and Brodsky forty years ago \cite{Lepage:1980fj}. 
Early systematic measurements by the CELLO collaboration \cite{CELLO:1990klc} 
investigated the process $e^+ e^- \to e^+e^- \left[ \gamma \gamma^\ast \to \right] {\cal P}$ for pseudoscalar mesons ${\cal P} = \pi^0, \eta, \eta^\prime$. 
At low momentum transfers ($Q^2 < 2.7$ GeV$^2$ for $\pi^0$ TFF and $Q^2 < 3.4$ GeV$^2$ for $\eta^{(\prime)}$ TFFs), 
the data were successfully described by both the vector meson dominance model with a simple $\rho$-pole [Landsberg:1985gaz] 
and the QCD-inspired Brodsky-Lepage model \cite{Brodsky:1981rp}. 
Particularly significant was the observed agreement between neutral and charged pion TFFs, 
which resolved the long-standing confusion regarding the slope sign of the neutral form factor. 
These pioneering measurements were later extended to higher momentum transfers by the CLEO-II collaboration \cite{CLEO:1997fho} 
($1.9 \leqslant Q^2 \leqslant 9$ GeV$^2$ for $\pi^0$, and $1.5 \leqslant Q^2 \leqslant 20 (30)$ GeV$^2$ for $\eta (\eta^\prime)$ TFFs), 
providing crucial data for testing QCD predictions across a broader kinematic range.

The perturbative QCD prediction of TFFs is expressed as a convolution of hard scattering kernel and the LCDAs of mesons, 
and the QCD asymptotic behavior reads \cite{Brodsky:1981rp}
\beq \lim_{Q^2 \to \infty} Q^2 {\cal F}_{{\cal P} \gamma \gamma^\ast}(Q^2) = \sqrt{2} f_{\cal P}. \label{eq:TFF-asy}\eeq
In parallel, the chiral limit of QCD shows the axial anomaly \cite{Adler:1969gk,Melikhov:2012bg}
\beq \lim_{Q^2 \to 0} {\cal F}_{ {\cal P} \gamma \gamma^\ast}(Q^2) = \frac{1}{4\pi^2 f_{\cal P}}. \label{eq:TFF-chiral-limit}\eeq
The CZ wave functions is proposed \cite{Chernyak:1983ej} to quantify the long-distance nonperturbative effects in a hard exclusive process. 
However, Kroll concluded that the CZ function of pseudoscalar mesons disagrees with the CLEO-II data. 
On the contrary, a similar pQCD analysis calculated by Cao, Huang and Ma \cite{Cao:1996di} yielded that 
neither the asymptotic nor the CZ functions can be excluded by the CLEO-II data. 
Kroll, Raulfs and Feldman then introduced the iTMD functions, supplementing to the sudakov suppression mechanism, 
in the pQCD calculation of light pseudoscalar TFFs \cite{Kroll:1996jx,Feldmann:1997vc}. 
This improved the pQCD prediction down to an intermediate transfer momentum, saying a few GeV$^2$. 

In parallel developments, Musatov and Radyushkin pioneered a sum rules approach for calculating the pion TFF at low momentum transfers  \cite{Musatov:1997pu}. While this method combined elements from different theoretical frameworks, 
the absence of quantified theoretical uncertainties made it difficult to distinguish the asymptotic and Chernyak-Zhitnitsky (CZ) wave functions, 
despite their predictions producing clearly separated central curves.
A significant advancement came with the development of LCSRs \cite{Khodjamirian:1997tk}, 
which provided a unified framework where the long-distance dynamics of the transition are governed by LCDAs. 
The leading-order LCSR calculation correctly reproduces the asymptotic behavior and demonstrates particular predictive power 
in the intermediate-to-high $Q^2$ regime ($Q^2 \geqslant 1$ GeV$^2$), 
outperforming traditional QCD sum rules with vacuum condensates  \cite{Musatov:1997pu}.
Subsequent work incorporated radiative corrections and higher-twist effects \cite{Schmedding:1999ap,Bakulev:2002uc,Agaev:2005rc}, 
to extract the Gegenbauer coefficients in the leading-twist pion LCDAs. 
When confronted with CLEO-II data, the precision of LCSRs decisively ruled out both the asymptotic and CZ models, 
revealing instead a richer structure in the pion's quark-gluon wave function that goes beyond these simplified approximations.

In 2009, the BaBar collaboration reported groundbreaking measurements of the $\pi^0, \eta, \eta^\prime$ TFFs 
with the momentum transfers up to unprecedented momentum transfers of $40$ GeV$^2$ \cite{BaBar:2009rrj}. 
These results reached a dramatic climax in the high-$Q^2$ region ($Q^2 \geqslant 10$ GeV$^2$), 
where the measured $\pi^0$ TFF strikingly exceeded the asymptotic QCD prediction, 
generating significant excitement in the theoretical physics community.
The observed enhancement at large $Q^2$ was initially interpreted through pQCD approaches 
incorporating transverse momentum ($k_T$) dependence in the hard kernel combined with flat distribution amplitudes \cite{Radyushkin:2009zg,Polyakov:2009je}, building on earlier pQCD frameworks \cite{Stefanis:1998dg,Nandi:2007qx}. 
Meanwhile, light-cone sum rule analyses \cite{Agaev:2010aq} attempted to explain the anomaly through 
extended conformal expansions of the leading-twist pion LCDA and the inclusion of higher-twist corrections up to twist-six. 
However, these LCSR fits exhibited problematic convergence patterns, most notably yielding expansion coefficients with $a_4^\pi > a_2^\pi$.

The attractive pion TFF is heat off with subsequent measurements from the Belle collaboration \cite{Belle:2012wwz}.  
Their results in the low-to-intermediate momentum tranfers ($Q^2 \leqslant 9$ GeV$^2$) 
showed good agreement with previous data from CELLO, CLEO, and BaBar \cite{CELLO:1990klc,CLEO:1997fho,BaBar:2009rrj}. 
However, the high momentum transfers' behavior ($Q^2 \geqslant 9$ GeV$^2$) presented a striking contrast. 
Unlike BaBar's rapidly growing form factor\cite{BaBar:2009rrj}, Belle's measurements remained consistent with the asymptotic QCD prediction.
This discrepancy led to significant revisions in the extracted Gegenbauer coefficients $a_n^\pi$ from LCSRs analyses \cite{Agaev:2012tm}. 
While the overall difference between Belle and BaBar datasets remains within $1.5$-$2$ standard deviations, 
the scientific community has eagerly awaited independent verification. 
The BESIII collaboration's subsequent measurement \cite{Redmer:2018uew} provided important confirmation in the low $Q^2$ regime, 
showing excellent agreement with earlier CELLO and CLEO results.
Looking ahead, the high-precision measurements expected from Belle-II and future collider experiments 
promise to resolve the remaining controversy in the high-$Q^2$ region. 
Recent theoretical advances, including next-to-next-to-leading order (NNLO) corrections in collinear factorization \cite{Gao:2021iqq}, 
have demonstrated the crucial importance of two-loop perturbative effects for the pion TFF. 
These developments, combined with the anticipated experimental precision from Belle-II, 
will undoubtedly provide deeper insights into the structure of leading twist pion LCDAs.

The TFFs of isospin-zero $\eta$ and $\eta^\prime$ mesons have been most precisely measured 
by the CLEO \cite{CLEO:1997fho} and BaBar \cite{BaBar:2006ash,BaBar:2011nrp} collaborations. 
While the $\eta^\prime$ meson measurements show reasonable consistency between experiments 
despite increasing uncertainties at higher momentum transfers, 
the $\eta$ meson results reveal significant discrepancies, particularly at $Q^2 = 7$ and $13$ GeV$^2$ 
where BaBar's data lies approximately $3 \sigma$ below CLEO's measurements.
Analysis of the BaBar dataset within a simple $\eta-\eta^\prime$ mixing framework allows extraction of the flavor-basis TFFs 
${\cal F}_{\eta_n \gamma \gamma^\ast}$ and ${\cal F}_{\eta_s \gamma \gamma^\ast}$. 
Interestingly, the high-$Q^2$ behavior of ${\cal F}_{\eta_n \gamma \gamma^\ast}$ 
shows no evidence of the rapid growth seen in the $\pi^0$ TFF \cite{BaBar:2009rrj}, 
while ${\cal F}_{\eta_s \gamma \gamma^\ast}$ systematically falls below the pQCD asymptotic prediction of $2/3 f_s \sim 0.12$ GeV. 
Theoretical efforts to explain these anomalies have explored various models for $\eta_n, \eta_s$ LCDAs 
\cite{Bakulev:2001pa,Kroll:2010bf,Agaev:2014wna} as well as potential two-gluon component in the $\eta^\prime$ meson \cite{Ali:2000ci,Kroll:2012gsh,Agaev:2003kb}. 
Significant experimental progress came with BaBar's measurement of the double-virtual TFF 
${\cal F}_{\gamma^\ast \gamma^\ast \eta^\prime}$ in $e^+e^- \to e^+e^- \eta^\prime$ processes \cite{BaBar:2018zpn}, 
spanning an extensive kinematic range ($2 \leqslant Q_1^2, Q_2^2 \leqslant 60$ GeV$^2$) that provides rigorous tests of pQCD predictions \cite{Ji:2019som} and valuable insights into pseudoscalar meson structure at high momentum transfers.

\subsection{pQCD formulism}\label{subsec:pQCDamplitudes}

%-----------------------------------------------------------------------
\begin{figure}[h] \begin{center} 
\includegraphics[width=0.8\textwidth]{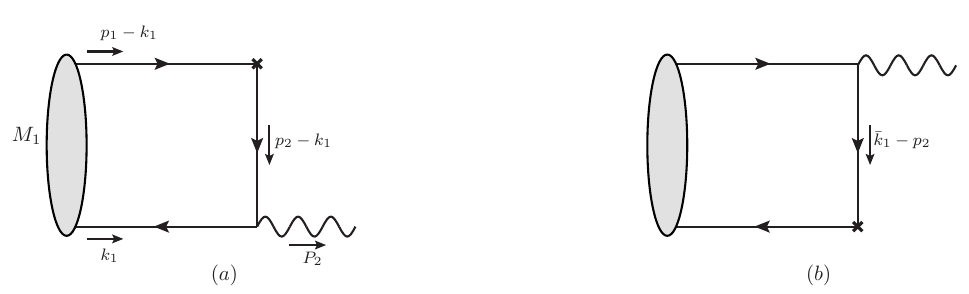}  
\end{center}\vspace{-8mm}
\caption{Leading-order feynman diagrams of photonic transition process.  }
\label{fig:tff-feynman} \end{figure} 
%-----------------------------------------------------------------------

In this work, we investigate TFFs incorporating soft iTMD effects, 
with the operator product expansion (OPE) for LCDAs extended to include twist-four contributions. 
Our analysis focuses on TFFs where one photon carries large virtuality while the other remains on-shell, 
described by the matrix element between a neutral meson state and a photon state with electromagnetic current.
\beq \langle \gamma(p_2, \varepsilon_2^\ast) \vert J_\mu^{\rm em} \vert {\cal P}^0 (p_1) \rangle 
= - i 4\pi \alpha \epsilon_{\mu\nu\rho\sigma} \varepsilon_2^{\ast \nu} p_1^\rho p_2^\sigma {\cal F}_{{\cal P}\gamma\gamma^\ast}(Q^2).
\label{eq:tff-definition} \eeq
Here $\alpha$ denotes the fine structure constant,  $\epsilon_{\mu\nu\rho\sigma} $ is the antisymmetric Levi-Civita tensor, 
and $\varepsilon^\mu$ represents the polarization vector of the virtual photon with momentum transfer $Q^2 = -q^2 = -(p_1-p_2)^2$. 
The real photon with momentum $p_2$ has polarization vector $\varepsilon^{\ast \mu}_{2}$. 
Figure \ref{fig:tff-feynman} shows the leading-order Feynman diagram for these meson-photon transitions, 
where a wavy line indicates the emitted real photon and a solid dot marks the electromagnetic vertex. 
For our kinematic setup, we choose $p_1 = \left( Q/\sqrt{2}, 0, 0 \right), p_2 = \left(0, Q/\sqrt{2},0 \right), 
k_1 = \left( x Q/\sqrt{2},0,k_{T} \right)$, where we neglect the light meson mass. 

Similar to the EMFFs of pion meson discussed in the last section, 
the meson-photon transition matrix element can also be written in the factorization formulization 
\beq \langle \gamma(p_2) \vert J_\mu^{\rm em} \vert {\cal P}^0 (p_1) \rangle 
= \varepsilon_2^{\ast \nu} \oint dz_1 H_{\alpha \delta}(z_1,0) \langle 0 \Big\vert {\bar q}_{\gamma}(z_1) [z_1, 0] q_{\delta}(0) 
\Big\vert {\cal P}^0(p_1) \rangle_{\mu_t}. \label{eq:tff-fact} \eeq
The hard function at leading order reads as 
\beq H^{(0)}_{\alpha\delta}(z_1,0) 
= (-1) \left[ \left(ie \gamma_\nu \right) S^{(0)}(z_1,0) \left( i e \gamma_\mu \right) \right]_{\alpha\delta}, \label{eq:tff-hk} \eeq
in which $S^{(0)}$ is the free quark propagator. 
In the pseudoscalar transition, only the chiral-even term ${\bar q} \gamma_\rho \gamma_5 q$ in the fierz identity 
Eq. (\ref{eq:fierz-indentity}) survives due to the conservation of chiral symmetry. 
In this sense, TFFs involve only the leading twist LCDA $\varphi(u)$ and two-particle twist four LCDAs $g_{1,2}(u)$, 
and it has nothing to do with the twist three LCDAs which brings large pollution to extract leading twist LCDAs 
in the pion EMFFs due to the chiral enhancement. 
Summing over the charge factor of electromagnetic currents and the locations of real photon emission shown in figure \ref{fig:tff-feynman}, 
we arrive at the pQCD prediction of TFFs at born level 
\beq {\cal F}^{2p}_{{\cal P} \gamma\gamma^\ast}(Q^2) &=& 
e_q^2 \int_0^1 du \int b db e^{-S(u,b,Q,\mu)} S_t(u,Q)  f_{\cal P}  \non 
&\cdot& \left\{ 2 \varphi(u) K_0(\kappa_1b) 
\left[ 1 -\frac{\alpha_s(\mu)}{4\pi} C_F \left( \ln\frac{\mu^2 b}{2 \kappa_1 }+\gamma_E + 2\ln u+3-\frac{\pi^2}{3} \right) \right] \right. \non 
&-& \left. 4g_1(u) \left[ \frac{b}{2 \kappa_1} K_1(\kappa_1 b) + \frac{b}{2 \kappa_2} K_1(\kappa_2 b) \right] \right. \non
&-& \left. 4 \tilde{g}_2(u) \left[ \frac{b}{2 \kappa_1} K_1(\kappa_1 b) - \frac{b}{2 \alpha_2} K_1(\kappa_2 b) \right] \right\}.
\label{eq:tff-P-2p} \eeq
In the above equation, $K_0$ and $K_1$ are the second kind of modified Bessel functions (Basset function) of order zero and one, respectively. 
The variables are $\kappa_1 = \sqrt{u}Q$ and $\kappa_2 = \sqrt{\bar u} Q$, 
and the auxiliary twist four LCDA is defined by $\tilde{g}_2(u) \equiv \int_0^{u} du' g_2(u)$. 

\begin{figure}[t] \begin{center} \vspace{-2mm}
\includegraphics[width=0.90\textwidth]{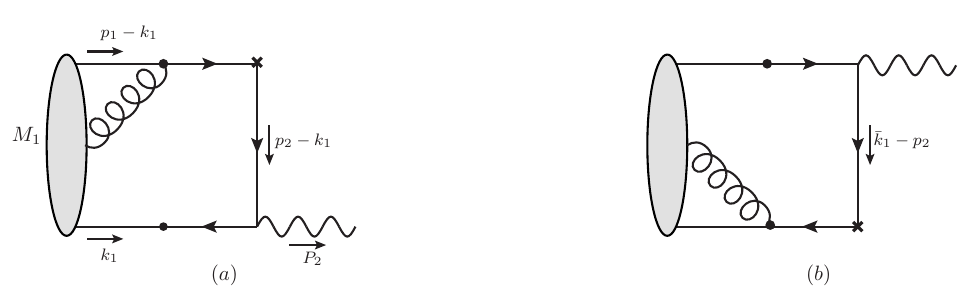}  \non
\includegraphics[width=0.90\textwidth]{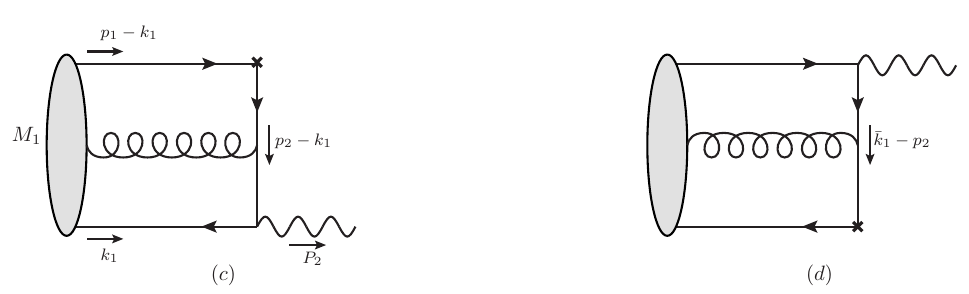}  
\end{center} \vspace{-8mm}
\caption{The soft gluon corrections to meson-photon TFFs.}
\label{fig:tff-feynman-soft} \end{figure}

The NLO hard gluon correction to the leading twist contribution had been calculated in 2009 by Li and Mishima \cite{Li:2009pr}. 
In order to take into account the soft gluon correction from the three-particle LCDAs as shown in figure \ref{fig:tff-feynman-soft}, 
we retain the quark propagator with a zero mass up to the ${\bar q}qg$ term. 
\beq S(z_1,z_2) &=& \frac{\zsl_1 - \zsl_2}{2\pi^2 \left(z_1-z_2\right)^2} \non 
&-& \frac{\int_0^1 dv G^{\mu\nu}(vz_1+{\bar v}z_2) 
\left[ \left(\zsl_1 - \zsl_2 \right) \sigma_{\mu\nu} - 4i v \left( z_1 - z_2 \right)_\mu \gamma_\nu \right]}{16\pi^2 \left(z_1-z_2\right)^2}.  
\label{eq:quark-propagator} \eeq
Here the gluon field $A_\mu$ is expressed in terms of field strength tensor $G_{\mu\nu}$, 
and the Fock-Schwinger gauge $z \cdot A(z) = 0$ is taken to guarantee the gauge invariance. 
The soft gluon corrections to the meson-photon transitions are depicted in figure \ref{fig:tff-feynman-soft}, 
where the bullets show the possible attaching points of soft gluon.
We note that the amplitudes of upper two diagrams with the soft gluon attaching to the external quark lines 
contributes at the power ${\cal O}(k_T/Q)$, this contribution could be absorbed into the transversal component of two-particle LCDAs. 
So the three-particle contribution corresponds to the diagrams where the soft gluon is attached to the internal quark propagator, 
as shown in the lower two diagrams. The pQCD result for figure \ref{fig:tff-feynman-soft} (c,d) is 
\beq {\cal F}^{3p}_{{\cal P} \gamma\gamma^\ast}(Q^2) &=& 
e_q^2 \int_0^1 du \, \int b db \, e^{-S(u,b,Q,\mu)} S_t(u,Q) \, \int_0^u d \alpha_1 \int_0^{\bar u} d \alpha_2  
\frac{2f_{\cal P} \, \varphi_\parallel(\alpha_i) }{\left(1-\alpha_1-\alpha_2\right)^2} \non
&\cdot& \left[ \left( u - \alpha_1 \right) \frac{b}{2 \kappa_3} K_1(\kappa_3 b) + \left( u - \alpha_2 \right) \frac{b}{2 \kappa_4} K_1(\kappa_4 b)\right]. 
\label{eq:tff-P-3p} \eeq
Again, only the chiral-even term ${\bar q} \gamma_\rho \gamma_5 G^{\rho\sigma} q$ survives in the three-particle LCDAs. 
The hard scales are $\kappa_3 = \sqrt{\left(\alpha_1+v \alpha_3\right)}Q, \kappa_4 = \sqrt{\left(\alpha_2 + v \alpha_3\right)}Q$ 
and $v = \left(u - \alpha_1\right)/(1-\alpha_1-\alpha_2)$.
Combing together the contributions from two particle and three particle LCDAs, the pQCD prediction of meson-photon TFFs is written by 
\beq {\cal F}_{{\cal P} \gamma\gamma^\ast}(Q^2) 
=  {\cal F}^{2p}_{{\cal P} \gamma\gamma^\ast}(Q^2) + {\cal F}^{3p}_{{\cal P} \gamma\gamma^\ast}(Q^2). \label{eq:tff-P-2p+3p} \eeq

%-----------------------------------------------------------------------
%-----------------------------------------------------------------------

\subsection{Transition form factor of pion}\label{subsec:pi-tff}

\begin{figure}[tb]\begin{center} 
\includegraphics[width=0.45\textwidth]{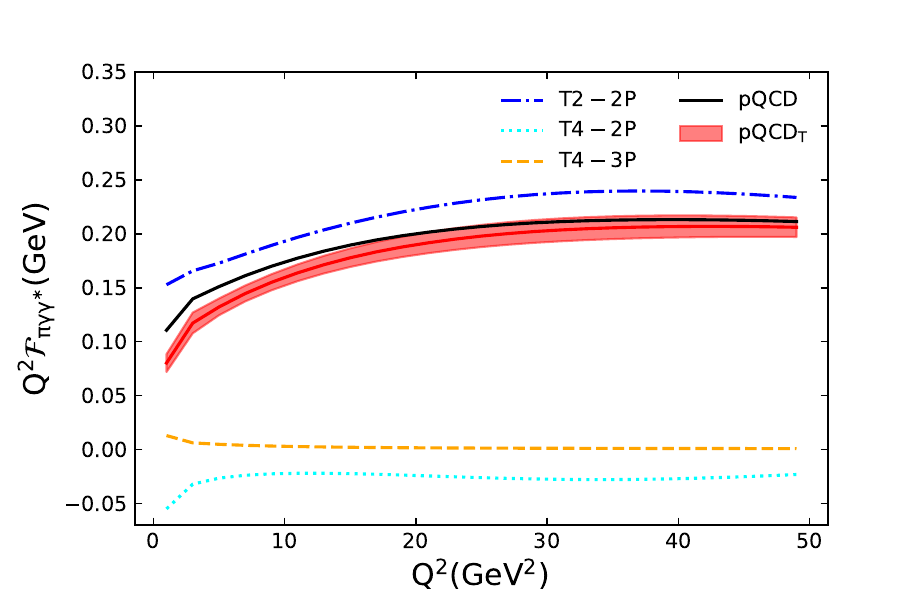} \hspace{8mm}
\includegraphics[width=0.45\textwidth]{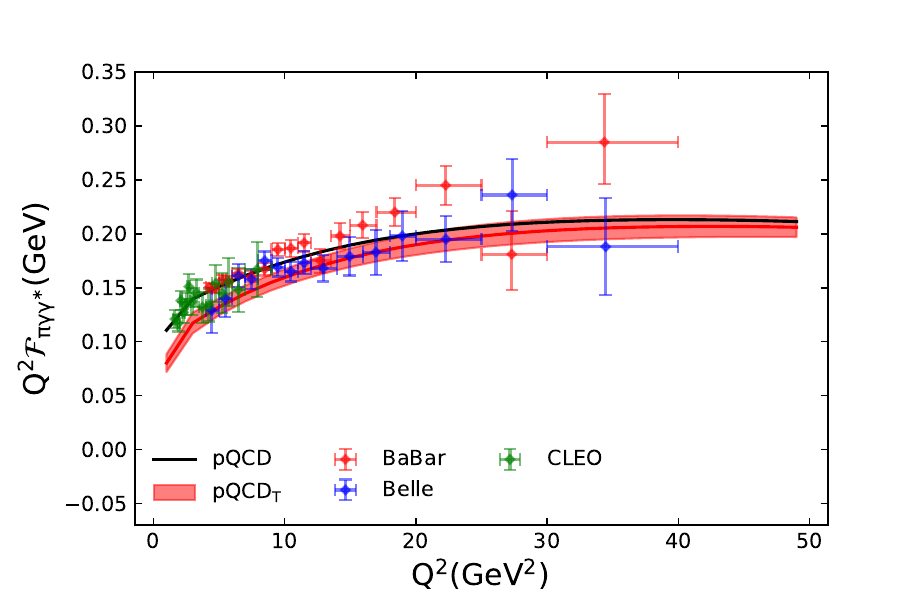}  \end{center}
\vspace{-6mm}
\caption{Left: Pion-photon TFF $Q^2{\cal F}_{\pi \gamma\gamma^\ast}$ at different twists, 
Right: The pQCD prediction of $Q^2{\cal F}_{\pi \gamma\gamma^\ast}$ in comparison to the measurements.}
\label{fig:pi-TFF}
\end{figure}

The pion-photon transition form factor (TFF) can be decomposed into contributions from various LCDAs 
\beq {\cal F}_{\pi \gamma\gamma^\ast}(Q^2) 
&=& \frac{e_u^2 - e_d^2}{\sqrt{2}} {\cal F}_{{\cal P} \gamma\gamma^\ast}(Q^2) \non
&=& {\cal F}^{t2}_{\pi \gamma\gamma^\ast}(Q^2) + {\cal F}^{t4-2p}_{\pi \gamma\gamma^\ast}(Q^2)+{\cal F}^{t4-3p}_{\pi \gamma\gamma^\ast}(Q^2), 
\label{eq:tff-pi-2p+3p} \eeq
where we explicitly separate the contributions from leading twist, two-particle and three-particle twist four LCDAs. 
The prefixal electron charge factor originates from the valence quark distribution in a neutral pion 
$\vert \pi^0 \rangle = \frac{1}{\sqrt{2}} \vert {\bar u}u - {\bar d}d \rangle$. 
Using the pion LCDA parameters listed in table \ref{tab:parameter_pi}, 
we present the individual contributions in figure \ref{fig:pi-TFF}(a). 
The results demonstrate the expected power hierarchy, confirming the theoretical expectation that 
higher-twist effects become increasingly suppressed at large momentum transfers.

%-----------------------------------------------------------------------
%-----------------------------------------------------------------------

To account for soft transverse momentum dynamics, 
we systematically incorporate iTMDs into the LCDAs of the factorization formula given in Eq. (\ref{eq:tff-fact}). 
For the leading-twist contribution, this is achieved directly through Eq. (\ref{eq:pi-soft-DA-bt}), 
where we employ the transverse-size parameter $\beta_\pi^2 = 0.51 \pm 0.04$ GeV$^{-2}$ determined from our analysis. 
The two-particle twist-four LCDAs $g_{1,2}(u)$ receive contributions from both genuine twist-four effects 
arising from three-particle Fock states and Wandzura-Wilczek-type mass corrections to the valence quark state's leading-twist DAs, 
as detailed in Appendix \ref{app:LCDAs-expression}. 
Given their proportionality to $m_\pi^2$, we safely neglect the numerically insignificant mass correction terms in our calculations. 
For the genuine twist-four components described by $\varphi_{\parallel,\perp}$, 
we implement iTMD effects via the three-particle Gaussian distribution specified in Eq. (\ref{eq:TMD-t3-bt}), 
thereby maintaining consistent treatment of transverse momentum effects across all relevant contributions. 
The pion-photon TFF is then modified to 
\beq {\cal F}_{\pi \gamma\gamma^\ast}(Q^2) &=& \frac{e_u^2 - e_d^2}{\sqrt{2}} 
\int_0^1 du \int b db \, e^{-S} S_t f_{\pi} \, \Big\{ 2 \varphi(u) \hat{\Sigma}(u,b) K_0(\kappa_1b) \non 
&~& \cdot \left[1 -\frac{\alpha_s(\mu)}{4\pi} C_F \left( \ln\frac{\mu^2 b}{2 \kappa_1 }+\gamma_E + 2\ln u+3-\frac{\pi^2}{3} \right) \right]  \non 
&~& - 2 g_1(u) \left[ \frac{b}{\kappa_1} K_1(\kappa_1 b) + \frac{b}{\kappa_2} K_1(\kappa_2 b) \right] \non
&~& - 2 \tilde{g}_2(u) \left[ \frac{b}{\kappa_1} K_1(\kappa_1 b) - \frac{b}{\kappa_2} K_1(\kappa_2 b) \right] \Big \} \non
&+& \frac{e_u^2 - e_d^2}{\sqrt{2}} \int_0^1 du \, \int b db \, e^{-S} S_t \, 
\int_0^u d \alpha_1 \int_0^{\bar u} d \alpha_2  \frac{f_\pi \, \varphi_\parallel(\alpha_i)}{\left(1-\alpha_1-\alpha_2\right)^2} \non
&~& \cdot  \left[ \left( u - \alpha_1 \right) \frac{b}{\kappa_3} K_1(\kappa_3 b) + \left( u - \alpha_2 \right) \frac{b}{\kappa_4} K_1(\kappa_4 b) \right]. 
\label{eq:pi-tff-amplitude}\eeq
We have set $b_1 = b_2 = b$ in the iTMD function associated to three-particle state, 
this is due to the same flavor of the valence quark and antiquark in the pion meson. 
Because the pion-photon TFF is predominated by the leading twist LCDAs, 
we do not consider the iTMD correction to the three particle contribution in the numerics. 

In figure \ref{fig:pi-TFF} (a), we also depict the result from the iTMDs-improved pQCD calculation (red band). 
In figure \ref{fig:pi-TFF} (b), we compare the pQCD predictions of pion-photon TFF 
to the experiment measurements from CLEO, BaBar and Belle collaborations. 
Our analysis reveals that incorporating iTMDs reduces the pion-photon TFF at small to intermediate momentum transfers ($Q^2 \geqslant 20$ GeV$^2$), while showing minimal impact on the TFF behavior in the high-$Q^2$ regime. 
This differential effect demonstrates how transverse momentum dynamics become increasingly negligible 
as the probing scale grows beyond $20$ GeV$^2$, where collinear factorization dominates.

\subsection{Transition form factor of $\eta,\eta'$ mesons}\label{subsec:eta-tff}

Within the framework of isospin symmetry, the valence quark structure of $\eta$ and $\eta^\prime$ mesons 
can be decomposed into their $SU(3)_F$ flavor singlet and octet components, says 
\beq \left(\begin{array}{c} \vert \eta \rangle \\ \vert \eta^\prime \rangle \end{array}\right) 
= \left(\begin{array}{cc} \psi_\eta^8 & \psi_\eta^1 \\ \psi_{\eta^{\prime}}^8 & \psi_{\eta^{\prime}}^1 \end{array}\right) 
\left(\begin{array}{c} \vert \eta_8 \rangle \\ \vert \eta_1 \rangle \end{array}\right) 
= \left(\begin{array}{cc} \cos \theta & -\sin \theta \\ \sin \theta & \cos \theta \end{array}\right) 
\left(\begin{array}{c} \frac{1}{\sqrt{6}} \vert {\bar u}u + {\bar d}d - 2 {\bar s}s \rangle \\ \frac{1}{\sqrt{3}} \vert {\bar u}u + {\bar d}d + {\bar s}s \rangle \end{array}\right). \label{eq:etap-81-decom}\eeq
Here $\psi_{\eta^{(\prime)}}^{i}$ describe the opportunity amplitude to find $\eta^{(\prime)}$ meson in the flavor singlet and octet states. 
The exact isospin symmetry implies that the amplitudes can be written in a mixing angle $\theta$. 

Decay constants of physical states $\eta^{(\prime)}$ and singlet-octet states $\eta_{i=q,s}$ are both defined 
via the local currents $J_{\mu 5}^{i=8} = \left( {\bar u}u + {\bar d}d - 2 {\bar s}s \right)/\sqrt{6}$ and 
$J_{\mu 5}^{i=1} = \left( {\bar u}u + {\bar d}d + {\bar s}s \right)/\sqrt{3}$, 
\beq \langle 0 \vert J_{\mu 5}^i \vert \eta^{(\prime)}(p) \rangle = i f_{\eta^{(\prime)}}^i p_\mu, \quad 
\langle 0 \vert J_{\mu 5}^i \vert \eta_i(p) \rangle = i f_{\eta_i} p_\mu.  \eeq
It can be read from Eqs. (\ref{eq:etap-81-decom}) that these two types decay constants satisfy the relations 
\beq \left(\begin{array}{cc}f_\eta^8 &  f_\eta^1 \\ f_{\eta^\prime}^8 &f_{\eta^\prime}^1 \end{array}\right) 
= \left(\begin{array}{cc} \cos \theta & -\sin \theta \\ \sin \theta & \cos \theta \end{array}\right) 
\left(\begin{array}{cc} f_{\eta_8} & 0 \\ 0 & f_{\eta_1} \end{array}\right). \label{eq:etap-81-decaycons}\eeq
A general parameterization is proposed \cite{Leutwyler:1997yr} by considering the $U(1)_A$ anomaly \cite{Witten:1978bc}, 
\beq \left(\begin{array}{cc}f_\eta^8 &  f_\eta^1 \\ f_{\eta^\prime}^8 &f_{\eta^\prime}^1 \end{array}\right) 
= \left(\begin{array}{cc} \cos \theta_8 & -\sin \theta_1 \\ \sin \theta_8 & \cos \theta_1 \end{array}\right) 
\left(\begin{array}{cc} f_{\eta_8} & 0 \\ 0 & f_{\eta_1} \end{array}\right).
\label{eq:etap-81-decaycons-1}\eeq
Here $\theta_1$ and $\theta_8$ are actually two different mixing angle. 

In order to investigate $SU(3)_F$ flavor symmetry breaking effects, 
Feldmann, Kroll, and Stech proposed an alternative approach using the orthogonal flavor basis for 
describing $\eta$ and $\eta^\prime$ meson states  \cite{Feldmann:1998vh}. 
Their framework decomposes the isoscalar mesons into pure quark flavor components, 
\beq \left(\begin{array}{c} \vert \eta \rangle \\ \vert \eta^\prime \rangle \end{array}\right) 
= \left(\begin{array}{cc} \psi_\eta^q & \psi_\eta^s \\ \psi_{\eta^{\prime}}^q & \psi_{\eta^{\prime}}^s \end{array}\right) 
\left(\begin{array}{c} \vert \eta_q \rangle \\ \vert \eta_s \rangle \end{array}\right) 
= \left(\begin{array}{cc} \cos \phi & -\sin \phi \\ \sin \phi & \cos \phi \end{array}\right) 
\left(\begin{array}{c}\frac{1}{\sqrt{2}} \vert {\bar u}u + {\bar d}d \rangle \\ \vert {\bar s}s \rangle \end{array}\right).
\label{eq:etap-flavor-1}\eeq
The mixing angle in the flavor basis is related to the angle in the singlet-octet basis by $\phi = \theta + \theta^\prime$, 
with $\theta^\prime$ being the rotation angle between the singlet-octet basis and the flavor basis 
( $\cos \theta^\prime = 1/\sqrt{3}, \sin \theta^\prime = \sqrt{2}/\sqrt{3}$). 
The decay constants of $\eta^{(\prime)}$ and $\eta_{q,s}$ defined via the orthogonal flavor currents 
$J_{\mu 5}^{i=q} = \left( {\bar u}u + {\bar d}d  \right)/\sqrt{2}$ and $J_{\mu 5}^{i=s} = {\bar s}s$ are related by the same mixing matrix, 
\beq \left(\begin{array}{cc}f_\eta^q &  f_\eta^s \\ f_{\eta^\prime}^q &f_{\eta^\prime}^s \end{array}\right) 
= \left(\begin{array}{cc} \cos \phi & -\sin \phi \\ \sin \phi & \cos \phi \end{array}\right) 
\left(\begin{array}{cc} f_{\eta_q} & 0 \\ 0 & f_{\eta_s} \end{array}\right).
\label{eq:etap-flavor-decaycons} \eeq
Transforming the flavor axial-vector currents into the single-octet currents, one obtain 
the relation between angles $\theta_1$ and $\theta_8$ proposed in Eq. (\ref{eq:etap-81-decaycons-1})
\beq \tan \left( \theta_1 - \theta_8 \right) = \frac{\sqrt{2}}{3} \left( \frac{f_{\eta_s}}{f_{\eta_q}} - \frac{f_{\eta_q}}{f_{\eta_s}} \right).
\label{eq:etap-81-decaycons-su3}\eeq
In this sense, the two angles mixing scenario shown in Eq. (\ref{eq:etap-81-decaycons-1}) reveals the breaking of $SU(3)_F$ symmetry. 

As the TFFs evolve to larger momentum transfers, 
the charmed quark component $\vert \eta_c \rangle = \vert {\bar c}c \rangle$ becomes increasingly relevant in the meson state decomposition. 
This necessitates extending the mixing framework to a three-flavor basis ($\eta_q$-$\eta_s$-$\eta_c$), 
where the physical states are described through a generalized rotation via 
\beq \left(\begin{array}{ccc} f_\eta^q &  f_\eta^s & f_\eta^c \\ f_{\eta^\prime}^q & f_{\eta^\prime}^s & f_{\eta^\prime}^c \\ 
f_{\eta_c}^q &  f_{\eta_c}^s & f_{\eta_c}^c \end{array}\right) 
= \left(\begin{array}{ccc} \cos \phi & -\sin \phi & - \theta_c \sin\theta_y \\ 
\sin \phi & \cos \phi & \theta_c \cos\theta_y \\ 
-\theta_c \sin \left( \phi-\theta_c \right) & -\theta_c \cos \left( \phi-\theta_c \right) & 1 \end{array}\right) 
\left(\begin{array}{ccc} f_{\eta_q} & 0 & 0 \\ 0 & f_{\eta_s} & 0 \\ 0 & 0 & f_{\eta_c} \end{array}\right).
\label{eq:etap-flavor-decaycons-1}\eeq
Two new angles are introduced to relate the decay constants by 
$f_\eta^c = - f_{\eta_c} \theta_c \sin \theta_y$ and $f_{\eta^\prime}^c = f_{\eta_c} \theta_c \cos \theta_y$, 
in additional to $f_{\eta_c}^c = f_{\eta_c}$. 
We note that the ${\cal O}(\theta_c^2)$ terms have been neglected in Eq. (\ref{eq:etap-flavor-decaycons-1}) 
since the mixing between light flavor basis and $\eta_c$ is at ${\cal O}(1/m_{\eta_c}^2)$. 
So strictly speaking, the mixing matrix shown in Eq. (\ref{eq:etap-flavor-decaycons-1}) is not a unitary matrix 
$UU^\dag = 1 + {\cal O}(\theta_c^2)$. 
Considering the axial vector anomaly at leading order, the three flavors mixing matrix is obtained in Ref. \cite{Feldmann:1998vh} as 
\beq U = \left(\begin{array}{ccc} \cos \phi & -\sin \phi & - 0.006 \\ \sin \phi & \cos \phi  & - 0.016 \\ 0.015 & 0.008 & 1 \end{array}\right).  
\label{eq:etap-flavor-mixing}\eeq
Here $\theta_c = - 1.0^\circ$ and $f_\eta^c = -2.4$ MeV, $f_{\eta^\prime}^c = -6.3$ MeV are indicated. 

%%-----------------------------------------------------------------------
\begin{table}[t] \begin{center}
\caption{Decay constants (in unit of GeV, $f_\pi = 0.13$ GeV) and mixing angles (in unit of degree) of flavor basis states. 
The default scale is $1$ GeV for $\eta_q$ and $\eta_s$, while $3$ GeV for $\eta_c$. } 
\label{tab:decaycons-eta} \vspace{2mm}
\begin{tabular}{c | c c c c } \hline
${\cal P}$ & $f_{\cal P}$-S1 \cite{Feldmann:1998vh} & \quad $f_{\cal P}$-S2 \cite{Escribano:2005qq,Escribano:2013kba} & 
\quad $f_{\cal P}$-S3 \cite{Cao:2012nj} & \quad $f_{\cal P}$-{\rm Lattice} \cite{Bali:2021qem} \\ \hline
$\eta_q$ & $(1.07 \pm 0.02) f_{\pi}$ & \quad $(1.09 \pm 0.03)f_{\pi}$ & \quad $(1.08 \pm 0.04) f_{\pi}$ & \quad 
$\left(1.02^{+0.02}_{-0.05}\right) f_\pi $ \\
$\eta_s$ & $(1.34 \pm 0.06) f_{\pi}$ & \quad $(1.66 \pm 0.06)f_{\pi}$ & \quad $(1.25 \pm 0.09) f_{\pi}$ & \quad 
$\left(1.37^{+0.04}_{-0.06}\right) f_\pi$ \\ 
$\phi$ &$39.3 \pm 1.0$ & \quad $40.3 \pm 1.8$ & \quad $37.7 \pm 0.7$ & \quad $39.6^{+1.2}_{-2.1}$ \\ \hline
$\eta_{c}$ & & & & \quad $0.40 \pm 0.01$ \cite{Hatton:2020qhk} \\
\hline \end{tabular}
\end{center} \end{table}
%%-----------------------------------------------------------------------
%%-----------------------------------------------------------------------
\begin{table}[t] \begin{center} 
\caption{Masses (in uint of GeV) and the first two gegenbauer coefficients of the flavor basis states. 
The default scale is $1$ GeV for $\eta_q$ and $\eta_s$, while $3$ GeV for $\eta_c$. }
\label{tab:mass-moment-eta}\vspace{2mm}
\begin{tabular}{c | c c c | c c} \hline
${\cal P}$ & $m_{\cal P}$-S1 \cite{Feldmann:1998vh} & \quad $m_{\cal P}$-S2 \cite{Escribano:2013kba} 
& \quad $m_{\cal P}$-S3 \cite{Cao:2012nj} \quad &  $a_2^{\cal P}$ \cite{RQCD:2019osh} & \quad $a_4^{\cal P}$ \\ \hline
$\eta_q$ & $0.11$ & \quad $0.31$ & \quad $0.21$ \quad & $0.14 \pm 0.02$ & \quad --- \\
$\eta_s$ & $0.71$ & \quad $0.72$ & \quad $0.71$ \quad &  $0.14 \pm 0.02$ & \quad --- \\ \hline
$\eta_{c}$ & & \quad $2.98$ \cite{Hatton:2020qhk} & &  $0.13$ \cite{Hatton:2020qhk} & \quad $0.04$ \cite{Hatton:2020qhk} \\ \hline
\end{tabular} \end{center} \end{table}
%%-----------------------------------------------------------------------

Within this framework, the physical $\eta^{(\prime)}$ TFFs can be expressed as linear combinations of the flavor basis TFFs,
\beq {\cal F}_{\eta \gamma \gamma^\ast} = 
\cos \phi \frac{e_u^2 + e_d^2}{\sqrt{2}} {\cal F}_{\eta_q\gamma \gamma^\ast} 
- \sin \phi \, e_s^2 \, {\cal F}_{\eta_s \gamma \gamma^\ast} - 0.006 \, e_c^2 \, {\cal F}_{\eta_c \gamma \gamma^\ast}, \non  
{\cal F}_{\eta^\prime \gamma \gamma^\ast} = 
\sin \phi \frac{e_u^2 + e_d^2}{\sqrt{2}} {\cal F}_{\eta_q \gamma \gamma^\ast} 
+ \cos \phi \, e_s^2 \, {\cal F}_{\eta_s \gamma \gamma^\ast} - 0.016 \, e_c^2 \, {\cal F}_{\eta_c \gamma \gamma^\ast}.
\label{eq:eta-tff-amplitude}\eeq
The first and second terms can be obtained directly by substituting the the flavor basis states $\eta_q$ and $\eta_s$ 
into ${\cal F}_{{\cal P} \gamma \gamma^\ast}$ expressed in Eq. (\ref{eq:tff-P-2p+3p}). 
In table \ref{tab:decaycons-eta}, we collect the decay constants for states in different flavor bases. 
The results in the first row, labeled S1, S2, and S3, are derived from distinct phenomenological analyses. 
We also include recent lattice QCD results \cite{Bali:2021qem}, which align more closely with the S1 predictions.
To determine the mixing angle ($\phi$) between the orthogonal flavor bases, 
we employ the simple relations connecting the singlet-octet scheme and the orthogonal flavor scheme, 
they read as $\theta_8 = \phi - \arctan \left( \sqrt{2} f_{\eta_s}/f_{\eta_q} \right)$ and $\theta_0 = \phi - \arctan \left( \sqrt{2} f_{\eta_q}/f_{\eta_s} \right)$. 
Using the lattice QCD results for the singlet-octet mixing angles $\theta_8 = -22.9^{+2.32}_{-3.79}$ degree and $\theta_0 = - 6.5^{+0.92}_{-1.85}$ degree, 
we can extract the angle $\phi$. 
For the charmonium state, we adopt the decay constant $f_{\eta_c} = 0.40 \pm 0.10$ GeV 
obtained from a lattice evaluation that incorporates both QCD and QED contributions \cite{Hatton:2020qhk}. 

In table \ref{tab:mass-moment-eta}, we list the masses and the first two Gegenbauer moments for the flavor basis states.
The Gegenbauer coefficients for the light-flavor basis states are taken from the RQCD collaboration\footnote{Specifically, 
their result for the  $SU(3)$ octet is $a_2^{\eta_8}(2 \, {\rm GeV}) = 0.11 \pm 0.02$ \cite{RQCD:2019osh}. 
Here, we assume that $SU(3)$ breaking effects manifest only in the decay constants.}. 
For the charmonium state, the Gegenbauer coefficient is adopted from the HPQCD collaboration \cite{Hatton:2020qhk}. 
We note that the default scales for the Gegenbauer moments are set at $1$ GeV for $a_2^{\eta_q,\eta_s} = 0.14 \pm 0.02$, 
while at $3$ GeV for $a_2^{\eta_c} = 0.13$ and $a_4^{\eta_c} = 0.04$. 
In our perturbative QCD calculations, the leading-twist LCDA is expected to exhibit similar behavior across these three quarkonium systems. 
However, at a hadronic scale (e.g., $1$ GeV), they display significant differences. 
Furthermore, the Gegenbauer polynomial expansion yields a piecewise convex-concave-convex leading-twist LCDA 
as a function of the momentum fraction $u \in [0,1]$. 
This behavior exhibits a departure from the broad, concave functions obtained through exponential \cite{Ding:2015rkn} 
or beta-function \cite{Blossier:2024wyx} parameterizations, which predict a narrower profile than the asymptotic form. 

The third term in Eq. (\ref{eq:eta-tff-amplitude}) comes from the $\eta_c$ TFF  
\beq {\cal F}_{\eta_c \gamma\gamma^\ast}(Q^2) &=& e_c^2 \int_0^1 du \int b db \, e^{-S} S_t  f_{\eta_c} \non
&\cdot& \Big\{  2 \varphi(u) \hat{\Sigma}(u,b) K_0(\kappa_1^\prime b)
+ %\int_0^u d \alpha_1 \int_0^{\bar u} d \alpha_2 \frac{\hat{\Sigma}^\prime(\alpha_i, b)}{1-\alpha_1-\alpha_2} 
g_2(u)  \left[\frac{2 u b}{ \kappa_1^\prime} K_1(\kappa_1^\prime b) - \frac{2 {\bar u} b}{\kappa_2^\prime} K_1(\kappa_2^\prime b) \right] \non
&~& - g_1(u) \left[ \frac{2 b}{\kappa_1^\prime} K_1(\kappa_1^\prime b) + \frac{2b}{\kappa_2^\prime} K_1(\kappa_2^\prime b) \right] \non
&~& - g_1(u) \left[ \frac{b^2 m_c^2}{\kappa_1^{\prime 2}} K_2(\kappa_1^\prime b) 
+ \frac{b^2 m_c^2}{\kappa_2^{\prime 2}} K_2(\kappa_2^\prime b) \right]  \non
&~& + \tilde{g}_2(u) \left[ \frac{2 b}{\kappa_1^\prime} K_1(\kappa_1^\prime b) - \frac{2b}{\kappa_2^\prime} K_1(\kappa_2^\prime b) \right] \non 
&~& + \tilde{g}_2(u) \left[ \frac{b^2 m_c^2}{\kappa_1^{\prime 2}} K_2(\kappa_1^\prime b) - \frac{b^2 m_c^2}{\kappa_2^{\prime 2}} K_2(\kappa_2^\prime b) \right] \Big\}. 
\label{eq:etac-tff-amplitude}\eeq
The auxiliary function $\tilde{g}_2(u) = \int_0^u du^\prime \, g_2(u^\prime)/ \left( p_1 \cdot z_1 \right)$ is called in the analysis. 
The virtualities are $\kappa_1^\prime = \left[ u Q^2 + m_c^2 \right]^{1/2}$ and $\kappa_s^\prime = \left[ {\bar u} Q^2 + m_c^2 \right]^{1/2}$. 
In addition to the virtualities, charm quark mass give another power correction terms ${\cal O}(b/\kappa_{1,2})$ 
to the hard kernel associated the LCDAs $g_1$ and $\tilde{g}_2$, here $b$ is a dimensionless transversal parameter. 
Eq. (\ref{eq:etac-tff-amplitude}) show the result accompanying with the valence quark state.  
In the numerics, we only take into account the leading twist contribution, since the high twist LCDAs are not understood well so far. 

\begin{figure}[t] \begin{center} \vspace{-6mm}
\includegraphics[width=0.75\textwidth]{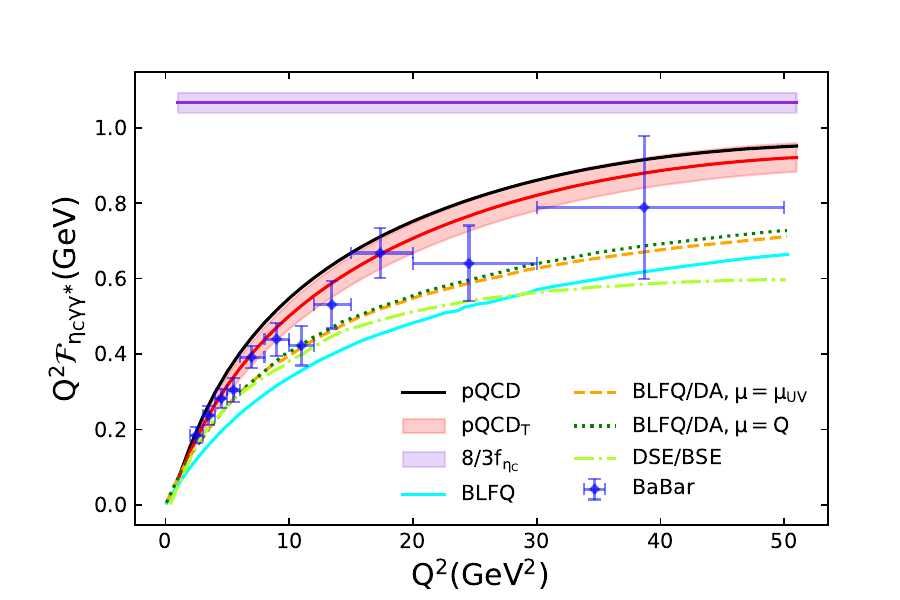}\end{center}
\vspace{-6mm}
\caption{The pQCD predictions of $Q^2 {\cal F}_{\eta_c \gamma \gamma^\ast}$ TFF in comparison to the measurement and the result obtained from other approaches. }
\label{fig:etac-tff}
\end{figure}

The transverse size parameter of $\eta_c$ was initially extracted from a phenomenological analysis of its two-photon decay width 
$\Gamma(\eta_c \to \gamma\gamma)$ in Ref. \cite{Kroll:2010bf}, yielding $\beta_{\eta_c}^2 = 0.19$ GeV$^{-2}$, 
wherein the perturbative QCD (pQCD) adopted the decay constant $f_{\eta_c} = 0.42$ GeV and an asymptotic distribution amplitude.
In our analysis, we adopt an updated value $f_{\eta_c} = 0.42$ GeV and, more importantly, incorporate non-asymptotic contributions to the leading-twist LCDAs. 
The transverse size parameter is then determined by fitting to the BaBar data of ${\cal F}_{\eta_c\gamma\gamma^\ast}(Q^2)/{\cal F}_{\eta_c\gamma\gamma^\ast}(0)$ 
in the intermediate momentum-transfer region $2 \leqslant Q^2 \leqslant 10$ GeV$^2$, where the data exhibit better statistical precision.
We obtain a larger value $\beta_{\eta_c}^2 = 0.60$ GeV$^{-2}$, along with ${\cal F}_{\eta_c \gamma \gamma^\ast}(0) = 0.092$ GeV$^{-1}$.
This result yields a mean-square transverse momentum $\langle {\bf k}_T^2 \rangle = 0.13 $ GeV$^2$. 
Notably, the conjugated mean-square transverse charge radius, $\langle b_{\eta_c}^2 \rangle = 0.058 $ fm$^{2}$, 
is nearly identity to the three-dimension charge radius $\langle r_{\eta_c}^2 \rangle$ obtained from the lattice QCD evaluation \cite{Delaney:2023fsc}. 
In figure \ref{fig:etac-tff}, we compare the updated pQCD prediction of $\eta_c$-photon TFF $Q^2{\cal F}_{\eta_c\gamma\gamma^\ast}(q^2)$ with 
other theoretical approaches, including the basis light-front quantization (BLFQ) with different scale choices \cite{Li:2021ejv}, 
the Dyson-Schwinger equation (DSE) and Bethe-Salpeter equation (BSE) \cite{Chen:2016bpj}. 
The BaBar data \cite{BaBar:2010siw} are also shown for comparison.
We note that our iTMD-improved pQCD prediction, calculated at leading order with leading-twist LCDAs, favors a larger value of 
${\cal F}_{\eta_c\gamma\gamma^\ast}(0)$. 
This explains why our curve lies slightly above those from BLFQ and BSE, 
where the form factors were normalized to ${\cal F}_{\eta_c\gamma\gamma^\ast}(0) = 0.068$ GeV$^{-1}$ 
from the $\Gamma(\eta_c \to \gamma\gamma)$ data. 
  
%-----------------------------------------------------------------------
%-----------------------------------------------------------------------

\begin{figure}[b] \begin{center} \vspace{6mm}
\includegraphics[width=1.0\textwidth]{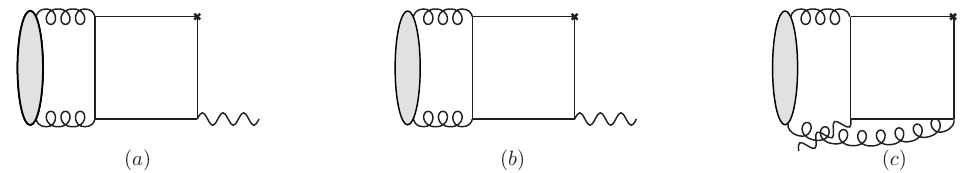}\end{center}
\vspace{-6mm}
\caption{Leading-order contribution from $\vert gg \rangle$ state to $\eta^{(\prime)}$ TFFs. }
\label{fig:tff-gg}
\end{figure}

We do not account for the $\vert gg \rangle$ component in $\eta^{(\prime)}$. 
While this component defines another leading-twist LCDA, the asymptotic term of the gluon LCDA vanishes due to charge conjugation symmetry. 
Its contribution arises solely from nonperturbative corrections proportional to Gegenbauer polynomials $C_n^{5/2}(2u-1)$. 
Furthermore, the $\vert gg \rangle$ state contributes to the TFFs at NLO with a hard coupling scale, as illustrated in figure \ref{fig:tff-gg}. 
Consequently, its contribution is formally of order ${\cal O}(\alpha_s)$. 
In summary, the valence gluon state's contribution to the $\eta^{(\prime)}$ TFF is doubly suppressed 
by the strong coupling constant and by the Gegenbauer expansion. 

%-----------------------------------------------------------------------
%-----------------------------------------------------------------------

%-----------------------------------------------------------------------
\begin{figure}[t] \begin{center} 
\includegraphics[width=0.45\textwidth]{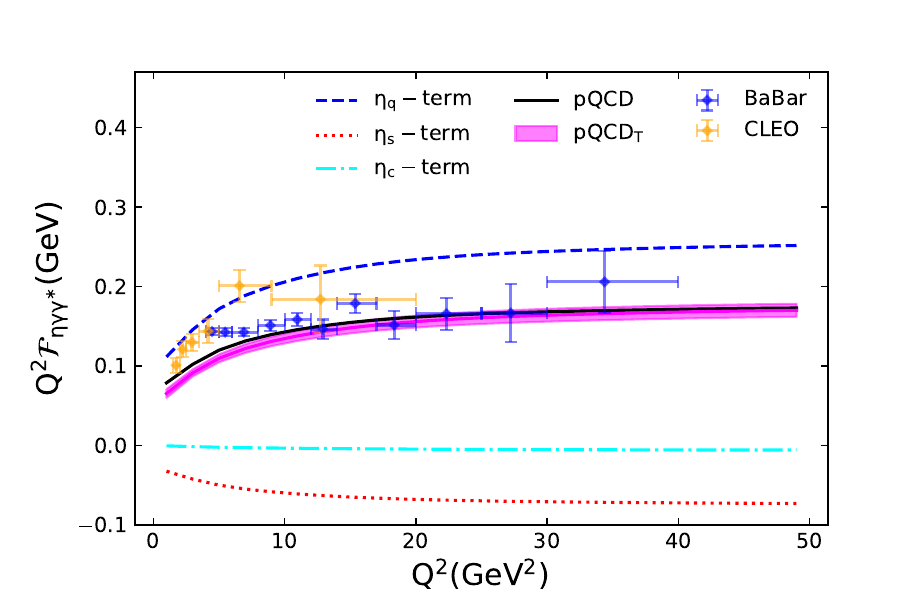} 
\includegraphics[width=0.45\textwidth]{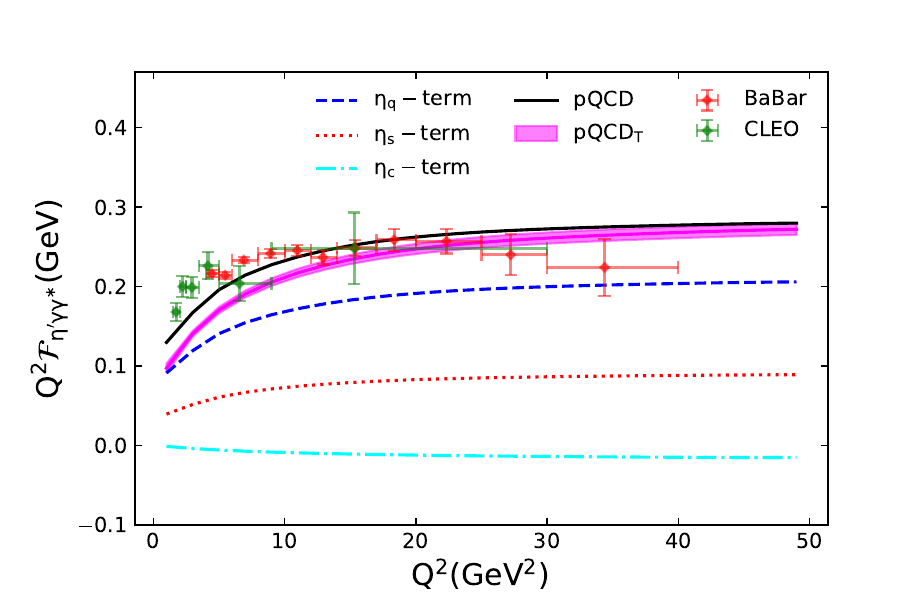} \non 
\includegraphics[width=0.45\textwidth]{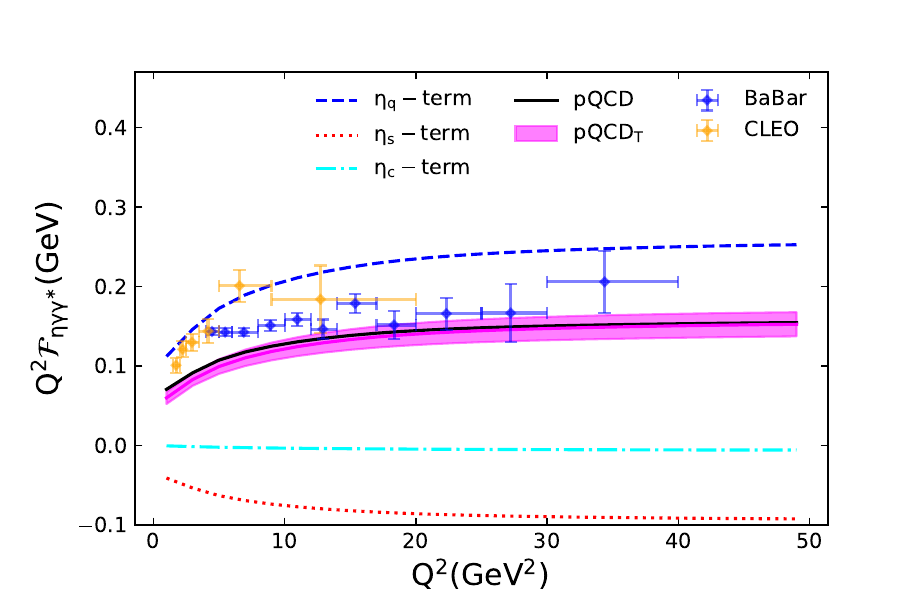} 
\includegraphics[width=0.45\textwidth]{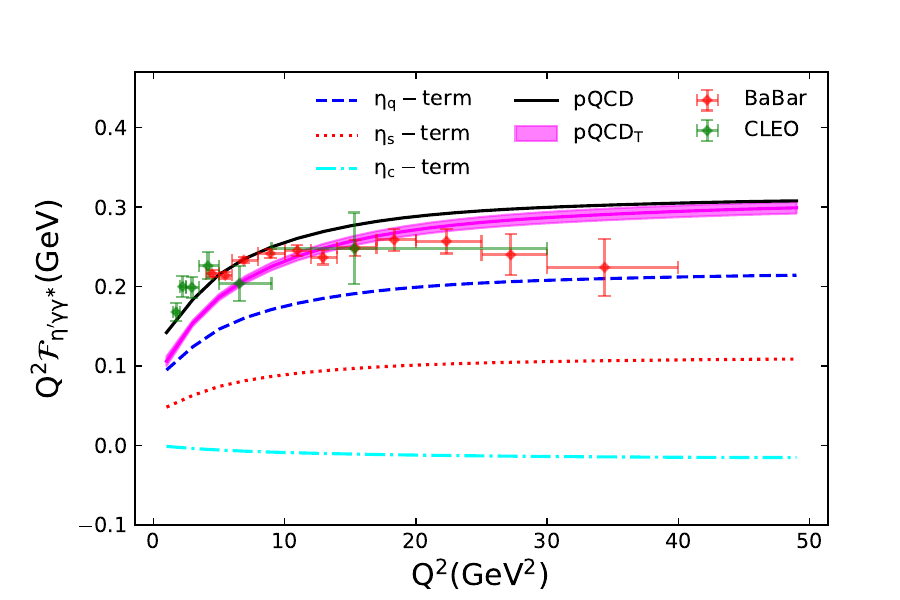}  \non 
\includegraphics[width=0.45\textwidth]{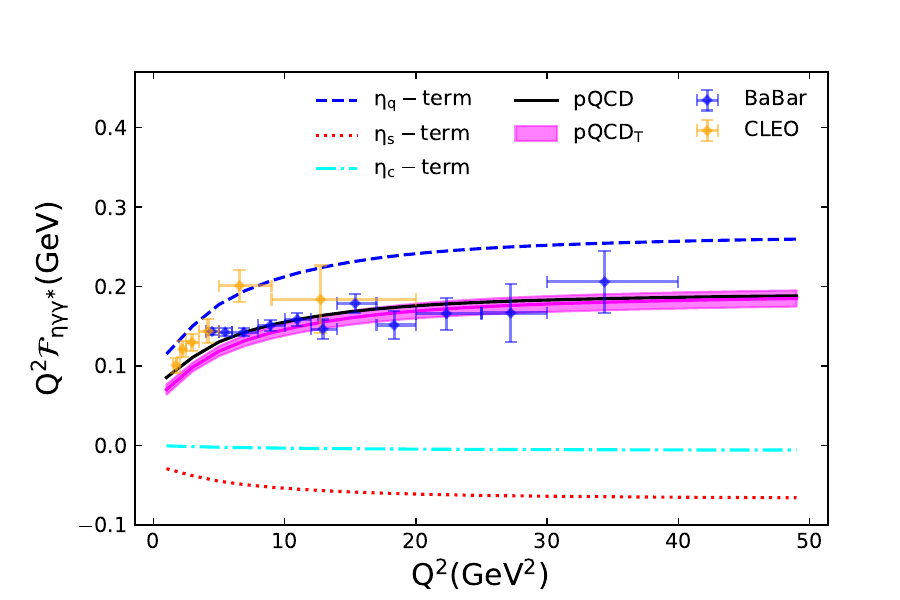} 
\includegraphics[width=0.45\textwidth]{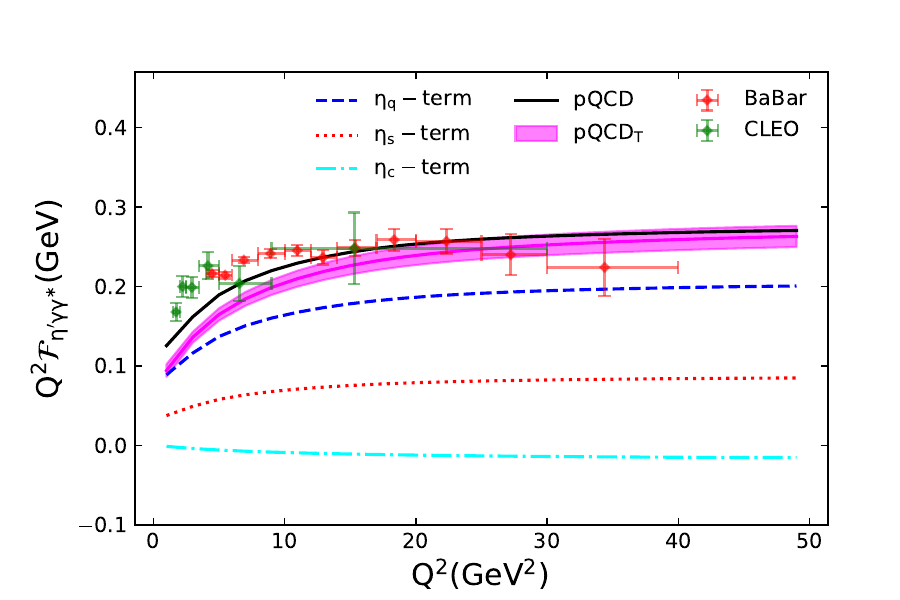}  \non 
\includegraphics[width=0.45\textwidth]{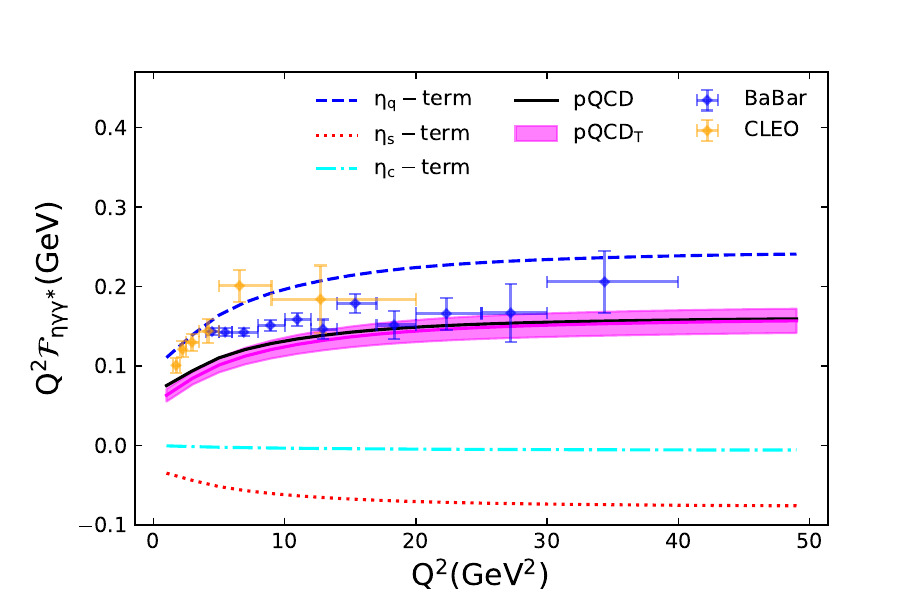} 
\includegraphics[width=0.45\textwidth]{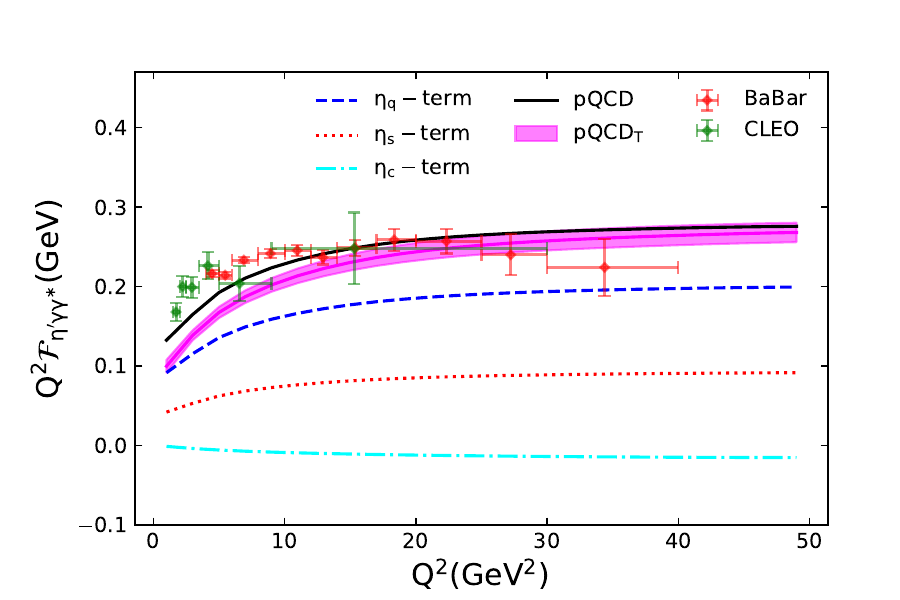}  \non 
\end{center}\vspace{-4mm}
\caption{The pQCD predictions of $\eta$ (left) and $\eta^\prime$ (right) TFFs in $Q^2 \leqslant 50$ GeV$^2$.}   
\label{fig:TFF-eta} \end{figure}
%-----------------------------------------------------------------------

We present the perturbative QCD (pQCD) predictions for the $\eta^{(\prime)}$ TFFs in figure \ref{fig:TFF-eta}. 
The individual contributions from the $\eta_q$, $\eta_s$ and $\eta_c$ components are shown as blue-dashed, red-dotted, 
and cyan-dash-dotted curves, respectively, while their sum is represented by the solid black curve. 
The iTMD-improved pQCD prediction is displayed as a magenta band. 
The TFFs are calculated using all four mixing schemes listed in tables \ref{tab:decaycons-eta} and \ref{tab:mass-moment-eta}. 
Our analysis reveals that
\begin{itemize} 
\item[(a)] The TFFs of both $\eta$ and $\eta^\prime$ mesons are predominantly determined by the $\eta_q$ component. 
However, the $\eta^\prime$ TFF also receives a significant contribution from the $\eta_s$ component.
\item[(b)] The $\eta_c$ component contribution is negligible in magnitude and therefore plays no significant role in explaining the experimental data.
\item[(c)] The iTMD-improved pQCD predictions of $\eta$ TFF favors the scenarios with smaller mixing angles (S1 and S3), 
characterized by larger decay constants and smaller masses of the flavor basis states. 
\end{itemize} 

In the perturbative QCD limit, the TFFs of all flavor basis states become identical 
${\cal F}_{\eta_q \gamma \gamma^\ast} = {\cal F}_{\eta_s \gamma \gamma^\ast} 
= {\cal F}_{\eta_c \gamma \gamma^\ast} = {\cal F}_{\pi \gamma \gamma^\ast}$. 
This leads to an asymptotic relation for the difference between $\eta$ and $\eta^\prime$ TFFs derived from Eqs. (\ref{eq:eta-tff-amplitude}), says, 
\beq \delta {\cal F} \equiv {\cal F}_{\eta \gamma \gamma^\ast} - {\cal F}_{\eta^\prime \gamma \gamma^\ast} \stackrel{Q^2 \to \infty }{\longrightarrow} 
\left( 0.071 \pm 0.032 \right) \sqrt{2} f_\pi = 0.013 \pm 0.006, \eeq
where the uncertainty primarily stems from the mixing angle $\phi = 39.6^\circ \pm 2.6^\circ$. 
This prediction can be compared with the BABAR measurement at $Q^2 = 112$ GeV$^2$ that 
$\delta {\cal F}(Q^2 = 112 \, {\rm GeV}^2) = 0.25^{+0.02}_{-0.02} - 0.23^{+0.03}_{-0.03} = 0.02 \pm 0.02$ \cite{BaBar:2011nrp}, 
though the experimental uncertainties should be noted.
Our analysis reveals that this observable exhibits significant sensitivity to the mixing angle. 
We therefore recommend precise measurements at ultraviolet momentum transfers $Q^2 \sim {\cal O}(10^2) \, {\rm GeV}^2$ 
to better constrain this fundamental parameter and elucidate the mixing mechanism.

\section{Summary}\label{sec:summary}

We present a systematic study of the EMFFs and TFFs of light pseudoscalar mesons ($\pi, K, \eta^{(\prime)}$) 
within the perturbative QCD approach based on $k_T$ factorization. 
Our predictions are derived at next-to-leading-order (NLO) QCD corrections, ${\cal O}(\alpha_s^2)$, 
for the leading and sub-leading twist LCDAs, and at leading order (${\cal O}(\alpha_s)$) for the twist four LCDAs. 
To account for the soft transverse momentum degrees of freedom in the parton distribution of hadrons, 
we introduce the iTMDs, which complement the LCDAs that describe the longitudinal momentum distribution. 
This is motivated by the fact that the initial and final hadrons are formed outside the hard interaction region, 
where the sudakov suppression of small transverse momenta is weak. 
Consequently, the partons inside the hadrons exhibit transverse oscillations orthogonal to the light-cone direction.

Due to chiral enhancement, the EMFFs of the pion and kaon are dominated by the twist-three LCDAs 
associated with the valence quark state, which are proportional to the chiral mass. 
Using the double-photon radiation relation and current constraints on Gegenbauer coefficients, 
we determine the transverse-size parameter of the pion to be $\beta_\pi^2 = 0.51 \pm 0.04$ GeV$^{-2}$.
By combining our iTMDs-improved pQCD calculation of the spacelike pion EMFF 
with the results from the modular dispersion relation which incorporates timelike experimental data and the pQCD high-energy tail, 
we extract the pion chiral mass, obtaining $m_0^\pi(1 \, {\rm GeV}) = 1.84 \pm 0.07$ GeV. 
This value is approximately $30\%$ larger than the previous pQCD estimates but agrees with ChPT predictions. 
The discrepancy suggests a significant suppression of the form factor due to soft transverse dynamics, 
particularly in the low-to-intermediate momentum transfer regions.

For the kaon EMFF, the abundance of timelike-region data at large momentum transfers, where the pQCD is reliably applicable, 
allows us to directly fit the iTMDs-improved pQCD calculation to measurements. 
Adopting the well-established chiral relation for the kaon chiral mass, $m_0^K(1 \, {\rm GeV}) = 1.90$ GeV, 
we extract the transverse-size parameter as $\beta_K^2 = 0.30 \pm 0.05$ GeV$^{-2}$. 
This value is notably smaller than that of the pion, indicating significant $SU(3)$ flavor symmetry breaking 
and reflecting stronger attractive dynamics between the strange and light quark system.
Our improved pQCD predictions for the spacelike form factors show good consistency with 
next-to-next-to-leading-order (NNLO) collinear factorization calculations, though both QCD-based results remain smaller than lattice QCD evaluations. 
We attribute this discrepancy primarily to the influence of the strange quark in the kaon LCDAs, 
which suppresses the pQCD predictions by approximately $30\%$.

The meson-photon TFFs demonstrate significant sensitivity to iTMDs in the low-to-intermediate momentum transfer regime. 
Our improved pQCD framework shows good agreement with Belle collaboration data for the pion-photon TFF, 
particularly reproducing the observed plateau in $Q^2 F_{\pi\gamma\gamma^\ast}$ at large $Q^2$. 
As this observable directly probes the leading-twist LCDAs, 
we expect forthcoming Belle-II measurements to provide crucial constraints on the pion's Gegenbauer coefficients, 
where current lattice QCD and QCD-based analyses show notable discrepancies.
For the $\eta$, $\eta^\prime$ TFFs, we implement a comprehensive pQCD analysis within the $\eta_q$-$\eta_s$-$\eta_c$ flavor mixing framework, 
testing four distinct parameter sets. 
The results reveal two key features. First, the charmonium ($\eta_c$) contribution remains negligible across all momentum transfers. 
Second, the $\eta_q$ component dominates both in $\eta$ and $\eta^\prime$ TFFs. 
The analysis strongly favors a scenario with relatively small mixing angles, larger decay constants, 
and reduced masses for the $\eta_q$ and $\eta_s$ flavor states. 
Of particular phenomenological importance is the predicted behavior of 
${\cal F}_{\eta\gamma\gamma^\ast}$ and ${\cal F}_{\eta^\prime\gamma\gamma^\ast}$ at ultraviolet scales, 
where their differential response provides exceptional sensitivity to the mixing angle, 
making this an ideal observable for future precision measurements.

\section{ACKNOWLEDGEMENTS}
We are grateful to Vladimir Braun, Heng-tong Ding, Jun Hua, Guang-shun Huang, Hsiang-nan Li, Wei Wang and Yu-ming Wang 
for their invaluable discussions and insights. We particularly thank Gunnar Bali for his careful reading and insightful corrections 
concerning the treatment of lattice results presented in table \ref{tab:mass-moment-eta}. 
This work is supported by the National Key R$\&$D Program of China under Contracts No. 2023YFA1606000 
and the National Science Foundation of China (NSFC) under Grant No. 11975112. 
J. C is also supported by the Launching Funding of Henan University of Technology (No.31401697). 

\begin{appendix}

\section{Definition of the light-cone distribution amplitudes}\label{app:LCDAs-definition}

Light-cone distribution amplitudes (LCDAs) are probability amplitudes 
to find the hadron in a state with certain number of fock constituents at small transversal separation. 
They are usually studied by applying the conformal symmetry in QCD \cite{Braun:2003rp}. 
The underlying idea is somewhat like the partial-wave expansion in quantum mechanism 
where the angular degree of freedom (dof) is detached from the radial one for a spherically symmetric potential. 
For the LCDAs, the transversal and longitudinal dofs are separated after considering the invariance of massless QCD under the conformal transformation. 
The transversal momentum dependence is governed by the renormalization group equations which show a scale dependence on the ultraviolet cutoff $\mu$. 
The dependence on the longitudinal momentum fractions is governed in terms of orthogonal Jacobi polynomials which deduces to the Gegenbauer polynomials representing the so-called collinear subgroup of the conformal group \cite{Ball:1998sk}.
In addition, the distributions with different conformal spins ($j=(l+s/2$ with $l$ being the canonical dimension and $s$ being the lorentz spin projection) 
have independently behaviors \cite{Balitsky:1987bk}. 

LCDAs of pseudoscalar meson ($\mathcal{P}= \pi$ and $K$) with the valence quark state is defined via the nonlocal matrix element \cite{BallWN,BallJE}.
\beq &~& \big\langle 0 \big\vert \overline{u}(z_2) (\gamma_\rho \gamma_5) q(z_1) \big\vert \mathcal{P}^-(p) \big\rangle 
= f_\mathcal{P} \int_0^1 du \, e^{-i upz_1 - i\overline{u}pz_2} \Big\{ i p_\rho \big[ \varphi(u,\mu) \non 
&~& \hspace{1.6cm} + (z_1-z_2)^2 g_{1}(u,\mu) \big] + \left[ (z_1-z_2)_\rho - \frac{p_\rho (z_1-z_2)^2}{p(z_1-z_2)} \right] g_{2}(u,\mu) \Big\}, 
\label{eq:DA-lt} \\ 
&&\big\langle 0 \big\vert \overline{u}(z_2) (\sigma_{\tau\tau'} \gamma_5) q(z_1) \big\vert \mathcal{P}^-(p) \big\rangle 
= f_\mathcal{P} m_0^\mathcal{P} \int_0^1 dx \,  e^{-i upz_1 - i\overline{u}pz_2} \left(1 - \frac{m_\mathcal{P}^2}{(m^0_\mathcal{P})^2}\right) \non
&~& \hspace{1.6cm} \cdot \left[ p^\tau (z_1-z_2)_{\tau'} - p^{\tau'}(z_1-z_2)_\tau \right] \, \varphi^\sigma(u,\mu), \label{eq:DAs-sigma} \\
&&\big\langle 0 \big\vert \overline{u}(z_2) (i \gamma_5) q(z_1) \big\vert \mathcal{P}^-(p) \big\rangle 
= f_\mathcal{P} m_0^\mathcal{P} \int_0^1 du \,  e^{-i upz_1 - i\overline{u}pz_2} \varphi^p(u,\mu). \label{eq:DAs-p} \eeq
Here $\varphi$, $\varphi^{P, \sigma}$ and $g_{1,2}$ corresponds to the leading twist, twist three and twist four LCDAs, respectively. 
The twist is defined by the minus between canonical dimension and spin projection $t=l-s$. 
$f_\mathcal{P}$ is the decay constant and $m_0^\mathcal{P} \equiv \frac{m_\mathcal{P}^2}{m_u+m_q}$ is the chiral mass. 

For the high fock state with ${\bar q}qg$ assignment, the LCDAs are defined via the matrix elements 
with the gluon field strength tensor operator $G_{\kappa\kappa'}=g_s G_{\kappa\kappa'}^a \lambda^a/2$,
\beq &~&p^+ \big\langle 0 \big\vert \overline{u}(z_2) (\sigma_{\tau\tau'} \gamma_5) G_{\kappa\kappa'}(z_0) q(z_1) \big\vert \mathcal{P}^-(p) \big\rangle 
= i f_{3\mathcal{P}} \int \mathcal{D}x_i \, e^{-i \alpha_1pz_1 - i\alpha_2pz_2 -i\alpha_3z_0} \non
&& \cdot \Big[ \left(p_{\kappa}p_{\tau} g_{\kappa'\tau'} - p_{\kappa'} p_{\tau} g_{\kappa\tau'} \right) 
- \left( p_{\kappa}p_{\tau'} g_{\kappa'\tau} - p_{\kappa'} p_{\tau'} g_{\kappa\tau}\right) \Big] \varphi_{3}(\alpha_i,\mu),  \label{eq:DAs-3p-t4-1} \\
&~&p^+ \big\langle 0 \big\vert \overline{u}(z_2) (\gamma_\rho \gamma_5) G_{\kappa\kappa'}(z_0) q(z_1) \big\vert \mathcal{P}^-(p) \big\rangle 
= f_{\mathcal{P}} \int \mathcal{D}x_i \, e^{-i \alpha_1pz_1 - i\alpha_2pz_2 -i\alpha_3z_0} \non
&~& \cdot \Big[ p_\rho \frac{p_{\kappa}(z_1-z_2)_{\kappa'} - p_{\kappa'}(z_1-z_2)_{\kappa}}{p(z_1-z_2)} \varphi_\parallel(\alpha_i,\mu) 
+ (g^{\perp}_{\rho\kappa}p_{\kappa'} - g^{\perp}_{\rho\kappa'} p_\kappa) \varphi_\perp(\alpha_i,\mu) \Big], \label{eq:DAs-3p-t4-2} \\
&~&p^+ \big\langle 0 \big\vert \overline{u}(z_2) (\gamma_\rho ) \tilde{G}_{\kappa\kappa'}(z_0) q(z_1) \big\vert \mathcal{P}^-(p) \big\rangle 
= f_{\mathcal{P}} \int \mathcal{D}x_i \, e^{-i \alpha_1pz_1 - i\alpha_2pz_2 -i\alpha_3z_0} \non
&~& \cdot \Big[ p_\rho \frac{p_{\kappa}(z_1-z_2)_{\kappa'} - p_{\kappa'}(z_1-z_2)_{\kappa}}{p(z_1-z_2)} \tilde{\varphi}_\parallel(\alpha_i,\mu) 
+ (g^{\perp}_{\rho\kappa}p_{\kappa'} - g^{\perp}_{\rho\kappa'} p_\kappa) \tilde{\varphi}_\perp(\alpha_i,\mu) \Big]. 
\label{eq:DAs-3p} \eeq
Here $\varphi_{3\mathcal{P}}$ is the twist-3 DA, and $\varphi_{\parallel,\perp}, \tilde{\varphi}_{\parallel,\perp}$ are twist-4 DAs, 
$\tilde{G}_{\kappa\kappa'}=1/2 \, \epsilon_{\kappa\kappa'\eta\eta'} G^{\eta\eta'}$ is the dual gluon-filed-strength tensor 
at the location $z_0=vz_1+\overline{v}z_2$ with the free variable $v \in [0,1]$. 
The integration measure is 
\beq \int {\cal D}\underline{\alpha} = \int_0^1 d\alpha_1 d\alpha_2 d \alpha_3 \delta(1-\alpha_1-\alpha_2-\alpha_3). \label{eq:int-measure} \eeq

\section{Expressions of the light-cone distribution amplitudes}\label{app:LCDAs-expression}

For the pseudoscalar meson without polarisation, the leading twist LCDA is written in terms of Gegenbauer polynomials as 
\beq \varphi(u, \mu) = 6u\bar{u} \sum_{n=0} \, a_n(\mu)  \, C_n^{3/2}(2u-1). \label{eq:DA-t2} \eeq
Two-particle twist-3 DAs are related to the three-particle DA $\varphi_{3}(\alpha_i)$ by the QCD equation of motion (EOM). 
The EOM relations contain the quark mass terms which are subsequently written by means of two 
dimensionless parameters $\rho_+^\mathcal{P} = (m_q+m_u)/m_0^\mathcal{P}$ and $\rho_-^\mathcal{P} = (m_q-m_u)/m_0^\mathcal{P}$.
We take into account the strange quark mass In the calculation, 
and neglect the $u, d$ quark masses unless in the chiral masses $m_0^\mathcal{P}$.
To the next-to-leading-order definition of conformal spin and to the second moments in the truncated conformal expansion, we get
\beq \varphi^p(u, \mu) &=& 1 + 3 \rho_+^\mathcal{P} \Big( 1+6a_2\Big) - 9 \rho_-^\mathcal{P} a_1
+ \left[ \frac{27}{2} \rho^\mathcal{P}_+ a_1 - \rho^\mathcal{P}_- \left( \frac{3}{2} + 27 a_2 \right) \right] C_1^{1/2}(2u-1) \non
&+&  \left[ 30 \eta_{3} +15 \rho_+^\mathcal{P} a_2 - 3 \rho_-^\mathcal{P} a_1 \right] C_2^{1/2}(2u-1)
+ \left[10 \eta_{3} \lambda_{3} - \frac{9}{2}\rho_-^\mathcal{P} a_2 \right] C_3^{1/2}(2u-1) \non
&-& 3 \eta_{3} \omega_{3} \, C_4^{1/2}(2u-1) + \frac{3}{2}(\rho_+^\mathcal{P}+\rho_-^\mathcal{P}) \left( 1 - 3 a_1 + 6 a_2 \right) \ln u \non 
&+& \frac{3}{2}(\rho_+^\mathcal{P} - \rho_-^\mathcal{P}) \left( 1 + 3 a_1 + 6 a_2 \right) \ln {\bar u},
\label{eq:DA-t3-p} \\
\varphi^\sigma(u, \mu) &=& 6u{\bar u} \Big\{ 1 + \frac{5}{2} \rho_+^\mathcal{P} + 15 \rho_+^{\cal P} a_2 - \frac{15}{2} \rho_-^{\cal P} a_1 
+ \left[ 3 \rho_+^\mathcal{P} a_1 - \frac{15}{2} \rho_-^{\cal P} a_2 \right] C_1^{3/2}(2u-1) \non
&+& \left[ 5 \eta_{3} - \frac{1}{2} \eta_3 \omega_3 + \frac{3}{2} \rho_+^\mathcal{P} a_2 \right] C_2^{3/2}(2u-1)
+ \eta_{3} \lambda_{3} C_3^{3/2}(2u-1) \non
&+& \frac{3}{2} \left( \rho_+^\mathcal{P} + \rho_-^{\cal P} \right) \left( 1- 3a_1 + 6 a_2 \right)  \ln u + 
\frac{3}{2} \left( \rho_+^\mathcal{P} - \rho_-^{\cal P} \right) \left( 1+ 3a_1 + 6 a_2 \right)  \ln {\bar u} \Big\} \,,
\label{eq:DA-t3-t} \\
\varphi_{3}(\alpha_i, \mu) &=& 360 \alpha_1 \alpha_2 \alpha_3^2 \Big\{ 1 + \lambda_{3} (\alpha_1-\alpha_2) + \omega_{3} \frac{1}{2} (7\alpha_3-3) \Big\}.
\label{eq:DA-t3-3p} \eeq
The asymptotic term, the terms from EOM and the ${\bar q}qg$ operators are clearly separated in the above expressions. 
The three-particle parameters $f_{3 {\cal P}}, \lambda_{3}, \omega_{3}$ can be defined by the matrix element of local twist-3 operators,
and their evolution have the mixing terms with the quark mass \cite{BallWN}.

For the two-particle twist-4 DAs, the gauge invariant and lorentz invariant definition in Eq. (\ref{eq:DA-lt}) 
is more convenient to be used in the QCD calculation. 
They are related to the invariant amplitudes $\psi_{4\mathcal{P}}, \phi_{4\mathcal{P}}$ by 
\beq g_{2}(u, \mu) = -\frac{1}{2}\int_0^u du' \psi_{4}(u'), \quad
g_{1}(u, \mu) = \frac{1}{16} \phi_{4}(u) + \int_0^u du' g_{2}(u'). \label{eq:DA-t4-2p} \eeq
The relations between different operators by EOM indicate that the lorentz invariant amplitudes 
are written in terms of the "genuine" twist-4 contribution from the three-particle DAs $\varphi_\parallel(\alpha_i), \varphi_{\perp}(\alpha_i)$
and the Wandzura-Wilczek-type mass corrections from the two-particle lower twist DAs, 
distinguishing by parameters $\delta_{\mathcal{P}}^2$ and $m_\mathcal{P}^2$, respectively \cite{KhodjamirianYS}.
\beq \psi_{4}(u, \mu) &=& \delta_{\mathcal{P}}^2 \Big[ \frac{20}{3} \, C_2^{1/2}(2u-1) + \frac{49}{2} a_1 \, C_3^{1/2}(2u-1) \Big] \non
&+& m_\mathcal{P}^2 \Big\{6 \rho^\mathcal{P} \Big( 1-3 a_1 +6a_2\Big) \,C_0^{1/2}(2u-1) \non
&-& \Big[\frac{18}{5}a_1 + 3\rho^\mathcal{P}\Big(1-9a_1+18a_2\Big) + 12\kappa_{4}\Big] \, C_1^{1/2}(2u-1) \non
&+&\Big[ 2 - 6 \rho^\mathcal{P} \Big( a_1 - 5 a_2 \Big) + 60 \eta_{3} \Big] \, C_2^{1/2}(2u-1) \non
&+&\Big(\frac{18}{5}a_1 - 9 \rho^\mathcal{P} a_2 + \frac{16}{3}\kappa_{4} + 20 \eta_{3} \lambda_{3} \Big) \, C_3^{1/2}(2u-1) \non
&+&\Big(\frac{9}{4} a_2 - 6 \eta_{3} \omega_{3} \Big) \,C_4^{1/2}(2u-1) \Big\} +6m_q^2 \Big(1-3a_1+6a_2\Big) \, \ln u,
\label{eq:DA-t4-t2-psi} \\ 
\phi_{4}(u, \mu) &=& \delta^2 \Big\{ \Big( \frac{200}{3} + 196(2x-1)a_1 \Big) u^2\bar{u}^2 \non
&+& 21 \omega_{4} \Big( u\bar{u} (2+13u\bar{u}) + [2u^3(6u^2-15u+10)\ln u] + [u \leftrightarrow \bar{u}] \Big) \non
&-& 14 a_1 \Big(u\bar{u}(2u-1)(2-3u\bar{u}) - [2u^3(u-2) \ln u] + [u \leftrightarrow \bar{u}] \Big)\Big\} \non
&+&m_\mathcal{P}^2 \Big\{
\frac{16}{3}\kappa_{4} \Big( u (2u-\bar{u}) (1-2u\bar{u}) + [5(u-2)u^3 \ln u] - [u \leftrightarrow \bar{u}] \Big) \non
&+& 4\eta_{3} u\bar{u} \Big[60\bar{u} + 10 \lambda_{3} \Big((2u-1)(1-u\bar{u}) - (1-5u\bar{u}) \Big) \non
&-& \omega_{3} \Big( 3-21u\bar{u}+28u^2\bar{u}^2+3(2u-1)(1-7u\bar{u})\Big) \Big] \non
&-& \frac{36}{5}a_2 \Big(\frac{1}{4} u\bar{u}(4-9u\bar{u}+110u^2\bar{u}^2) + [u^3(10-15u+6u^2) \ln u] + [u \leftrightarrow \bar{u}] \Big) \non
&+& 4 u\bar{u} (1+3u\bar{u}) \Big( 1 + \frac{9}{5}(2u-1)a_1 \Big)  \Big\}.
\label{eq:DA-t4-t2-phi} \eeq
Here $\eta_{3}=f_{3\mathcal{P}}/(f_\mathcal{P}m_0^\mathcal{P})$. 
It is noticed in Eq. (\ref{eq:DA-t4-t2-psi}) that $\psi_{4}(u)$ has a logarithm end-point singularity for the finite quark mass,
while this singularity is not existed in $\phi_{4}(u)$. Three-particle twist-4 DAs read as 
\beq \psi_\parallel(\alpha_i, \mu) &=& 120\alpha_1\alpha_2\alpha_3 \Big\{
\delta^2 \Big[ \frac{21}{8}(\alpha_1-\alpha_2)\omega_{4} + \frac{7}{20}a_1(1-3\alpha_3) \Big] \non
\hspace{1cm} &+& m^2 \Big[-\frac{9}{20}(\alpha_1-\alpha_2)a_2 + \frac{1}{3}\kappa_{4} \Big] \Big\},
\label{eq:DA-t2-3p-A-para} \\
\psi_\perp(\alpha_i, \mu) &=& 30\alpha^2_3 \Big\{ \delta^2 \Big[\frac{1}{3}(\alpha_1-\alpha_2)
+ \frac{7}{10}a_1 \Big( -\alpha_3(1-\alpha_3) + 3(\alpha_1-\alpha_2)^2\Big)\non
\hspace{1cm} &+& \frac{21}{4} \omega_{4}(\alpha_1-\alpha_2)(1-2\alpha_3) \Big] 
+ m^2 (1-\alpha_3) \Big[ \frac{9}{40}(\alpha_1-\alpha_2) - \frac{1}{3}\kappa_{4} \Big] \Big\}\,,
\label{eq:DA-t2-3p-A-perp} \\
\tilde{\psi}_\parallel(\alpha_i, \mu) &=& -120\alpha_1\alpha_2\alpha_3 \delta^2 \Big\{
\frac{1}{3} + \frac{7}{4}a_1(\alpha_1-\alpha_2) + \frac{21}{8}\omega_{4}(1-3\alpha_3) \Big\},
\label{eq:DA-t2-3p-V-perp} \\
\tilde{\psi}_\perp(\alpha_i, \mu) &=& 30\alpha_3^2 \Big\{\delta^2 \Big[ \frac{1}{3}(1-\alpha_3) - \frac{7}{10}a_1(\alpha_1-\alpha_2)(4\alpha_3-3) +
\frac{21}{4}\omega_{4}(1-\alpha_3)(1-2\alpha_3) \Big] \non
\hspace{1cm} &+& m^2 \Big[ \frac{9}{40}a_2 (\alpha_1^2- 4\alpha_1\alpha_2+\alpha_2^2) - \frac{1}{3}(\alpha_1-\alpha_2)\kappa_{4} \Big] \Big\}.
\label{eq:DA-t2-3p-V-perp} \eeq
Three nonperturbative parameters $\delta^2, \omega_{4}$ are introduced. 
We mark that all parameters in the conformal expansion of DAs have the scale dependence
and the behaviours of their evolutions can be found in Ref. \cite{BallWN}.

\section{Derivation of the modulus squared dispersion relation}\label{app:dr}

In order to derive the modulus squared dispersion relation shown in Eq. (\ref{eq:Fpi_DR2}), we firstly introduce an auxiliary $g_\pi$ function
\beq g_\pi (q^2) \equiv \frac{\ln F_\pi (q^2)}{q^2 \sqrt{s_0 - q^2}}. \label{eq:g-function} \eeq
The only assumption in our derivation is that the form factor $F_\pi(q^2)$ is free of zeros in the complex $q^2$ plane, 
then the $g_\pi$ function has no additional singularities in the $q^2$ plane, apart from the region $q^2=s>s_0$ on the real axis.
The normalization condition $F_\pi(0) = 1$ indicates that $g_\pi(0)$ is finite. 
What's more, the power asymptotics of pion form factor $F_\pi(q^2) \sim 1/q^2$ implies 
$g_\pi(q^2) \sim 1/(q^2)^\alpha$ at $\vert q^2 \vert \to \infty$ with the parameter $\alpha > 1$, 
which enables a dispersion relation for the $g_\pi$ function 
\beq g_\pi (q^2) = \frac{1}{\pi} \int\limits_{s_0}^\infty d s \frac{{\rm Im} \, g_\pi (s)}{s - q^2- i \epsilon}. \label{eq:DR-gpi} \eeq
The derivation of Eq. (\ref{eq:DR-gpi}) is in the same way as for the standard dispersion relation of pion form factor shown in Eq. (\ref{eq:Fpi_DR1}).

At $s>s_0$ on the real axis, the imaginary part of $g_\pi$ reads as 
\beq &~&{\rm Im} \, g_\pi(s) = {\rm Im}  \left[\frac{\ln (|F_\pi (s)| e^{i \delta_\pi (s)})}{- i s \sqrt{s - s_0} } \right]  
=  \frac{\ln |F_\pi (s)|}{s \sqrt{s -s_0}}, \label{eq:img-gpi} \eeq
where the pion form factor is written by $F_\pi (s) = |F_{\pi} (s)| e^{i \delta_\pi (s)}$ 
and the branch point is chosen at $\sqrt{s_0 - (s+i\epsilon)} \stackrel{\epsilon \to 0}{\longrightarrow} - i \sqrt{s - s_0}$. 
We note that the other branch point $+i \sqrt{s - s_0}$ would lead to an unphysical divergence of the pion form factor, 
saying $F_\pi(q^2) = \infty$ at $q^2 \to -\infty$.

Substituting Eq. (\ref{eq:g-function}) and Eq. (\ref{eq:img-gpi}) into the dispersion relation Eq. (\ref{eq:DR-gpi}), we get 
\beq \frac{\ln F_\pi(q^2)}{q^2 \sqrt{s_0 - q^2}} = \frac{1}{2 \pi} \int\limits_{s_0}^\infty \frac{d s \, \ln |F_\pi (s)|^2}{s\,\sqrt{s -s_0}  \, (s -q^2)}, 
\qquad q^2 < s_0.  \label{eq:DR-Fpi-log} \eeq
Taking exponent to the both sides, we arrive at the modulus squared dispersion relation 
\beq F_\pi (q^2) = \exp \left[ \frac{q^2 \sqrt{s_0 - q^2}}{2 \pi} 
\int\limits_{s_0}^\infty \frac{d s \, \ln |F_\pi (s)|^2}{s\,\sqrt{s - s_0}  \, (s -q^2)} \right].  \label{eq:DR-Fpi-2} \eeq

\end{appendix} 

%%---------------------------------------------------------------------------------------

\end{document}